\RequirePackage{snapshot}
\documentclass[usenatbib,fleqn]{mnras}

\makeatletter
\newlength{\abovecaptionskip}%
\makeatother

\usepackage{mathtools}
\usepackage{threeparttable}
\usepackage{makecell}
\usepackage{paralist}
\usepackage{verbatim}

\usepackage{amsmath,amssymb}
\usepackage{cases}
\usepackage{mathrsfs}
\usepackage{graphicx}
\usepackage{epstopdf}
\usepackage{hyperref}
\graphicspath{{./figures/}}
\usepackage{url}
\usepackage{aas_macros}

\newcommand\lsim{\mathrel{\rlap{\lower4pt\hbox{\hskip1pt$\sim$}}
    \raise1pt\hbox{$<$}}}
\newcommand\gsim{\mathrel{\rlap{\lower4pt\hbox{\hskip1pt$\sim$}}
    \raise1pt\hbox{$>$}}}
\newcommand{\Mbh}[1][]{M_{\bullet#1}}
\newcommand{\Msun}{{\rm M_\odot}}

\title[An Overabundance of BH-XRBs in the GC from Tidal Captures]{An Overabundance of Black Hole X-Ray Binaries in the Galactic Center from Tidal Captures}  
\author[Generozov et al.]{A. Generozov\thanks{ag@astro.columbia.edu}, N.~C.~Stone$^{1}$, B.~D.~Metzger, J.~P.~Ostriker\\
  Columbia Astrophysics Laboratory, Pupin Hall, Columbia University, 550 W.~120th Street, New York, NY 10027, USA\\
  $^{1}$Einstein Fellow
  }

\begin{document} \maketitle \begin{abstract} A large population of X-ray
binaries (XRBs) was recently discovered within the central parsec of the
Galaxy by Hailey et al.  While the presence of compact objects on this scale
due to radial mass segregation is, in itself, unsurprising, the fraction of
{\it binaries} would naively be expected to be small because of how easily
primordial binaries are dissociated in the dynamically hot environment of the
nuclear star cluster (NSC).  We  propose that the formation of XRBs in the
central parsec is dominated by the {\it tidal capture} of stars by black holes
(BHs) and neutron stars (NSs).  We model the time-dependent radial density
profiles of stars and compact objects in the NSC with a Fokker-Planck
approach, using the present-day stellar population and rate of {\it in situ}
massive star (and thus compact object) formation as observational constraints.
Of the $\sim 1-4\times 10^4$ BHs that accumulate in the central parsec over
the age of the Galaxy, we predict that $\sim 60 - 200$ currently exist as
BH-XRBs formed from tidal capture, consistent with the population seen by
Hailey et al.  A somewhat lower number of tidal capture NS-XRBs is also
predicted.  We also use our observationally calibrated models for the NSC to
predict rates of other exotic dynamical processes, such as the tidal
disruption of stars by the central supermassive black hole ($\sim 10^{-4}$ per
year at z=0). \end{abstract}


\section{Introduction} 
A large fraction of low- to moderate-mass galaxies host nuclear star clusters
(NSCs). The large mean stellar densities in these clusters, typically $\sim
10^2-10^6 \Msun$ pc$^{-3}$ (e.g.~\citealt{georgiev+2014}), result in
correspondingly high rates of collisional interactions \citep{leigh+2016}.  Stellar-mass compact
objects, particularly black holes (BHs) and neutron stars (NSs), play an
important role in these environments; for example, they form sources of {\it
LIGO} and {\it LISA}-band gravitational waves
(e.g.~\citealt{quinlan&shapiro1987, oleary+2009,tsang2013,baror&alexander2016,
antonini&rasio2016,stone+2017, bartos+2017}), serve as probes of the
relativistic spacetime near the central supermassive BH (SMBH;
\citealt{paczynski&trimble1979, pfahl&loeb2004}), and potentially contribute
to the $\gamma$-ray excess observed in our own Galactic Center (GC;
\citealt{brandt+2015}).  Compact objects in NSCs will also induce strong tidal
interactions during close flybys with stars.  A sequence of weak tidal
encounters will stochastically spin up GC stars \citep{alexander&kumar2001,
sazonov+2012}, while a single very strong tidal encounter may disrupt the
victim star and produce a luminous transient \citep{perets+2016}, but a tidal
encounter of intermediate strength will bind the star to the compact object in
a ``tidal capture'' \citep{fabian+1975}, as is the focus of this paper.

There is strong evidence of a population of NSs and stellar-mass BHs in the
Milky Way (MW) GC.  The hundreds of O/B stars currently located in the central
parsec indicate a high rate of {\it in situ} NS/BH formation in this region
(e.g.~\citealt{levin&beloborodov03,genzel+03}).  The discovery of even a
single magnetar within $\lesssim 0.1$ pc of Sgr A* \citep{mori+2013}, given
their short active lifetimes, also demands a high current rate of NS
formation.  The X-ray point sources in the GC also directly indicate a
population of {\it binaries} containing compact objects. There are a total of
six known X-ray transients in the central parsec
\citep{muno+2005,hailey&mori2017}. Of these six, three are strong BH X-ray
binary (BH-XRB) candidates based on their spectral and timing properties. For
example, one of these has a radio jet \citep{bower+2005}, while the other two
show a broadened Fe-K$\alpha$ line and a 13 mHz QPO (see Mori et al. in prep
for details and other arguments). The identity of the remaining transients is
unknown, but they may be NS-XRBs. In addition to these transient sources,
\citet{hailey+2018} recently discovered 12 quiescent non-thermal X-ray sources
within the central parsec. These sources are spectrally consistent with
quiescent XRBs and distinct from the magnetic CVs that make up most of the
X-ray sources outside of the central parsec. Additionally, their luminosity
function is consistent with that of dynamically confirmed BH XRBs in the
field, while NSs are on average brighter in quiescence
\citep{armas-padilla+2014,hailey+2018}. Other confirmed NS XRBs with
comparable X-ray luminosities in the Galactic Center region and globular
clusters show bright outbursts with a characteristic cadence of 5-10 years
\citep{degenaar+2010,bahramian+2014}; the exceptions are almost all thermal
sources, and therefore not directly comparable to the GC population. Thus, if
the quiescent population were NS-XRBs, there would be many more outbursts than
are observed. It is possible that there is an undiscovered class of NS-XRBs
that mimic the properties of BH-XRBs. Nonetheless, the most likely
identification for this new population is quiescent BH XRBs, though an
admixture of up to six milisecond pulsars cannot be ruled out. Reasonable
extrapolation of the point source luminosity function below the instrumental
detection threshold implies hundreds of BH-XRBs inside the central parsec.

The number of NS XRBs per stellar mass in the GC is three orders
of magnitude greater than in the field, and comparable to the number in
Globular clusters (see Table~\ref{table:Nick}). The number of BH XRBs per
stellar mass in the GC is also three orders of magnitude greater
than in the field, and an order of magnitude greater than in any existing globular,
suggesting the GC BH XRBs are not brought in via globular cluster
in-fall. Also, any BH XRBs brought in by globulars  are unlikely to survive to
the present day, as the lifetime of BH XRBs is at most a few$\times 10^9$
years (see Fig.~\ref{fig:lifetime}).

This indicates that the unusual environment of the GC dynamically - and
efficiently - assembles BH-XRBs, in a manner analogous to the dynamical
overproduction of NS-XRBs in globular clusters \citep{katz1975,
benacquista+2013}. Although a high concentration of compact objects in the GC
is itself unsurprising \citep{alexander&hopman2009}, an overabundance of
mass-transferring {\it binaries} is more challenging to understand.  In other
dense stellar systems like globular clusters, exchange interactions that swap
compact objects into binaries can explain the overabundance of NS-XRBs and
their MSP progeny (e.g.~\citealt{ivanova+2008}). However, this channel is
strongly suppressed in NSCs because nearly all primordial stellar binaries
would be evaporated by three-body encounters, and those that survive would be
so hard as to present a minimal cross-section for exchange interactions (see
\citealt{leigh+2018} and appendix~\ref{app:binFrac}).\footnote{Note that the
maximum semi-major axis above which binaries are evaporated scales as velocity
dispersion $\sigma^{-2}$, whereas the maximum pericenter for tidal capture scales
$\sigma^{-0.2}$; see appendix~\ref{app:binFrac} for more details.}

This paper instead focuses on an alternative channel of XRB formation: the
tidal capture of main sequence stars by compact objects
\citep{press&teukolsky1977,lee&ostriker1986}. Stars that pass sufficiently
close to a compact object$-$approximately, within its tidal radius $r_{\rm
t}-$are completely torn apart by tidal forces (e.g.~\citealt{rees1988}).
However, for pericenter radii somewhat larger than $r_{\rm t}$, tidal forces
are not necessarily destructive; instead, they transfer orbital energy into
internal oscillations of the star, binding it to the compact object. Following
a complex and potentially violent process of circularization, the
newly-created binary settles into a tight orbit.  The necessarily small
orbital separation of the tidal capture binary guarantees that subsequent
gravitational wave emission will drive the star into Roche Lobe overflow in
less than a Hubble time, forming a mass-transferring X-ray source. The high
density of compact objects and stars in the GC inevitably lead to a
significant rate of tidal captures, representing a promising explanation for
the observed overabundance of BH- and NS-XRBs.

This paper is organized as follows.  In $\S$~\ref{sec:nsc} we describe our
model for the dynamical evolution of stars and compact remnants in the GC.  In
$\S$~\ref{sec:coll} we use the time-dependent density profiles of the stars
and compact objects from our NSC models to calculate the rates of collisions
and tidal capture of stars by compact objects, and make predictions for the
present-day BH and NS-XRB population. In $\S$~\ref{sec:obs} we compare our
predictions to observations of the XRB populations in the GC measured by
\citet{hailey+2018}. In $\S$~\ref{sec:discuss} we describe several auxiliary
predictions of our model for the rates of stellar interactions and exotic
transients.  In $\S$~\ref{sec:conc} we briefly summarize our results and
conclude.

\begin{table*}
\begin{threeparttable}
\begin{minipage}{15cm}
\centering
\caption{\label{table:Nick} Estimated abundances of NS-XRB/BH-XRB/MSP per unit
stellar mass in different environments. Q=Quiescent LMXBs ($L_x\lsim 10^{33}$
erg s$^{-1}$); T=Transient LMXBs; B=Persistently bright or transient LMXBs
(only persistently bright sources are NSs). No outbursts have been seen from
globular BH XRB candidates, while only one quiescent BH XRB candidate has been
identified in the field \citep{tetarenko+2016}. The abundances in the ``transient'' BH-XRB column estimate the total population of BH-XRBs prone to periodic outburst, computed by assuming that the observed BH-XRB transients represent a larger population with a characteristic recurrence time of 100 years.}
\begin{tabular}{l|l|l|l|l|l|l}
  Environment & $N_{\rm NS-XRB}$ (B) & $N_{\rm NS-XRB}$ (Q) & $N_{\rm BH-XRB} $ (T) & $N_{\rm BH-XRB} $ (Q) & $N_{\rm MSP}$ & References \\ 
              & [$M_{\odot}^{-1}$] & [$M_{\odot}^{-1}$]  & [$M_{\odot}^{-1}$] & [$M_{\odot}^{-1}$] & \\
  \hline
  Field          & 2$\times 10^{-9}$ & $-$ & $2\times 10^{-8}$ & $-$ & 5$\times 10^{-9}$ & 1-3\\
  \hline
  Globular clusters (all)  & $3\times 10^{-7}$ & 6$\times 10^{-7}$ &  $-$ & $10^{-7}$ & $7\times 10^{-6}$ & 5-7\\
  \hline 
  47 Tuc  & $-$ & 7$\times 10^{-6}$  & $-$ & $10^{-6}$ & 3$\times 10^{-5}$ & 8-11\\
  \hline 
  Terzan 5 & 1.5$\times 10^{-6}$ & 6$\times 10^{-6}$ & $-$ & $-$ & $2\times 10^{-5}$ & 12-13 \\
  \hline 
  Galactic Center (central parsec)  & $1-3 \times 10^{-6}$ & $-$ & $2\times 10^{-5}$ & $10^{-5}$ & $\lesssim 1.3\times 10^{-4}$ & 14-18
\end{tabular}
\begin{tablenotes}
\item \textbf{References} \begin{inparaenum}[(1)] \newcounter{field} \item \citet{corralsantana+2016} \stepcounter{field} \item \citet{tetarenko+2016} \stepcounter{field} \item \href{http://astro.phys.wvu.edu/GalacticMSPs/GalacticMSPs.txt}{Galactic MSPs} \stepcounter{field} 
\item \citet{benacquista+2013} \item \citet{strader+2012a} \item \citet{strader+2012} \item \href{http://www.naic.edu/7Epfreire/GCpsr.html}{Globular MSPs} \item \citet{miller-jones+2015} \item \citet{bahramian+2017}
\item \citet{heinke+2005} \item \citet{heinke+2005a} \item \citet{heinke+2006} \item \citet{bahramian+2014}
\item \citet{degenaar+2015} \item \citet{muno+2005} \item \citet{hailey&mori2017} \item \citet{hailey+2018} \item \citet{perez+2015}
\end{inparaenum}
\end{tablenotes}
\end{minipage},
\end{threeparttable}
\end{table*}

\section{Galactic NSC Model} \label{sec:nsc} 

The number of BH-XRBs that form in the GC clearly depends on the number of
stellar-mass BHs that reside there.  Previous works have predicted that
$\gtrsim 10^3-10^4$ BHs accumulate within the central parsec over timescales
of several Gyr due to radial mass segregation from the stellar population on
larger scales (e.g.~\citealt{morris1993,miralda-escude&gould2000,freitag+2006,
hopman&alexander2006a,oleary+2009,dale+2009,merritt2010}).
Most previous models assume that the BHs are distributed at birth in the same
way as the lower mass stars, and neglect ongoing star formation (though see
\citealt{aharon&perets2015,baumgardt+2018}).

In fact, much of our NSC's total stellar population was likely deposited by
the infall of globular clusters early in its history
\citep{tremaine+1975,antonini+2012,gnedin+2014,arca-sedda+2014,abbate+2018,arca-sedda+2017}.
Historically, globular clusters (at least of the kind which survive to the
present day) were predicted to lose all but a few of their BHs due to strong
kicks in multi-body interactions during a core collapse or Spitzer instability
phase (e.g. \citealt{spitzer1987,kulkarni+1993,banerjee+2010}).  Such
lossiness would limit the ability of globular infall to seed the GC with BHs,
although modern studies have challenged this conventional wisdom by showing
that $\sim 10^2-10^3$ BHs could be retained in globulars due to three-body
processes reversing core collapse
\citep{mackey+2008,morscher+2013,morscher+2015,askar+2018,kremer+2018}.\footnote{Physically,
this is because the thermodynamics of the subcluster is regulated by the
longer relaxation time of the bulk cluster
\citep{breen&heggie2012,breen&heggie2013}.}  While a few candidate globular
cluster BH-XRBs have been identified in the MW
\citep{strader+2012,bahramian+2017}, and other BH-XRBs have been seen in
extragalactic globulars \citep{maccarone+2007}, the total inventory of
globular BHs is challenging to infer from observations because the number of
BH-XRBs is likely a weak function of the number of retained BHs
\citep{kremer+2018a}. Overall, if globulars retain an order unity fraction of
their BHs they can deliver $\sim 2\times10^4$ BHs to the Galactic Center
\citep{arca-sedda+2017}.

A potentially larger population of BHs is formed in the GC by {\it in
situ} star formation. A disk of young stars of age $\approx 4$ Myr is observed
to extend between $\sim 0.03-0.3$ pc of SgrA*
\citep{krabbe+1995,paumard+2006,lu+2013}, containing a total of $\sim$100 WR/O
stars with a top-heavy initial mass function (IMF; e.g.~\citealt{bartko+2010,
lu+2013}). If the average formation rate of massive stars is comparable to the
rate in the last few million years, a total of $\sim 10^5$ compact objects
would be injected into the Galactic Center over 10 Gyr. However, there is no
evidence for multiple bursts of star formation over the last few$\times 10^7$
years \citep{habibi+2017}\footnote{The existing stellar population would not
probe the star formation history on longer time-scales if the IMF is truncated
below $\sim 10\Msun$}, and feedback from stellar winds and Supernovae would
suppress star formation on this time-scale. This suggests star-bursts occur
with a cadence of at least $4\times 10^7$ years. In principle, the cadence may
be much longer, but this would mean we are observing the GC at a very atypical
time. In this paper, we focus on models with bursts of star formation every
$4\times 10^6$-$4\times 10^7$ years.

This section describes our prescription for how the 1D radial density profiles
of stars and compact remnants in the GC evolve in time.  Our goal is to create
a small set of simple but physically-motivated models for building up the NSC,
that are consistent with both the present-day stellar density profile and the
observed rate of compact object formation.  Motivated by the above discussion,
our model consists of two stellar populations: (1) stars injected in the
distant past, near the formation time of the NSC; and (2) a continuously
forming in- situ population with a top heavy IMF concentrated within the
central parsec, as is motivated by the observed disks of young stars
(\citealt{lu+2013}). All of our models assume spherical symmetry and isotropic
velocities.

In $\S \ref{sec:populations}$ we motivate the parameters of our models using
the observed stellar populations and constraints on the star formation history
in the GC.  In $\S$~\ref{sec:num} we describe our numerical procedure for
evolving the density profiles of stars and compact objects through two-body
relaxation.  To build up physical intuition, we first calculate how the
compact objects evolve in isolation in $\S$~\ref{sec:compact}, before adding
the effects of the stellar background in $\S$~\ref{sec:fiducial}. In
$\S$~\ref{sec:nonfiducial} we discuss several additional hypothetical
scenarios for building the NSC, in order to assess the uncertainty in our work
and to make contact with previous papers in the literature (which generally
neglect centrally-concentrated compact object formation).

\subsection{Stellar and Compact Object Populations}
\label{sec:populations}
Eighty percent of the stars in the GC are older than 5 Gyr \citep{pfuhl+2011},
consistent with the bulk of the NSC's growth being due to the infall of
globular clusters via dynamical friction over a period of $\sim 1$ Gyr in the
early history of the Galaxy \citep{gnedin+2014}. The observed diffuse stellar
light\footnote{This profile differs from the observed giant density profile,
which has a core inside of $\sim 0.5$ pc (the so-called ``missing giants"
problem; e.g. \citealt{buchholz+2009, do+2009}). The diffuse light tracks
emission from early G and late F main sequence and sub-giant stars, and is
likely a better probe of the underlying stellar density. One solution to the
missing giants problem is mass-stripping by collisions between the giants with
other stars and compact objects \citep{dale+2009} or with a clumpy gas disc
\citep{amaro-seoane&chen2014,kieffer&bogdanovic2016}.} is well fit by the parameterization (\citealt{schodel+2017}; their Tables 2 and 3),
\begin{align}
\rho_\star(r)= 2^{(\beta-\gamma)/\alpha}\rho_0
\left(\frac{r}{r_0}\right)^{-\gamma}
\left(1+\left(\frac{r}{r_0}\right)^{\alpha}\right)^{(\gamma-\beta)/\alpha},
\label{eq:schodel}
\end{align}
with best-fit parameters of $\gamma=1.16 \pm 0.02$, $\beta=3.2 \pm 0.3$,
$r_0=3.2\pm 0.2$ pc for fixed $\alpha=10$. The density normalization at 1 pc
is $0.8-1.7\times10^5$ pc$^{-3}.$\footnote{We use the values from the
first version of \citet{schodel+2017}. The best fit parameters are
slightly different in the published version, but consistent within
uncertainties.}

Compact objects are also deposited at early times if they arrive with the
globular clusters. \citet{ivanova+2008} estimate that a typical globular
cluster retains 1 NS per $10^{3}M_{\odot}$ of other stars. Although globulars
may also bring in a sizeable BH population, this is less certain because, as
discussed above, BHs may be ejected from globulars by binary-single
interactions (e.g.~\citealt{kulkarni+1993,banerjee+2010}; see, however,
\citealt{morscher+2015}).  Given this uncertainty, and because the current BH
population is anyways likely to be dominated by {\it in situ} star formation
(see below), we neglect BHs deposited with the old stellar population.

Our fiducial models include a population of compact objects formed {\it in
situ}, motivated by the sub-parsec disk of young stars observed in the GC.
The K-band luminosity function of the young stars is consistent with a single
starburst that occurred $2.5-5.8$ Myr ago \citep{lu+2013,habibi+2017}. The
burst produced a total of $\sim$250 stars of mass $\gtrsim 8 \Msun$ with an
IMF of the form $dN/dm
\propto m^{-\beta}$ with $\beta \approx 1.7$.  If stars with masses in the range $m \approx
8-25 \Msun$ form NSs, while those with $m \gtrsim 25M_{\odot}$ form BHs, then
a total of $N_{\rm ns} \sim 160$ NSs and $N_{\rm bh} \sim 90$ BHs
were, or will be, formed from the disk stars.  If the time since the last
star formation episode of $\approx$ 4 Myr is comparable to the typical interval between starbursts,
then the implied average formation rates of NSs and BHs in the central parsec
are $\dot{N}_{\rm ns} \sim 4 \times 10^{-5}$ yr$^{-1}$
and $\dot{N}_{\rm bh} \sim 2 \times 10^{-5}$ yr$^{-1}$, respectively. 

The above estimates assume that the current epoch is a representative snapshot
of the central parsec's average star formation history.  In possible tension
with this, \citet{pfuhl+2011} find that the star formation rate $\sim1-5$ Gyr
ago was $\sim 1-2$ orders magnitude smaller than the present-day rate
(their Figure 14), in which case the average star formation rate is $\lesssim
10\%$ of its recent value. However these observations probe only low mass
stars ($\lesssim 2 \Msun$), and thus do not constrain the rate of NS/BH
formation within the star-forming disks if the top-heavy disk IMF is truncated below a
few solar masses.  Other nearby galactic nuclei such as M31 possess disks of A
stars, but no O and B stars \citep{leigh+2016}; in these NSCs at least, the
last major episode of star formation occurred $\gtrsim 100$ Myr ago.

Motivated by the above, we construct our fiducial models for the GC using the following three populations:
\begin{enumerate}[a.]

\item ``Primordial" stars, which are assumed to form impulsively at $t = 0$
($10$ Gyr ago) with an initial density profile following
eq.~\eqref{eq:schodel}.  We model all the stars as being of a single mass
0.3$M_{\odot}$, which represents the root-mean-square mass of main sequence
and sub-stellar objects in an evolved Kroupa IMF.  For simplicity, the
parameters of the stellar profile ($\alpha, \beta, \gamma$) are fixed to the
best-fit values from \citet{schodel+2017}, except for the scale radius $r_o$
and normalization $\rho_o$.  The cluster expands radially over time, so we
chose smaller initial values of $r_o = 0.5,1.5$ pc in order to match the {\it
present-day} stellar density at 1 pc (though we note that the functional form
of the density profile is not exactly preserved in the evolution). A
normalization of $\rho_{\star}(1\, {\rm pc})=1.1\times 10^5$ $M_\odot ~$
pc$^{-3}$ is chosen to fix the total stellar mass to $5.7\times 10^7
\Msun$.\footnote{The true total mass, $2.5 \pm 0.4 \times 10^7 \Msun$, is
somewhat lower. The difference comes from the fact that we assume the stellar
density profile extends to infinity (with an $r^{-3.2}$ profile), but in
reality the stellar density steepens at $\sim$10 pc.}

\item  ``Primordial" NSs of mass $1.5M_{\odot}$, which are deposited
impulsively at $t = 0$ with the same density profile as the stars.  The total
number of NSs is normalized to a fraction $10^{-3}$ of the number of stars,
motivated by their expected abundance in globular clusters \citep{ivanova+2008}. 
\item  Compact objects from {\it in situ} star formation (NSs and BHs of
masses $1.5$ and $10M_{\odot}$, respectively) that are continuously injected
near the present-day disk of young stars. The source term is narrowly peaked
at the potential energy at $0.3$ pc (the outer edge of the disk). In physical
space, star formation is concentrated inside of this radius with an $r^{-0.5}$
density profile. We found our results do not change if the star formation is
instead concentrated inside of 0.03 pc (the inner edge of the star forming
disks). In our {\bf ``Fiducial"} model, we adopt conservative formation rates
of $\dot{N}_{\rm ns} = 4 \times 10^{-6}$ yr$^{-1}$ and $\dot{N}_{\rm bh} =
2\times 10^{-6}$ yr$^{-1}$, respectively. We also consider a model (``{\bf
Fiducial $\times$ 10}") in which $\dot{N}_{\rm ns}$ and $\dot{N}_{\rm bh}$ are
ten times larger, corresponding to the present day formation rate of massive
stars.
\end{enumerate}

The parameters of our fiducial models are summarized in Table~\ref{tab:models}.  Several hypothetical (non-fiducial) models are introduced in $\S\ref{sec:nonfiducial}$ in order to assess the robustness of our conclusions.

\subsection{Numerical Method: Fokker-Planck}
\label{sec:num}
The radial distribution of stars and compact objects evolves over time due to
two-body relaxation.  We follow this evolution using the \textsc{PhaseFlow}
code \citep{vasiliev2017}, which solves the time-dependent, isotropic
Fokker-Planck equation for the energy-space distribution function
$f(\epsilon,t)$.  This equation can be written in flux-conserving form as
\begin{equation}
\frac{\partial f(\epsilon, t)}{\partial t} = -\frac{\partial}{\partial
  \epsilon} \underbrace{\left[ D_{\epsilon \epsilon} \frac{\partial f(\epsilon,
    t)}{\partial \epsilon} + D_{\epsilon} f(\epsilon, t) \right]}_{F(\epsilon)}
-\frac{f(\epsilon, t)}{\tau_{\rm LC} (\epsilon, t)}+S(\epsilon, t),
\label{eq:fp}
\end{equation}
where $\epsilon$ is the binding energy, $D_{\epsilon}$ and $D_{\epsilon
\epsilon}$ are the first and second order energy diffusion coefficients,
$F(\epsilon)$ is the mass flux, and the last two terms account for the
draining of stars into the loss cone of the SMBH (see eq. 13 in
\citealt{vasiliev2017}), and injection of stars due to star formation. The
diffusion coefficients can be expressed as integrals over the distribution
function, which for a single species mass $m$ are given by
\begin{align}
 &D_{\epsilon \epsilon}=16 \pi^2 G^2 m \ln \Lambda \left[h(\epsilon)
 \int_{0}^{\epsilon} f(\epsilon') d\epsilon' + \int_{\epsilon}^{\epsilon_{\rm
 max}} f(\epsilon') h(\epsilon') d\epsilon' \right]\\ 
 &D_{\epsilon} = -16
 \pi^2 G^2 m  \ln \Lambda \int_{\epsilon}^{\epsilon_{\rm max}} f(\epsilon')
 g(\epsilon') d\epsilon',
\label{eq:diffusion}
\end{align}
where $h(\epsilon)$ is the phase volume and $g(\epsilon)=d
h(\epsilon)/d\epsilon$ is the density of states (see e.g. \citealt{merritt2013}).  
For a Keplerian potential, $h \propto  \epsilon^{-3/2}$ and $g \propto \epsilon^{-5/2}$.

The one-dimensional Fokker-Planck approach is computationally efficient and
reproduces the results from two-dimensional Fokker-Planck
\citep{cohn1985,merritt2015a} as well as Monte-Carlo and N-body calculations
\citep{vasiliev2017} reasonably well. A key assumption of this equation is
spherical symmetry, which is in tension with the physical motivation for our
source term $S(\epsilon, t)$: disk-mode star formation.  However, it is
reasonable to assume that compact remnants will become isotropic over time.
First of all, there is likely no preferred plane for disk mode star formation
in the GC. The current disk of young stars is not aligned with either the
Galactic disk or the circumnuclear ring of molecular gas
\citep{mccourt&madigan2016}. Thus, the injected remnants from many different
episodes of {\it in situ} star formation would naturally form with a
quasi-isotropic angular distribution. Recent work has found that vector
resonant relaxation can lead to a disk configuration for BHs and heavy stars
even if they are drawn from sixteen randomly oriented disks
\citep{szoelgyen&kocsis2018}. However, as the number of star formation
episodes increases, the disk would thicken and would approach an isotropic
distribution (Bence Kocsis, personal communication). In principle, resonant
relaxation \citep{rauch&tremaine1996,hopman&alexander2006} may flatten the
radial stellar density profile by causing stars to diffuse more rapidly into
the loss cone. In practice, this effect only becomes important on
small radial scales $\lesssim 0.1$ pc \citealt{baror&alexander2016}), interior
to where most tidal captures occur.

Finally, strong gravitational scatterings by BHs can lead to significant
evaporation of low mass stars and remnants from the cusp. This effect is not
included in our models, but a post-hoc calculation shows it changes
the stellar density profile by $\lesssim 40\%$ (see
$\S~\ref{sec:strong}$).

\subsection{Evolution with Compact Remnants Only}
\label{sec:compact}
Compact objects which are injected near the present disk of massive stars at
$\sim 0.3$ pc will diffuse outwards via two-body scattering. To
study this process, we solve eq. (\ref{eq:fp}) with a constant source
function of injected BHs, $\dot{N}_{\rm bh} =
2\times 10^{-5}$ yr$^{-1}$, corresponding to our
``Fiducial$\times 10$"  model. To whet our intuition in a controlled setting, we initially
neglect contributions to the gravitational potential or diffusion coefficients
from the background of NSs and stars.

\begin{figure}
  \includegraphics[width=8.5cm]{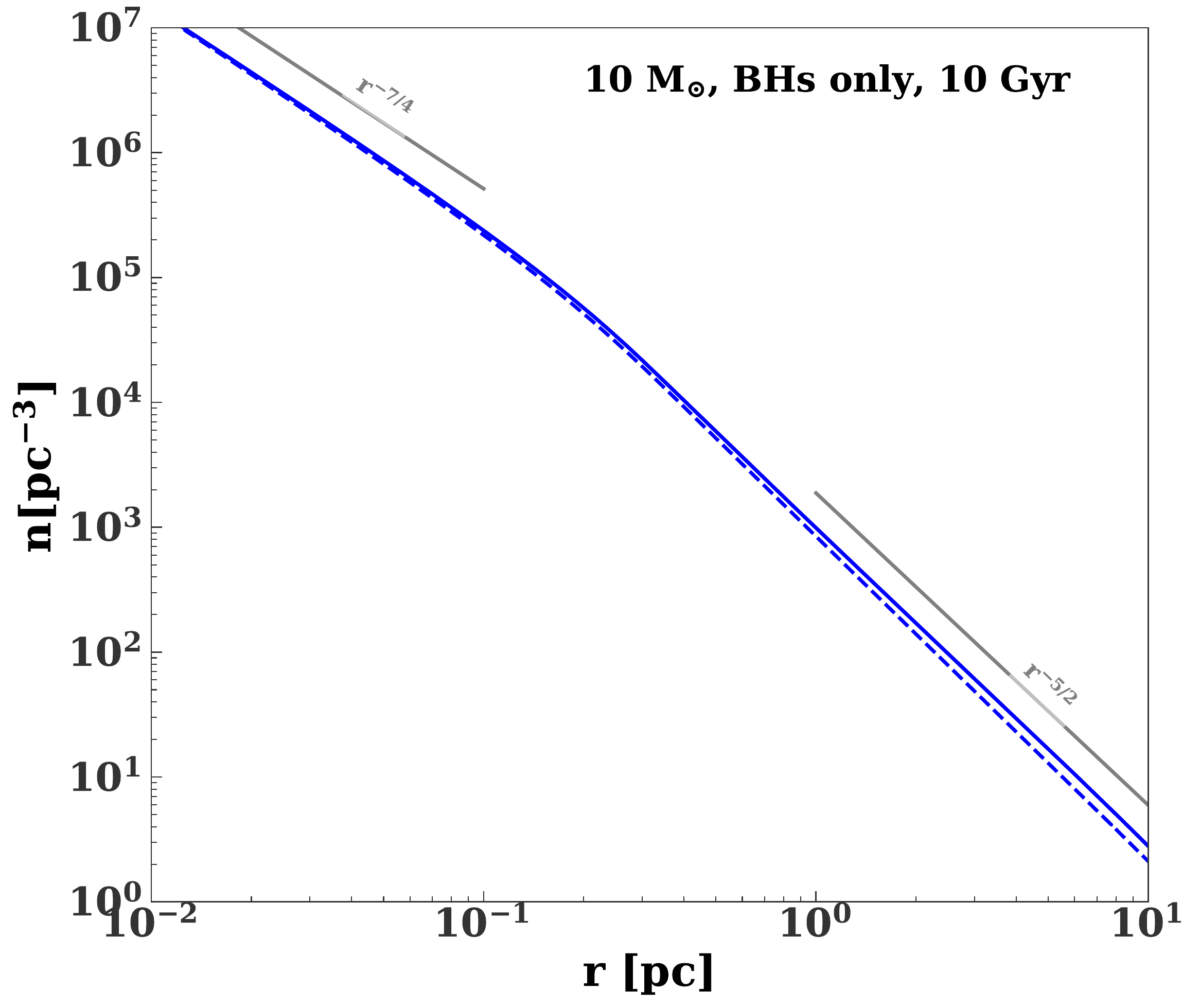}
  \includegraphics[width=8.5cm]{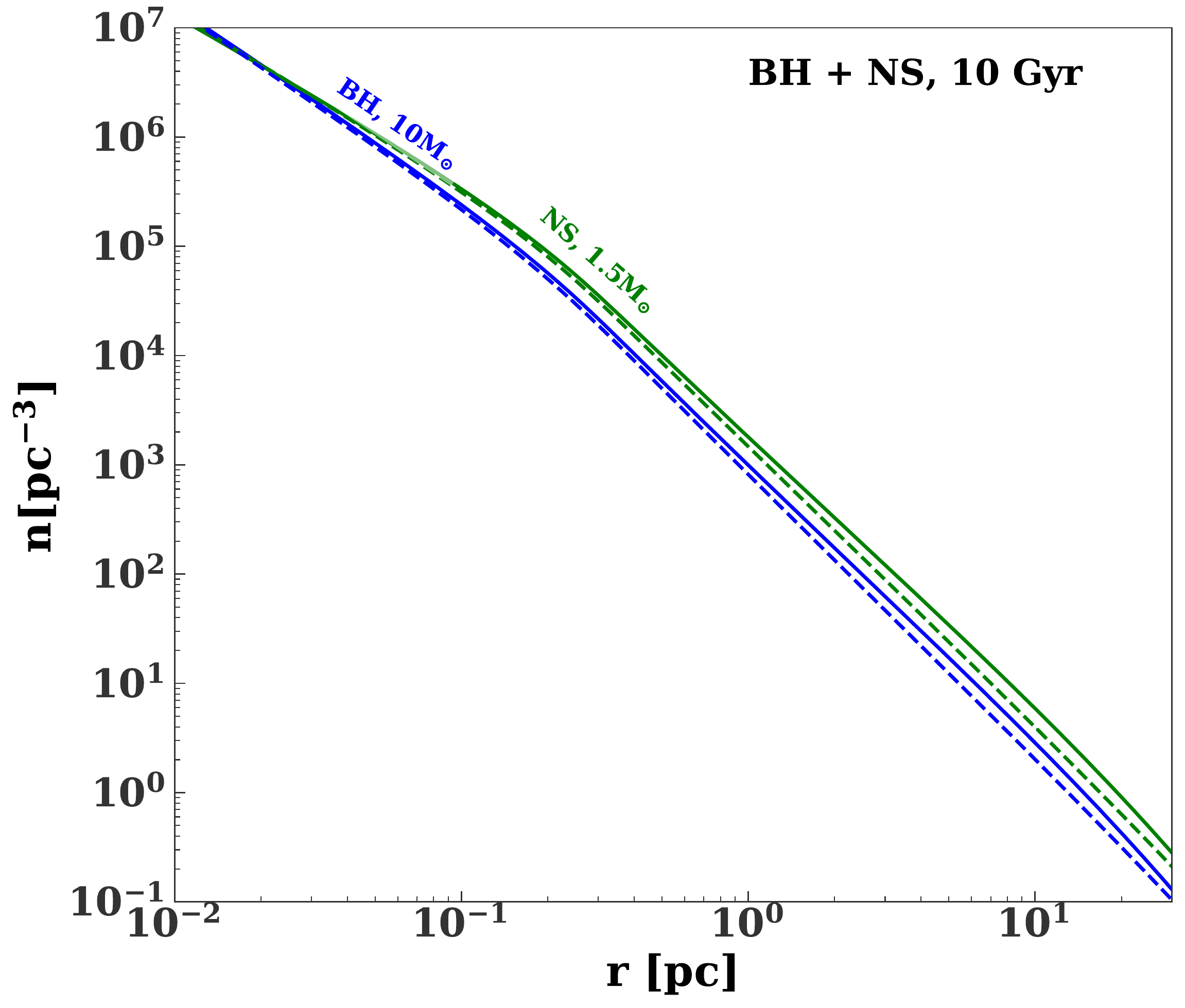}
  \caption{\label{fig:profiles} \emph{Top panel:} Number density of BHs in the
GC as a function of radius $r$ after 10 Gyr of evolution, in the case of
continuous BH injection (\emph{solid line}).  BHs are injected at a constant
rate $\dot{N}_{\rm bh} = 2 \times 10^{-5}$yr$^{-1}$ at $r_{\rm in} \approx
0.3$ pc. \emph{Bottom panel:} Density profiles of NSs and BHs after 10 Gyr of
evolution for injection rates at $r_{i}$ corresponding to our Fiducial$\times 10$ model
($\dot{N}_{\rm ns} = 4\times 10^{-5}$, $\dot{N}_{\rm bh} = 2 \times
10^{-5}$yr$^{-1}$).  Dashed lines show how the results change if the
gravitational potential of the compact objects, and the sink term due to SMBH
loss cone, are neglected.}
\end{figure}

Figure \ref{fig:profiles} (top panel) shows the resulting BH number density
profile $n(r)$ after 10 Gyr of evolution, over which a quasi-steady state is
achieved on small radial scales. For comparison, a dashed line shows how
little the solution changes if one neglects the gravitational potential of the
stellar mass BHs and the loss cone sink (the final term in eq.~\ref{eq:fp}).
The steady-state BH profile is well described by a broken power law, with $n
\propto r^{-7/4}$ at small radii $r \ll r_{\rm i}$ and $n \propto r^{-5/2}$ at
$r \gg r_{\rm i}$.  These power-law slopes and the normalization of the BH
profile can be understood through basic analytic arguments.

Compact objects injected at $r_{\rm i}$ diffuse outwards on the two-body
relaxation timescale, which for a single mass population $m_{\rm c}$ is
approximately given by\footnote{For a Keplerian potential, the pre-factor of
eq.~\eqref{eq:trx} varies from 0.2$-$0.4, depending on the density profile
power-law slope $\gamma = 0.5-3$, where $n \propto r^{-\gamma}$.},
\begin{align}
\tau_{\rm rx} \approx \frac{0.34}{\log \Lambda}\frac{ \sigma^3(r)}{G^2 n(r) m_c^2},
\label{eq:trx}
\end{align}
where $\sigma(r)$ is the one dimensional velocity dispersion, $n(r)$ is the
number density, and $\log \Lambda \approx 15$ is the Coulomb logarithm.

In a steady state, the formation rate of compact objects per unit volume
equals the rate of outwards diffusion, i.e.
\begin{align}
\frac{\dot{N}}{r_i^3} \sim \frac{n_i}{\tau_{\rm rx}(r_i)}.
\end{align}
This implies a steady-state density at the injection radius of 
\begin{align}
n(r_{i}) \propto r_i^{-9/4} \dot{N}^{1/2} M^{3/4} m_c^{-1}, 
\label{eq:densinj}
\end{align}
where in taking $\sigma \propto (GM/r)^{1/2}$ we have assumed that the SMBH of mass M dominates the gravitational potential.  Normalizing equation (\ref{eq:densinj}) using results from our numerical solutions, we find
\begin{align}
  n_i= &1.6\times 10^{4} \left(\frac{r_i}{0.3{\rm pc}}\right)^{-9/4}
  \left(\frac{\dot{N}}{2\times 10^{-5} {\rm yr^{-1}}}\right)^{1/2}\nonumber\\
  &\left(\frac{M}{4\times 10^6 \Msun}\right)^{3/4}
  \left(\frac{m_c}{10 \Msun}\right)^{-1} {\rm pc^{-3}},
\label{eq:densinj1}
\end{align}
What sets the power-law slopes of the BH density profile?  For a distribution
function $f \propto \epsilon^p$ which extends to a maximum energy
$\epsilon_{\rm max}$, the flux of mass through energy space (see eq.~\ref{eq:fp}) is
\begin{align}
&F(\epsilon) \propto \epsilon^{2 p-3/2}\left[ a_o(p)  + a_1(p) \left(\frac{\epsilon_{\rm
max}}{\epsilon}\right)^{p-3/2} + a_2(p) \left(\frac{\epsilon_{\rm
max}}{\epsilon}\right)^{p-1/2} \right]\nonumber\\ & a_o(p) = \frac{3(4 p
-1)}{(1+p)(2 p-1)(2 p -3)},
\label{eq:F}
\end{align}
where $a_1$, $a_2$, and $a_3$ are dimensionless functions of $p$.

\begin{figure} \includegraphics[width=8.5cm]{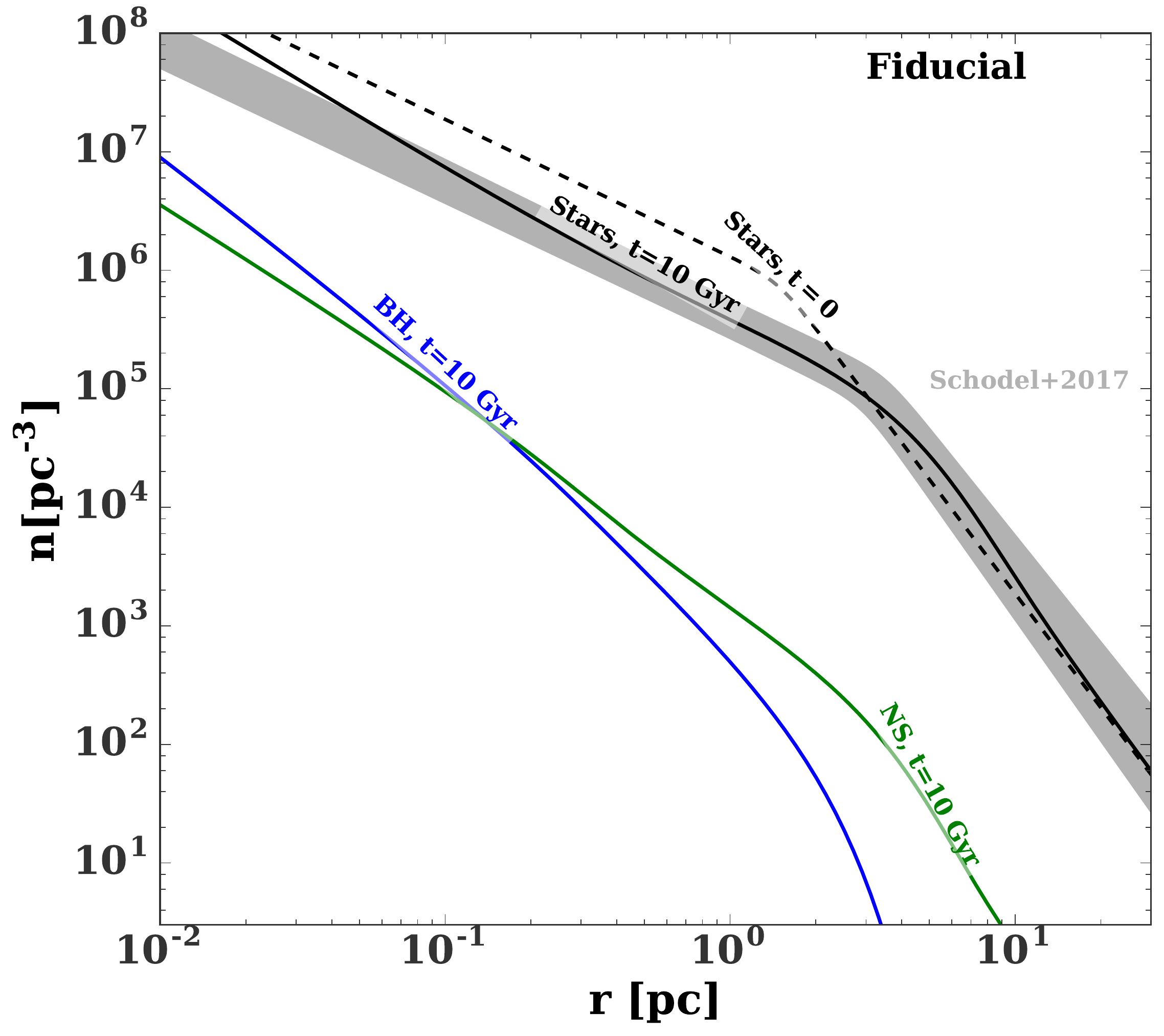}
\includegraphics[width=8.5cm]{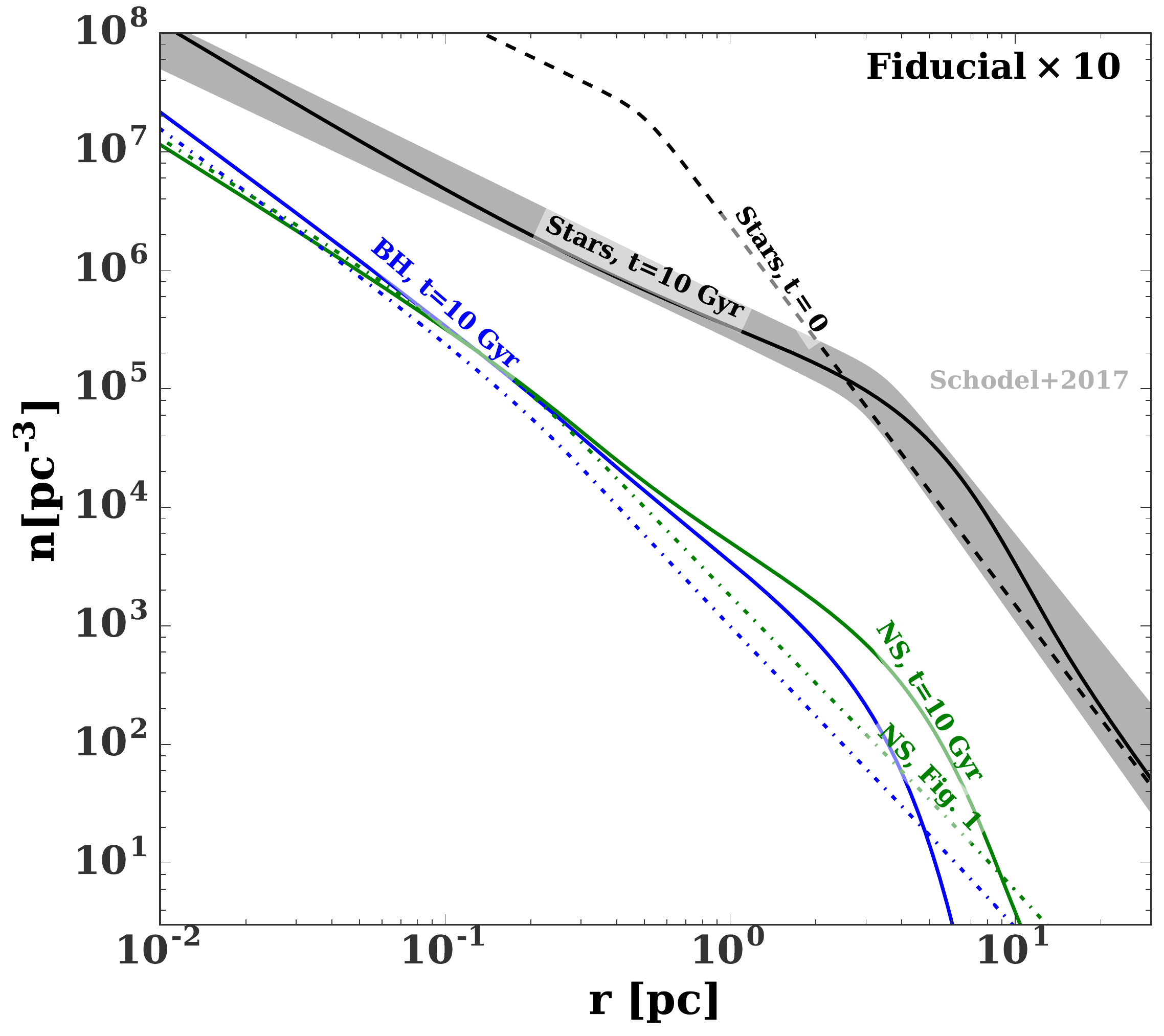}
\caption{\label{fig:stellarBackground2} {\emph{Top panel:}} Fiducial density profiles
of stars and compact remnants at $t = 10$ Gyr (solid lines). Compact remnants
are continuously injected near $\sim 0.3$ pc. The initial profile of stars is
shown as a dashed black line, while the present-day distribution of low mass
stars (with uncertainties) from \citet{schodel+2017} is shown as a shaded
region. \emph{Bottom panel:} Density profiles of stars and compact objects in
our Fiducial$\times 10$ model. For comparison, dash-dotted blue and green
lines show, respectively the profiles of BHs and NSs, neglecting the
pre-existing background of low mass stars/NSs (Fig.~\ref{fig:profiles}).}
\end{figure}

In steady state, the flux through energy space is constant.  For $p<1/2$, one
finds $F \approx a_o(p) \epsilon^{2 p-3/2}$ in the limit that $\epsilon_{\rm
max} \rightarrow \infty$ and thus $F$ will be zero for $p=1/4$; this is the
classical ``Bahcall-Wolf" (BW) solution \citep{bahcall&wolf1976}.  If $p>1/2$,
the $a_{2}$ term in equation (\ref{eq:F}) dominates over the first two
terms. A $p=1$ profile corresponds to the steady-state solution with a
constant, non-zero (outwards) flux. In this case $\epsilon_{\rm max}$ must
have a finite value, as otherwise the flux would diverge; in our case, this
maximum energy corresponds to the location of the source function of injected
BHs at $r_{i}$. These two steady state solutions (zero flux at small radii and
constant outward flux at large radii) correspond to density profiles $n\propto
r^{-7/4}$ and $\propto r^{-5/2}$, respectively. In this solution, energy is
transferred from the injection radius to larger scales by stars on eccentric
orbits. However, \citet{fragione&sari2018} have argued that these stars cannot
effectively transfer energy to the bulk of the stellar population and evolve
in a decoupled way (an effect which the isotropic Fokker-Planck solver in
\textsc{PhaseFlow} cannot capture). In this case, energy relaxation at any
radius would be dominated by the local stellar population, and by mass
conservation the density has a slightly shallower $r^{-9/4}$ profile (see also
\citealt{peebles1972}). Ultimately, the two-dimensional Fokker-Planck
simulations are necessary to determine the correct outer density profile.

The enclosed mass $\propto n r^{3} \propto r^{1/2}$ is dominated by the
largest radius to which the BHs have time to diffuse over the system age $t$.
The half-mass radius $r_{1/2}$, interior to which the above steady-state
profile is established, can be estimated by equating the system age with the
relaxation time; using equation (\ref{eq:trx}), this gives
\begin{align}
  &r_{\rm 1/2} \approx 3.3 \,\,{\rm pc}\,\, \left(\frac{t}{\rm 3\,
      Gyr}\right)^2 \left(\frac{r_i}{0.3{\rm pc}}\right)^{-1/2}
  \left(\frac{\dot{N}}{2\times 10^{-5} {\rm yr^{-1}}}\right)\nonumber\\
  &\left(\frac{M}{4\times 10^6 \Msun}\right)^{-3/2}
  \left(\frac{m_c}{10 \Msun}\right)^{2}.
\end{align}
BHs injected in the GC therefore have sufficient time to establish a steady-state profile 
within the central parsec by the present age ($t = 10$ Gyr).

The bottom panel of Fig.~\ref{fig:profiles} shows the present-day density
profiles for a calculation otherwise identical to the BH-only case, but
including the evolution of {\it both} the BHs and NSs, assuming each are
injected at $r_{\rm i}$ at rates of $\dot{N}_{\rm bh} = 2\times 10^{-5}$
yr$^{-1}$ and $\dot{N}_{\rm ns} = 4\times 10^{-5}$ yr$^{-1}$, respectively
(the old population of stars and their associated NSs are still neglected).
The addition of NSs has little effect on the BH profile compared to the
BH-only case.  Outside of the injection radius, the slope of the NS density
profile is similar to the BH one, but with a greater overall normalization
reflecting the relatively higher NS injection rate.  At radii $\lesssim r_i$,
the BH density profile approaches the BW shape $n \propto r^{-7/4}$ (as in the
BH-only case), while the NSs achieve a shallower profile $\propto r^{-3/2}$.
A shallower profile for the lighter species is expected for a two-component
model in which the heavy species dominates the diffusion coefficients
\citep{bahcall&wolf1977,alexander&hopman2009}.

The analytic arguments presented above are readily extended to the
multi-species case. In particular, the BH density at $r_i$ can again be
estimated by replacing the single remnant mass $m_c$ in eq.~\eqref{eq:densinj}
with a weighted generalization
\begin{equation}
\tilde{m}=\left(\frac{\dot{N}_{\rm bh} m_{\rm bh}^2 +\dot{N}_{\rm ns}
    m_{\rm ms}^2}{\dot{N}_{\rm bh}+ \dot{N}_{\rm ns}}\right)^{1/2}.
\end{equation}

\subsection{Effects of stellar background and potential}
\label{sec:fiducial}

We now explore the effects of including the old population of low mass
stars and NSs on the NSC evolution, the final step in constructing our
models. Fig.~\ref{fig:stellarBackground2} shows the profile of stars, NSs, and
BHs at $t = 10$ Gyr in our Fiducial (top panel) and Fiducial$\times 10$
(bottom panel) models. BHs and NSs dominate the mass density inside of 0.03
(0.4) pc in the Fiducial (Fiducial$\times 10$) model.

The cusp of compact remnants causes the star cluster to expand radially over
time, motivating our choice of a more compact initial stellar profile (black
dashed line) than the currently-observed one (shaded gray region;
\citealt{schodel+2017}). By contrast, the compact objects become slightly more
centrally concentrated than in the previous models where the stars were
neglected (Fig.~\ref{fig:profiles}). There are two reasons for this: (i) stars
tend to scatter the higher mass compact objects to larger binding energies
(ii) the gravitational potential of the stars suppresses outward diffusions of
compact objects.

\subsection{Non-fiducial NSC Models}
\label{sec:nonfiducial}

This section explores other (``non-fiducial'') scenarios for creating the GC's
NSC, in which all of the stars and compact objects instead form as a single
population with a common density profile and standard IMF. For one set of
models, we assume all stars formed impulsively 10 Gyr ago. Such models allow
us to compare our results to those of past work (e.g.~\citealt{morris1993,
miralda-escude&gould2000, freitag+2006}), and are a useful limiting case if
bursts of massive star formation occur with a $\gsim$100 Myr cadence. We also
consider models in which stars form at a constant rate with the present day
observed profile. This model is unrealistic for the GC, but may be useful for
other galactic nuclei with different star formation histories
\citep{leigh+2016}. These models are summarized in Table~\ref{tab:models}.

Our models assume that stars are accreted in the distant past or form in-situ,
neglecting on-going exchange of stars with the surrounding galaxy. We expect
the exchange of low mass stars to be negligible as the energy relaxation time
becomes longer than a Hubble time outside of a few parsecs. Ten solar mass BHs
within ten parsecs of the center would sink to smaller radii within a Hubble
time. Sinking from this scale would be captured by our impulsive models.
On-going star cluster in-fall can bring additional stars to the center.
However, the majority of the stellar mass brought in via globulars is accreted
within $\sim$1 Gyr \citep{arca-sedda+2014,gnedin+2014}.


\begin{table*} \begin{threeparttable} \centering
\caption{\label{tab:models} Summary of models for assembling the populations
of stars and compact remnants in the GC.  The top two rows summarize our
``fiducial" models, in which low mass stars and NSs are initialized at $t = 0$
following a radial profile given by eq.~\eqref{eq:schodel} with the scale
radius $r_{0}$ and NS-to-star number ratio $N_{\rm ns}/N_{\star}$. BHs and NSs
are  also continuously injected inside of $\sim 0.3$ pc, near the outer edge
of the observed young stellar disks, with rates $\dot{N}_{\rm bh}^\star$ and
$\dot{N}_{\rm ns}^\star$, respectively. The bottom three rows summarize other,
non-fiducial scenarios, in which all compact remnants and stars form with the
same radial distribution, either impulsively in the distant past or
continuously. The masses of the stars, NSs, and BHs, are taken to be
$0.3M_{\odot}, 1.5M_{\odot}$, and $10\Msun$, respectively.}
\begin{tabular}{l|l|l|l|l|l|l|l|l|l}
  Scenario & $r_o$ & ${N}_{\star}(t=0)$  & $\dot{N}_{\star}$ & $N_{\rm bh}/N_{\star}$ (t=0) & 
  $N_{\rm ns}/N_{\star}$ (t=0) &  $\dot{N}_{\rm bh}^\star$ & $\dot{N}_{\rm ns}^\star/\dot{N}_{\rm
      \rm bh}^\star$  & $\Mbh$ & Fig. \\
  & [pc] & & [yr$^{-1}$ ] &  &  & [yr$^{-1}$] & & & \\
  \hline
  Fiducial &  1.5 & $1.9 \times 10^{8}$ & $0$ & $0$ & $10^{-3}$ & $2\times 10^{-6}$ & 2 & $4 \times 10^6 
  \Msun$ & \ref{fig:stellarBackground2} \\
  \hline
  Fiducial$\times 10$ &  0.5 & $1.9 \times 10^{8}$  & $0$ & $0$ & $10^{-3}$ & $2\times 10^{-5}$ & 2 & $4 \times 10^6 
  \Msun$ & \ref{fig:stellarBackground2} \\
  \hline
  Impulsive & 1 & $1.9 \times 10^{8}$ & $0$ &
  ${3\times 10^{-4} , 10^{-3} , 10^{-2}}$ & $4 N_{\rm bh}/N_\star $ & $0$ & $-$ & $4 \times 10^6 
  \Msun$ & \ref{fig:compare} \\
  \hline
  \makecell{Continuous SF\\Existing SMBH} & 3  & $0$ & $1.9\times 10^{-2}$  &   ${3\times 10^{-4} , 10^{-3} , 10^{-2}}$ & $4 N_{\rm bh}/N_\star $ & $0$ & $-$ & $4 \times 10^6 
  \Msun$ & \ref{fig:compareInsitu} \\
  \hline
  \makecell{Continuous SF\\Growing SMBH} & 3 & $0$ & $1.9 \times 10^{-2}$ &   ${3\times 10^{-4} , 10^{-3} , 10^{-2}}$ & $4 N_{\rm bh}/N_\star $ & $0$ & $-$ & 7\% $M_{\rm tot}$ & \ref{fig:compareInsitu2} \\
\end{tabular}
\end{threeparttable}
\end{table*}

\begin{figure} \includegraphics[width=8.5cm]{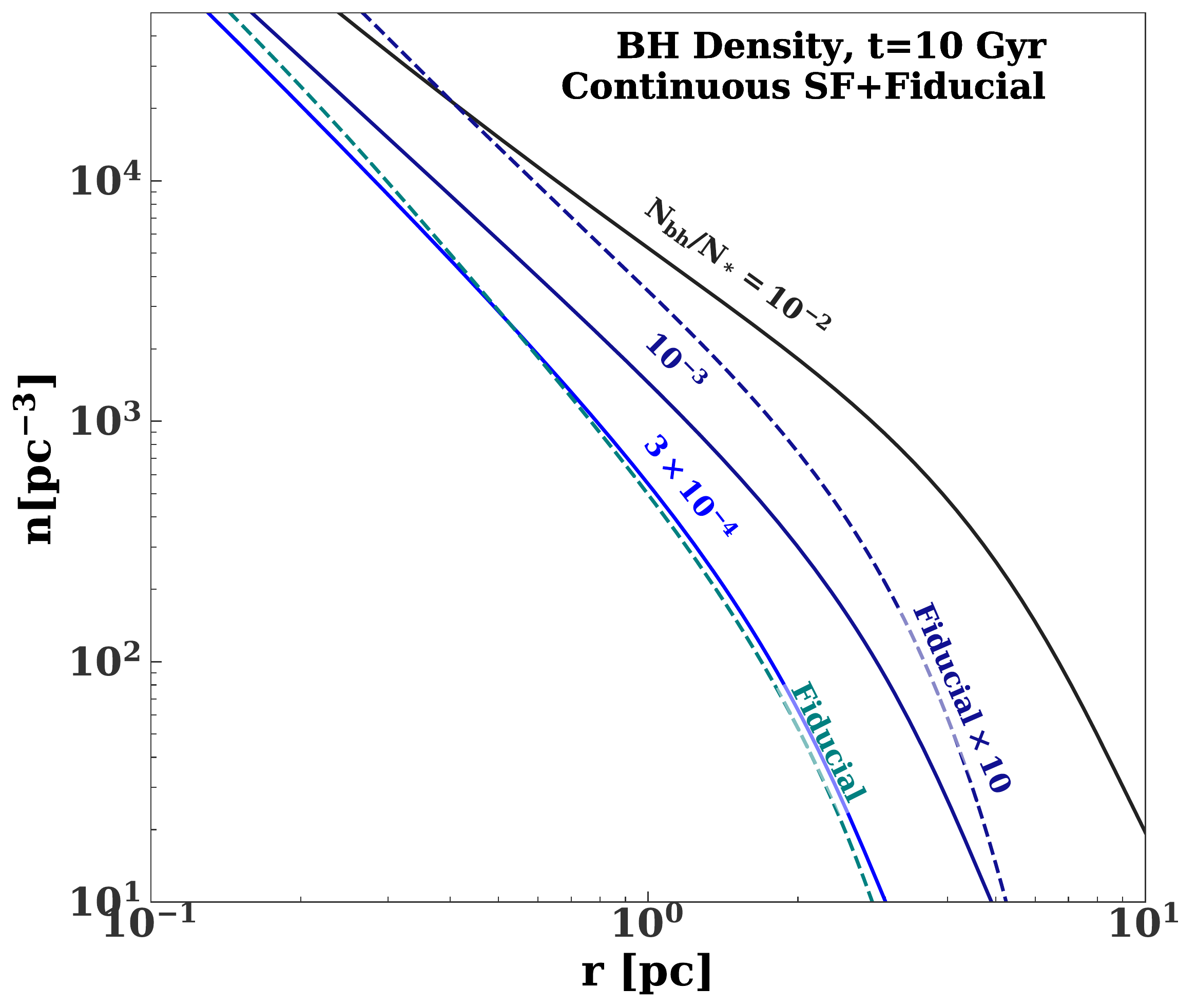}
\includegraphics[width=8.5cm]{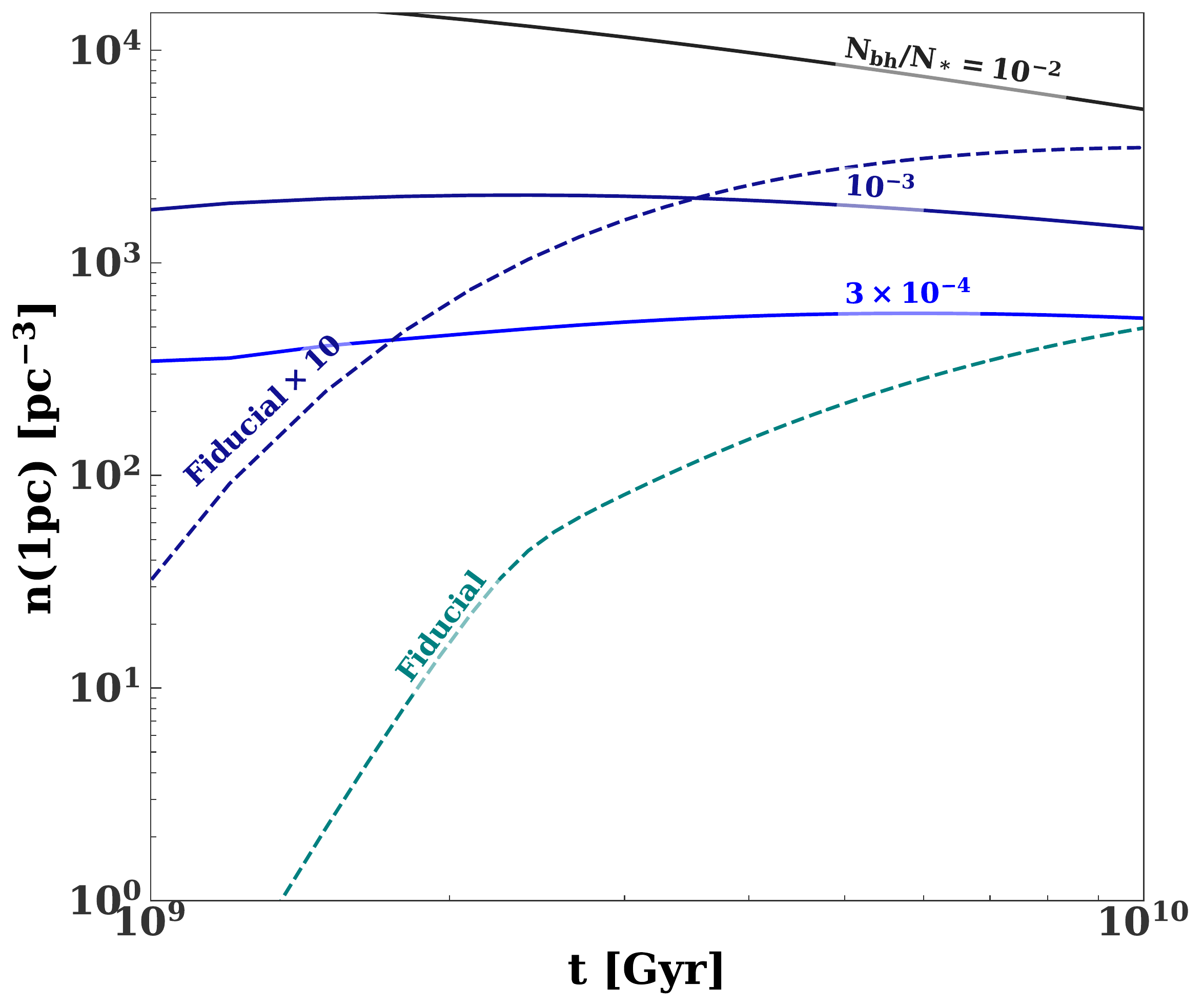} \caption{\label{fig:compare}
\emph{Top panel}: Density profile of BHs at $10$ Gyr for  different star
formation histories  (Table~\ref{tab:models}).  Solid lines show non-fiducial
models in which the BHs form implusively at $t=0$ with the same profile as the
stars (eq.~\ref{eq:schodel}), with colors labeling the ratio of BHs to stars.
For comparison, dashed lines show our Fiducial and Fiducial$\times$10 models,
in which the BHs are instead injected continuously at small radii (see
Fig.~\ref{fig:stellarBackground2} and surrounding discussion).  \emph{Bottom
panel:} Time evolution of the BH density at $r = 1$ pc for each of the
formation histories shown in the top panel.}
\end{figure}

Fig.~\ref{fig:compare} shows the BH profile at $t = 10$ Gyr which results if both stars and compact remnants
are formed impulsively at $t = 0$ (initial scale radius of $r_o = 1$ pc) and assuming no subsequent star formation.  We show results for a range of models which assume different ratios for the number of stars to
compact objects, $N_{\rm bh}/N_{\star}$.  For a Kroupa IMF in which BHs
originate from stars of mass $\gtrsim 25M_{\odot}$, we expect $N_{\rm
bh}/N_{\star} \sim 10^{-3}$, which (coincidentally) coincides with the number of BHs
injected by the present day in our Fiducial$\times 10$ model.  However, because the BHs in this case are injected directly at small radii, their density at radii $\lesssim 1$ pc in our Fiducial$\times$10 models exceeds the primordial model with $N_{\rm bh}/N_\star=10^{-3}$ by a factor of a few.

\begin{figure} \includegraphics[width=8.5cm]{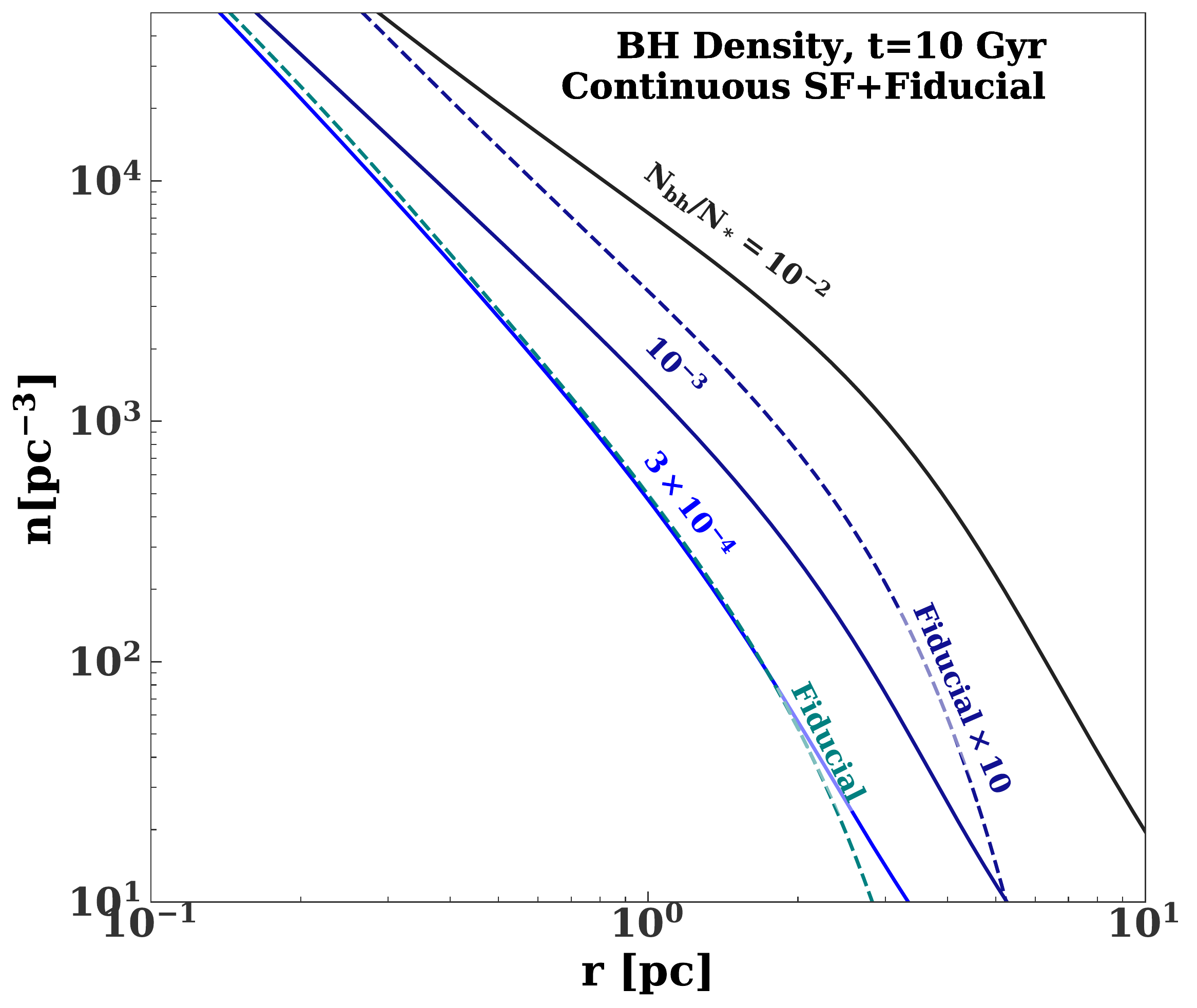}
\includegraphics[width=8.5cm]{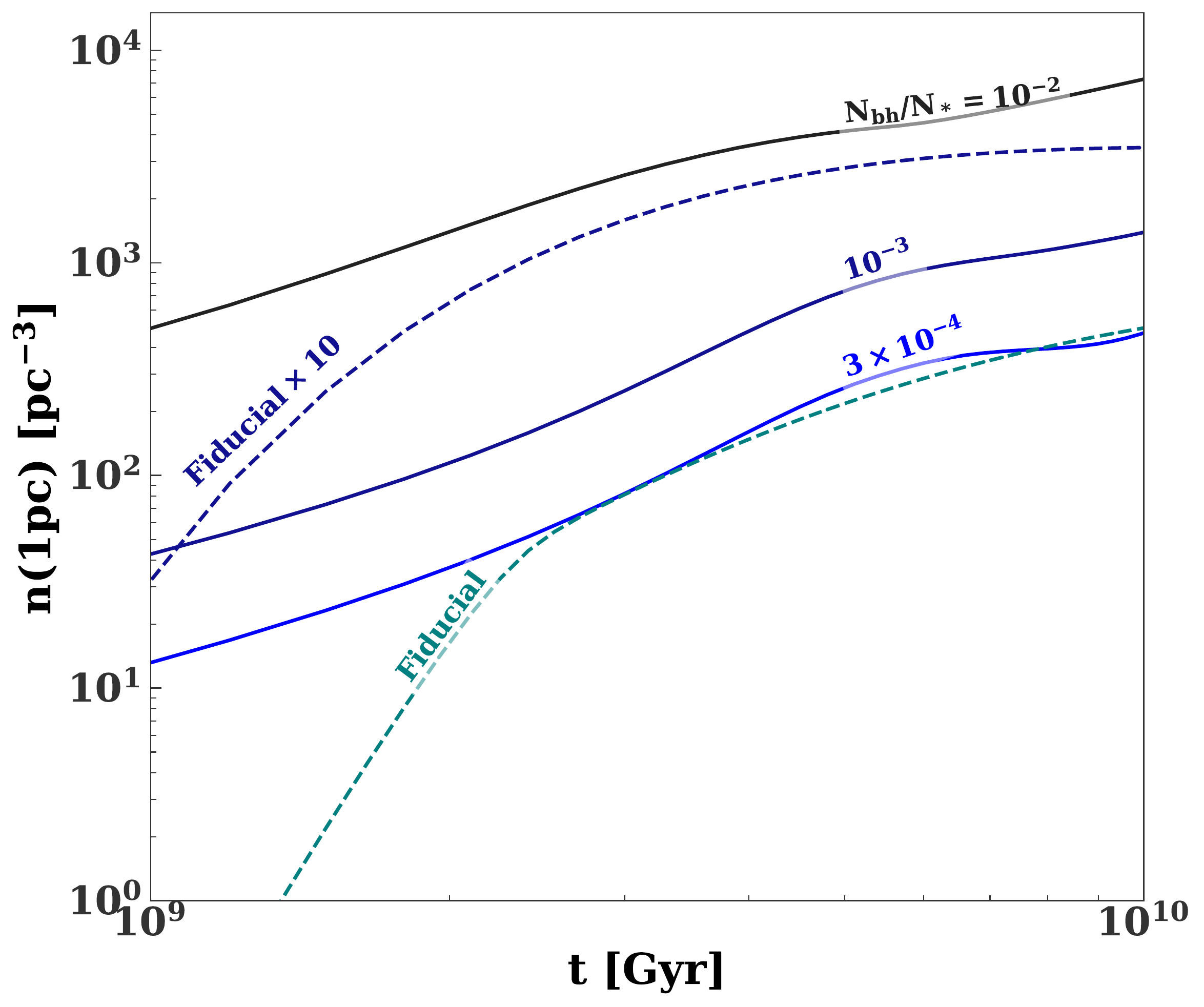}
\caption{\label{fig:compareInsitu} \emph{Top panel}: Density profile of BHs at
10 Gyr under the assumption that the NSC is built up by continuous star formation
at a constant rate with a spatial profile identical to the present-day stellar
population; colors denote different ratios of BHs to
stars, $N_{\rm bh}/N_\star$.  For comparison, dashed lines show the BH profile
in our Fiducial and Fiducial$\times 10$ scenarios
(Fig.~\ref{fig:stellarBackground2}). \emph{Bottom panel:} Time evolution
of the BH density at $1$ pc, for each of the formation histories in the top
panel.}  \end{figure}

Fig.~\ref{fig:compareInsitu} shows the BH density profile under the assumption
that they form continuously at a constant rate over the age of the NSC, with a
spatial profile identical to the present-day stellar population.  The BH
density at small radii $\lesssim 1$ pc evolves significantly over time, taking
several Gyr to reach a quasi-steady state.

\begin{figure}
\includegraphics[width=8.5cm]{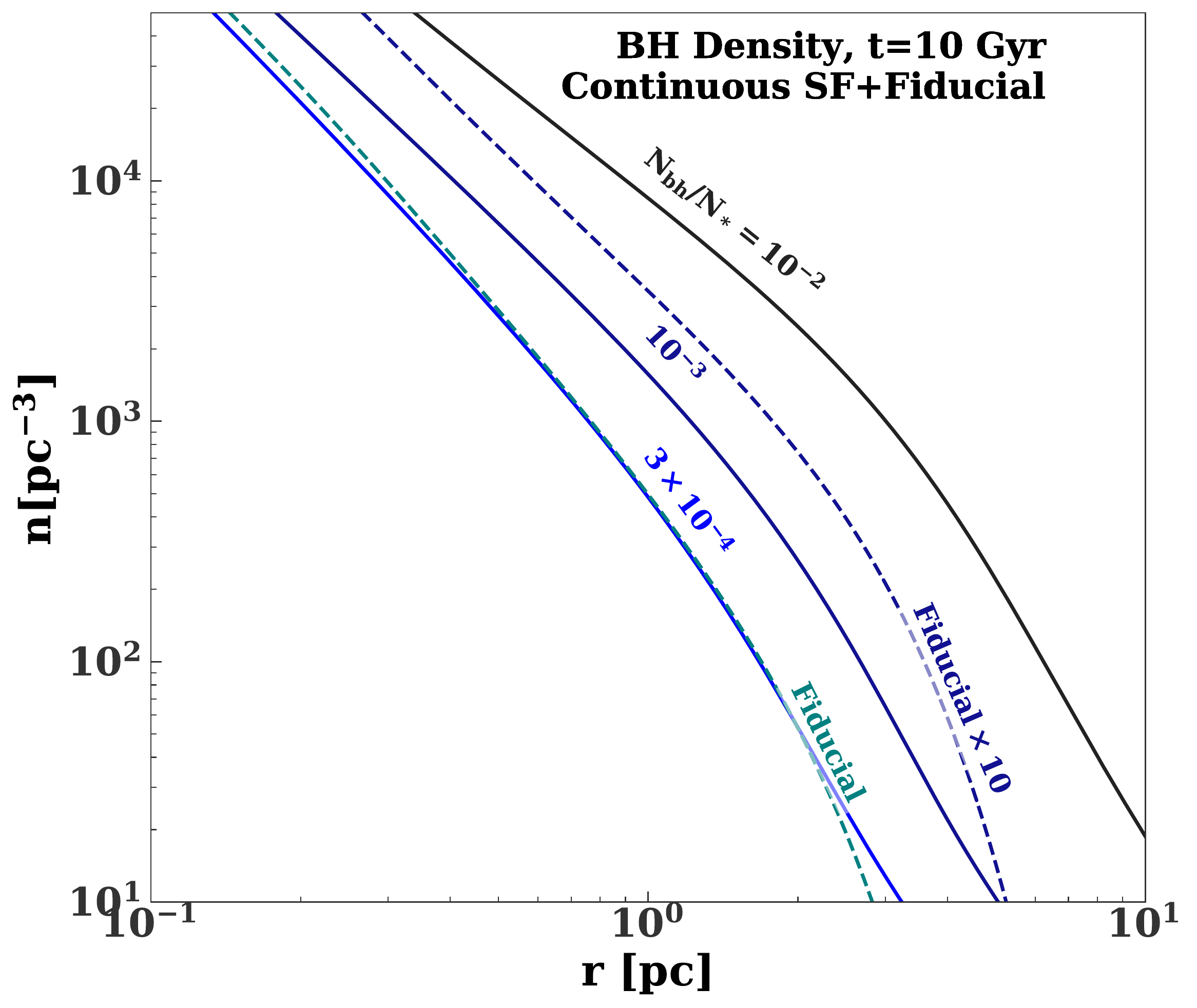}
\includegraphics[width=8.5cm]{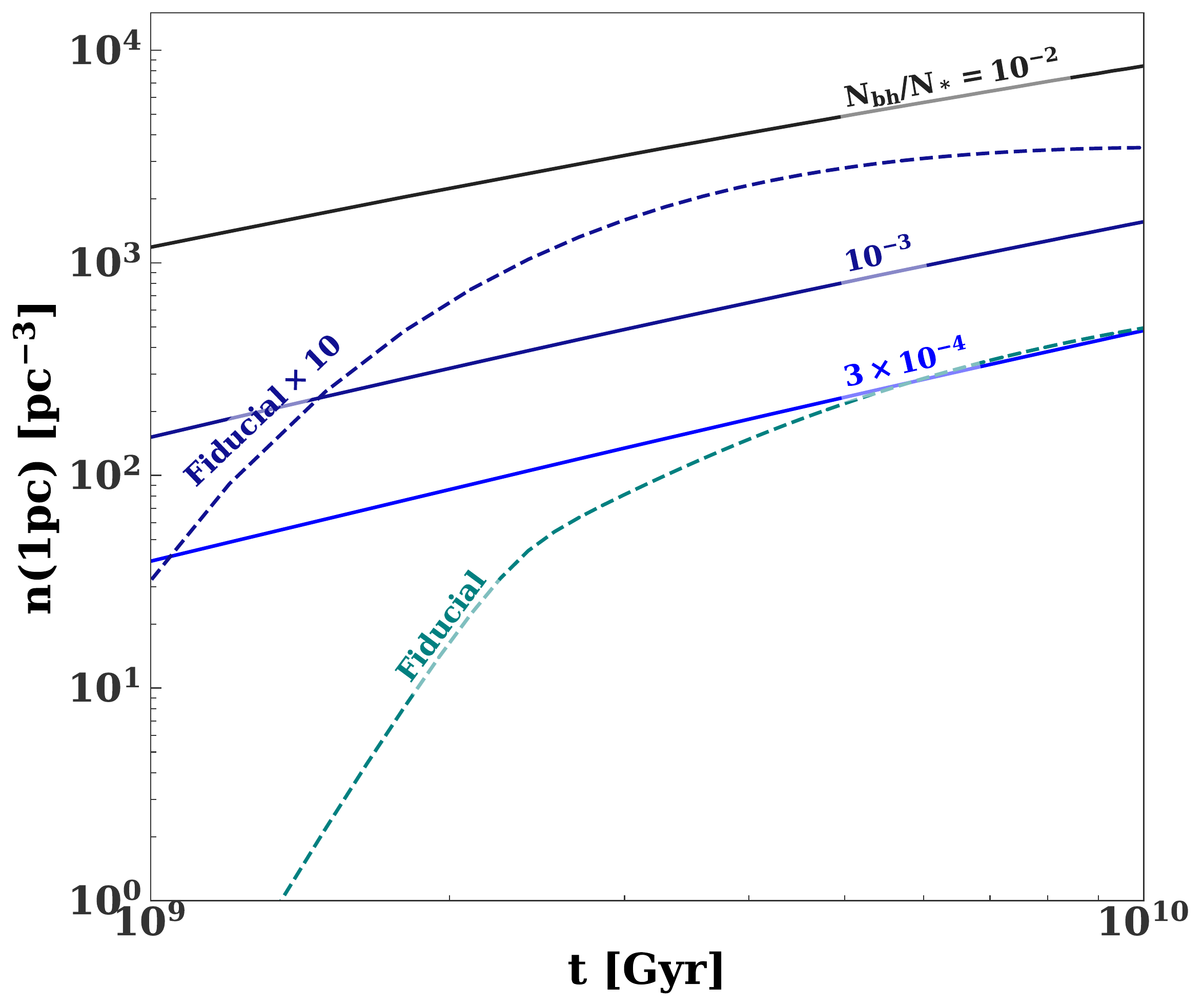}
\caption{\label{fig:compareInsitu2} Same as
  Fig.~\ref{fig:compareInsitu}, except the central SMBH is fixed to be 7\% of
  the total cluster mass at all times, so that it grows with the cluster.}
\end{figure}

Our previous scenarios assumed the central SMBH possesses a fixed mass, $4\times
10^6 \Msun$.  However, if the NSC is built up by continuous star formation,
then the SMBH may grow in concert with the cluster through gaseous accretion or star capture.
Fig.~\ref{fig:compareInsitu2} shows the density profiles of stars and compact
objects at $t = 10$ Gyr if the SMBH mass is artificially fixed at all times to be $7 \%$ of the mass of the NSC. This speeds up the evolution because the velocity
dispersion, and hence the cluster relaxation timescale (eq.~\ref{eq:trx}) is
smaller at early times.  Nevertheless, the final distribution of BHs is
similar to the previous cases (cf.~Fig.~\ref{fig:compareInsitu}).

Overall, we find that the final distribution of remnants after $\sim$10 Gyr is
mostly sensitive to the overall rate of production of BHs versus stars, and is rather
insensitive to the details of the star formation history, or its precise radial
distribution within the NSC.

\subsection{Effects of strong scattering}
\label{sec:strong}
So far we have neglected BHs ejecting stars via strong scatterings in our
models. In this section we quantify this effect, which under some
circumstances can be important for bulk cluster evolution
\citep{lin&tremaine1980}. The volumetric ejection rate at radius $r$ is
\begin{equation}
\dot{n}_{\rm ej}(r)= n_{\rm bh} (r) n_\star (r) \left<\Sigma(v_{\infty}) v_{\infty}\right>
\end{equation}
where $n_{\rm bh}(r)$ is the number density of BHs, $n_\star(r)$ is the 
stellar density, $v_\infty$ is relative velocity at infinity,
$\Sigma(v_\infty)$ is the cross-section for ejection, and the angle brackets
denote an average over the relative velocity distribution. This expression may
be rewritten as

\begin{align}
&\dot{n}_{\rm ej}= I(m_c/m_*) n_{\rm bh} (r) n_\star (r) \sigma (r) \pi b_o^2 \nonumber\\
&b_o= \frac{G (m_{c}+m_*)}{\sigma (r)^2}
\label{eq:ndotEj}
\end{align}
where $m_c$ and $m_\star$ are the masses of the compact object and star
respectively, $\sigma(r)$ is the (1D) velocity dispersion of the compact
objects. The likelihood of ejection increases with the mass of the compact
object, as quantified by the dimensionless number $I$. In an encounter, the
change in the star's velocity is given by
\begin{align}
&\Delta v_{\parallel}=\frac{-2 v_\infty}{1+ x^2}  \frac{m_c}{m_c+m_\star}\\ 
&\Delta v_{\perp}=\frac{2 v_\infty x}{1+ x^2}  \frac{m_c}{m_c+m_\star}\\
&x=\frac{b v_\infty^2}{G (m_c+m_\star)}
\end{align} 
where the first and second lines are the components parallel and perpendicular to the initial relative velocity. The change in the star's specific energy is 
\begin{align}
\Delta E= \frac{1}{2}\Delta v^2+ \mathbf{\Delta v} \cdot \mathbf{v_\star}
\end{align}
where $\mathbf{v_\star}$ is the star's initial velocity. For a star to be ejected $\Delta E$
should at least exceed the specific binding energy of the central SMBH, viz.
\begin{align}
\Delta E \geq \frac{G M}{2 r} =\frac{(1+\delta)}{2} \sigma (r)^2,  \label{eq:ejectCriterion}
\end{align}
where $\delta$ is the logarithmic BH density slope. To determine the
normalization of the ejection rate, we compute a Monte Carlo ensemble of encounters with
different relative velocities, approach angles, and impact parameters.
Assuming a Maxwellian velocity distribution for the stars and black holes, a
uniform distribution of the cosine of the approach angle, and $\delta=1.75$,
the numerical pre-factor ($I$) in equation~\eqref{eq:ndotEj} is 0.1, 1, and
1.3 for $m_c/m_\star=$1, 10, and 50 respectively (see also \citealt{henon1969}).

The total ejection rate may be dominated by stars on eccentric orbits. For a
thermal eccentricity distribution and an $r^{-1.75}$ BH perturber profile, the
ejection rate increases by a factor of $\sim 3.7$ (relative to purely circular
orbits). Then, the ejection rate from strong scatterings is
\begin{align}
\dot{n}_{\rm ej}&\approx 3.7 \pi I(m_c/m_*) n_{\star} (r) n_{\rm bh} (r) \sigma (r)^{-3} G^2 m_c^2 \nonumber\\
&\approx 4 n_\star(r) \ln \Lambda^{-1} \tau_{\rm rx, bh}(r)^{-1},
\label{eq:ndotEj1}
\end{align}
where $\ln \Lambda\approx 15$ is the Coulomb logarithm and we take
$m_c/m_\star=10$. At any radius the time-scale for a star to be unbound from
the central SMBH is approximately four times the local relaxation time of the
BHs. To test how this effect would modify the stellar density, we add an
additional sink term into \textsc{PhaseFlow}. We find that the stellar density
is modified by $\lsim 25\%$ (40\%) outside of 0.01 pc in our Fiducial
(Fiducial$\times$ 10) models. The total number of stars ejected from the cusp
is $\sim$2.7$\times 10^{6}$ in the Fiducial model and $7\times 10^6$ in the
Fiducial$\times 10$ model.  These likely represent upper limits on the
uncertainty caused by our neglect of strong scatterings, as Eq.
\ref{eq:ejectCriterion} represents a generous ejection criterion.

\section{Tidal Capture Binary Formation}
\label{sec:coll}
A close encounter between a star of mass $m_{\star}$ and a compact object of
mass $m_{\rm c}$ can lead to the formation of an XRB through tidal capture.
During pericenter passage, tidal forces transfer orbital energy into stellar
oscillations, capturing the star into an elliptical orbit.

The maximum initial pericenter distance that results in tidal capture,
$r_{\rm capt}$, can be estimated by equating the hyperbolic
orbital energy with the
energy deposited in tides (see Appendix D of \citealt{stone+2017b}, and Appendix \ref{app:coupling} in this paper).  This
condition can be expressed as

\begin{equation}
\mu \frac{v_{\infty}^2}{2}= \frac{G m_\star^2}{r_\star}
\left(\frac{m_c}{m_\star}\right)^2 \left(\frac{r_\star}{r_{\rm capt}}
\right)^{6} T_{2}(r_{\rm capt}, m_\star/m_c), 
\label{eq:rcapt}
\end{equation}
where $r_\star$ is the stellar radius, $\mu$ is the reduced mass, and $T_2$ is
the tidal coupling constant (we only include the dominant $l=2$ modes, as even
for an equal mass binary including the $l=3$ modes only increases
the maximum pericenter resulting in tidal capture by 5\%). For distant
pericenters this may be estimated using the linear theory (see
Appendix~\ref{app:coupling} and \citealt{lee&ostriker1986}). However, for the
closest pericenters relevant for capture, linear theory  underestimates the
tidal coupling constant by a factor of a few. The magnitude of non-linear
effects has been estimated for polytropic models by
\citet{ivanov&novikov2001}, and we adopt their prescriptions for close
pericenters, as discussed in Appendix~\ref{app:nonlinear}.

Fig.~\ref{fig:rcapt} shows the maximum pericenter distance for tidal capture
as a function of the relative velocity at infinity, $v_{\infty}$, normalized to
the stellar escape speed, $v_{\rm esc} =
\sqrt{2Gm_{\star}/r_{\star}}$. The capture radius $r_{\rm capt}$ is typically
$\lesssim 2$ times greater than the characteristic tidal radius $r_{\rm t} \equiv r_{\star}(m_{\rm
c}/m_{\star})^{1/3}$.  Note that tidal capture cannot occur if $v_{\infty}
\gtrsim v_{\rm esc}$ and thus is suppressed at small
radii $r \lesssim 0.1-1$ pc where the velocity dispersion is large; the same
considerations virtually prohibit the tidal capture of giant stars in
the GC. For even closer pericenter passages, inside of the so-called
\textit{disruption radius} $r_{\rm dis} \approx 0.5-1.1 r_t$ (depending on
stellar structure; \citealt{guillochon&ramirezruiz2013}), stars are tidally
disrupted rather than captured.

\begin{figure}
  \includegraphics[width=8.5cm]{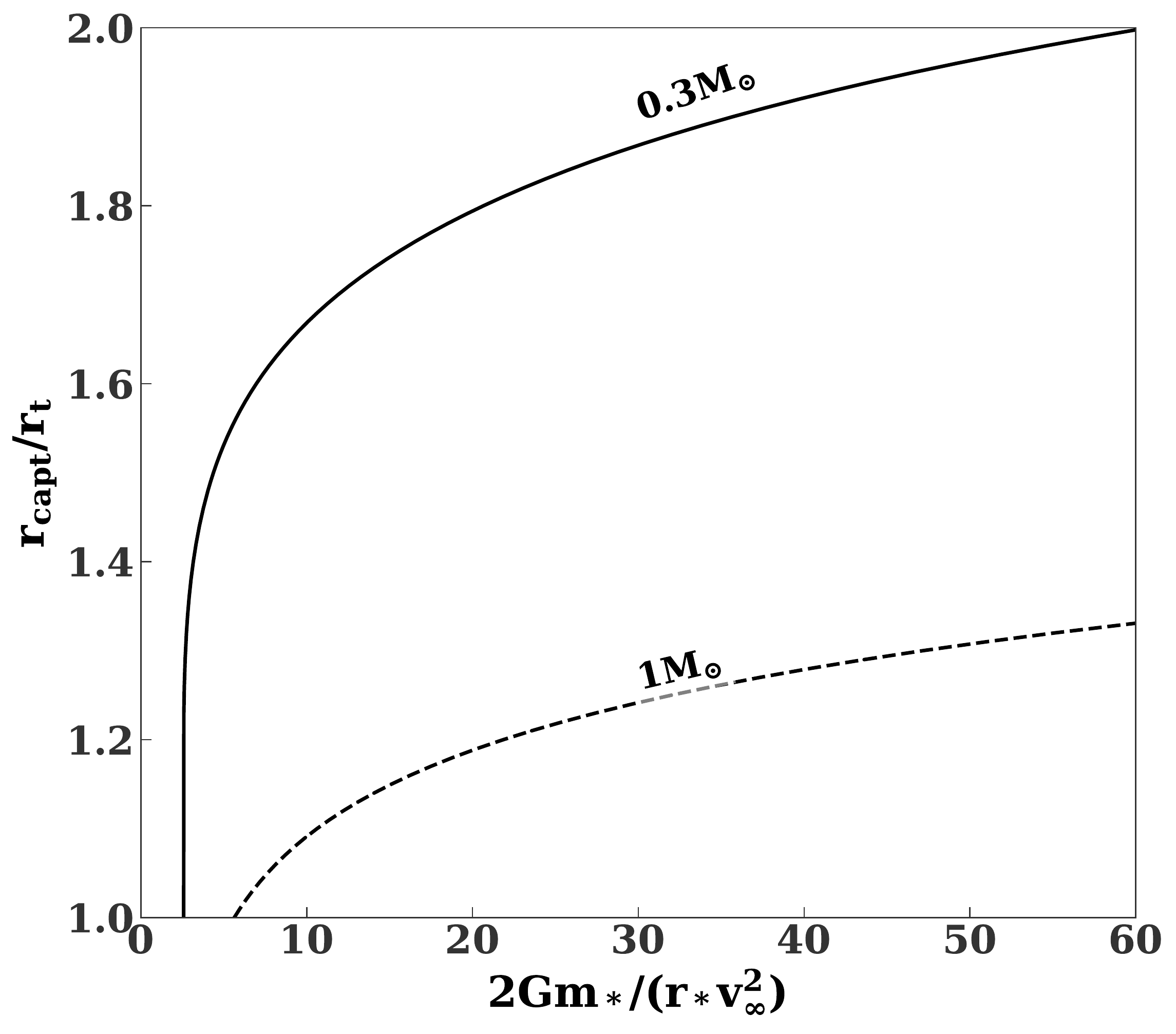}
  \caption{\label{fig:rcapt} Maximum pericenter distance $r_{\rm capt}$
(normalized to the BH tidal radius $r_{t}$) at which a main sequence star can
be tidally captured by a BH of mass $10M_{\odot}$, as a function of the
stellar escape speed (normalized to the relative velocity at infinity,
$v_{\infty}$).  Results are shown for two stellar masses, 0.3 $\Msun$
(\emph{solid line}) and 1 $\Msun$ (\emph{dashed line}). The former is modeled
as an $n=3/2$ polytrope and the latter is modeled as an $n=3$ polytrope.}
\end{figure}

The combined volumetric rate of tidal captures ($r_{\rm dis} \leq r_{\rm p}
\leq r_{\rm capt}$) and disruptions ($r_{\rm p} \leq r_{\rm dis}$) at
Galactocentric radius $r$ is given by
\begin{align}
  &\Gamma(r,t)= \int_{0}^{v_{\rm max} (m_\star)} n_c(r, t) n_{\star}(r, t) v_\infty \pi r_o^2 \nonumber\\
   &\times \left[1+\frac{2 G (m_c+m_\star)}{r_o v_\infty^2 }\right]
   f(v_\infty, r, t) d v_\infty, \nonumber\\
   &r_o={\rm max}[r_{\rm capt} (v_\infty, m_c/m_\star), r_{\rm dis}(m_\star)]
\label{eq:rate1}
\end{align}
where $n_c(r, t)$ is the number density of compact objects, $n_\star(r, t)$ is
the number density of stars, and $f(v_\infty)$ is the distribution of relative
velocities. A hard upper limit to the value of $v_{\rm max}(m_\star)$ is the
stellar escape velocity (for faster relative velocities most of star would
remain unbound from the compact object in any tidal interaction), but in practice $v_{\rm max} (m_\star)$ is the relative velocity such that $r_{\rm capt}=r_{\rm dis}$ in
eq.~\eqref{eq:rcapt}. This may be smaller than stellar escape speed by a
factor of $\sim$2.  We approximate the velocity distribution as a Maxwellian, with a scale
parameter equal to the local velocity dispersions of the two species added in
quadrature.\footnote{In detail the velocity distribution in the Keplerian
potential of the SMBH is not Maxwellian, but this is a good
approximation for our model stellar density profiles (see e.g.
\citealt{alexander&kumar2001}).}  The term in brackets allows for
gravitational focusing, which exceeds the geometric cross section for $r
\gtrsim 0.01$ pc.

\begin{figure} \includegraphics[width=8.5cm]{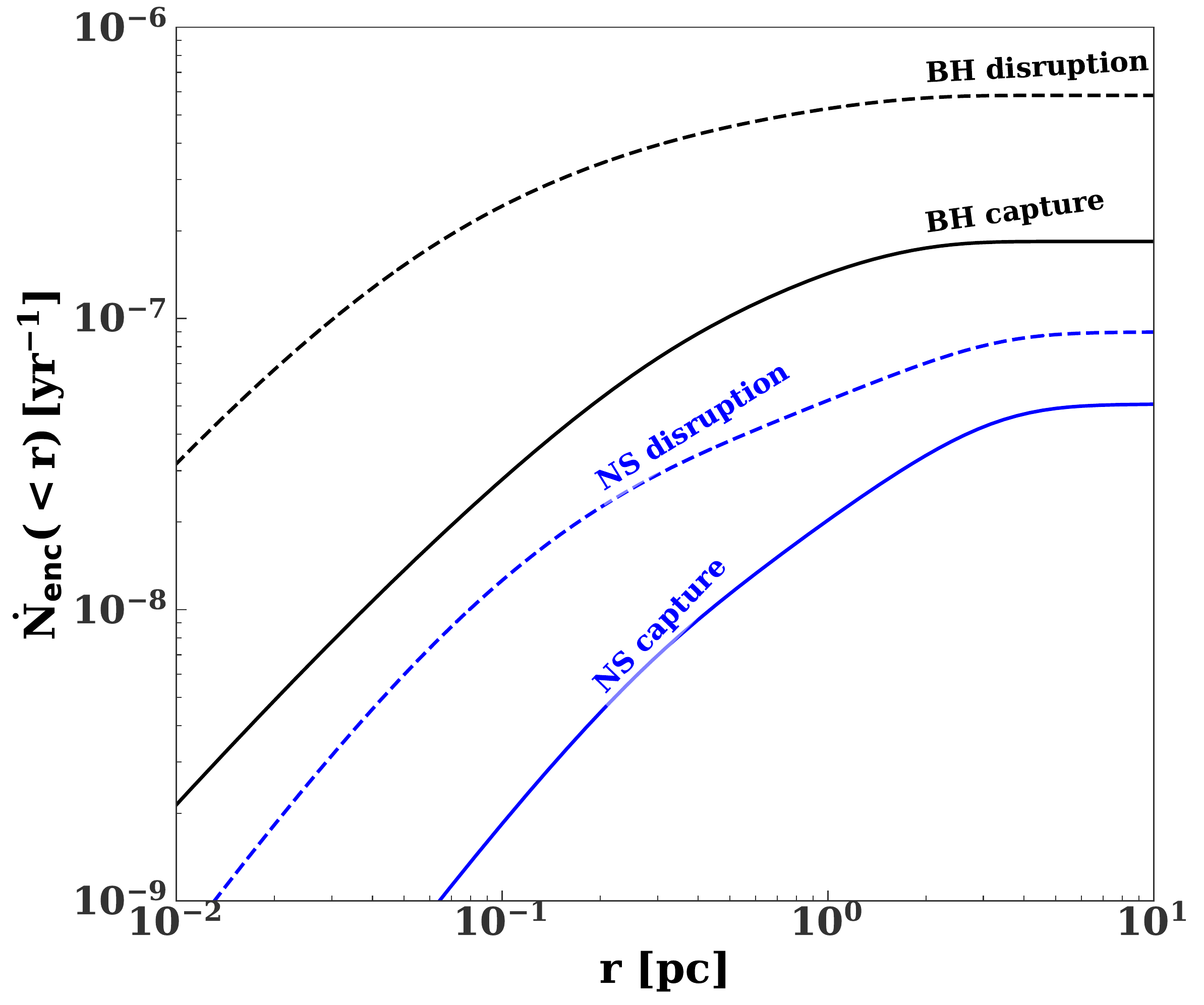}
\caption{\label{fig:colRateDist} Present-day cumulative rate of tidal
disruptions (\emph{dashed lines}) and tidal captures (\emph{solid lines}) inside of 
radius $r$ for our Fiducial scenario.  A total of $\approx 9\times 10^{-7}$
strong tidal encounters occur per year.} \end{figure}

Fig.~\ref{fig:colRateDist} shows our calculation of the present-day rate of
total stellar tidal disruptions (dashed lines) and tidal captures
(solid lines) by BHs and NSs, as calculated using the predictions of our
Fiducial model for $n_{\star}$ and $n_{\rm c}$.  The capture/disruption rate
by BHs exceeds that of NSs by a factor of $\gtrsim 3-10$ across most radii of
interest; this is partially because in the limit of gravitationally-focused
collisions, the rate of captures/disruptions obeys $\Gamma
\propto m_{\rm c}^{4/3}$. The rate of tidal captures is somewhat smaller than the rate
of disruptions, since for typical relative velocities capture occurs over a
narrower range of pericenter distance than does disruption. Some tidal
captures may lead to a series of partial disruptions instead of the formation
of a stable binary, even if the initial pericenter is outside of the
disruption radius. Specifically, significant mass loss from the star is likely
to lead to run-away heating that disrupts the star. In this paper we assume
the star is eventually disrupted if it loses more than $\sim 10\%$ of its mass
after its first pericenter passage (see the discussion in
$\S$~\ref{sec:xray}).

\begin{figure}
\includegraphics[width=8.5cm]{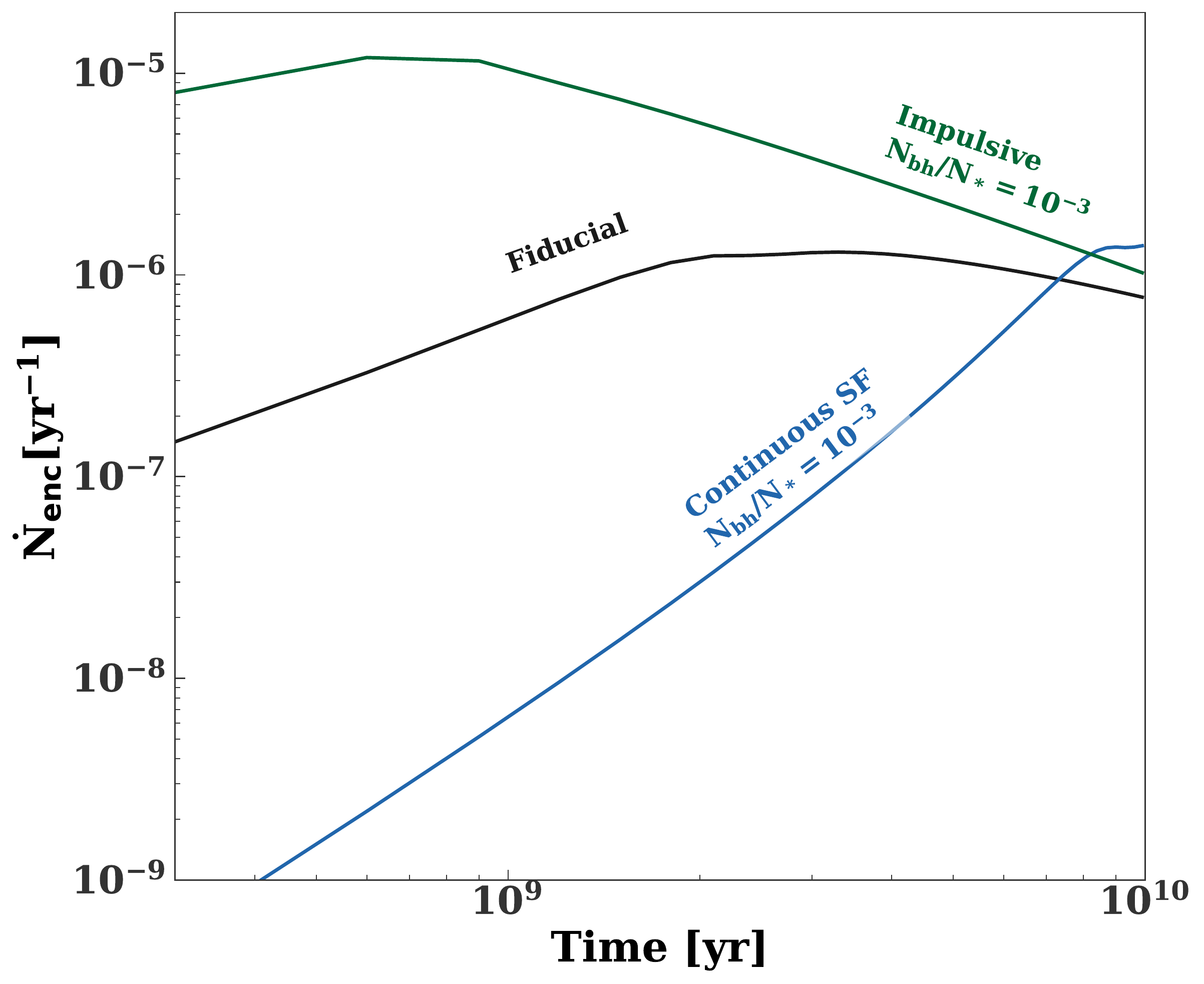}
\caption{\label{fig:colRate} Rate of strong tidal encounters (disruptions plus
captures) as a function of time in our Fiducial model ($\S$~\ref{sec:fiducial})
as well as two of the non-fiducial scenarios ($\S$~\ref{sec:nonfiducial}).}
\end{figure}

Fig.~\ref{fig:colRate} shows the tidal capture and disruption rate of stars by
BHs as a function of time for different star formation histories corresponding to
our Fiducial ($\S\ref{sec:fiducial}$) and non-fiducial scenarios
($\S\ref{sec:nonfiducial}$). In the Fiducial scenario, with compact remnant
injection inside of $\approx 0.3$ pc, the encounter rate increases for the
first $\sim$3 Gyr, as the number of compact objects increases. The rate then
declines slightly as the compact objects reach a steady state on small scales, while
the population of pre-existing low mass stars are pushed outwards to larger
radii.  The non-fiducial scenario with impulsive injection of compact remnants
and stars shows qualitatively similar behavior, but with the encounter rate
peaking much earlier in time.  Finally, in the non-fiducial scenario of
continuous star formation, the encounter rate monotonically increases.

\subsection{Tidal Capture and Circularization}
\label{sec:circ} 
As described by \citet{stone+2017b}, there are three possible outcomes of a tidal capture: (1)
the star continues to lose energy at each pericenter passage, until its orbit
is circular; (2) the binary is perturbed by another star or compact object
before circularization is complete; (3) the star inflates due to tidal
heating, and is destroyed in a series of partial tidal disruptions.

Circularization of the binary can be interrupted (option 2) if the initial
pericenter of the encounter is sufficiently large, in which case the tidal
energy transfer is weak and the star barely captures into a highly elliptical
orbit. In the limit of very large post-capture apocenter, an encounter with another star
will perturb the orbital angular momentum faster than circularization can
occur. Such encounters generally increase the angular momentum of the
binary (since there is more phase volume at larger angular momenta), derailing
the circularization process. However, comparing the time-scales for
circularization and angular momentum diffusion (\citealt{stone+2017b}; their
eqs.~21, 25), we find that circularization is slower than the outwards angular
momentum diffusion time from stellar interactions for only a extremely narrow
range of pericenters, within $10^{-3}$ of the maximum value for capture.  Only
a tiny fraction of tidally captured binaries will be perturbed by a third star
before they circularize.

\begin{figure}   
\includegraphics[width=8.5cm]{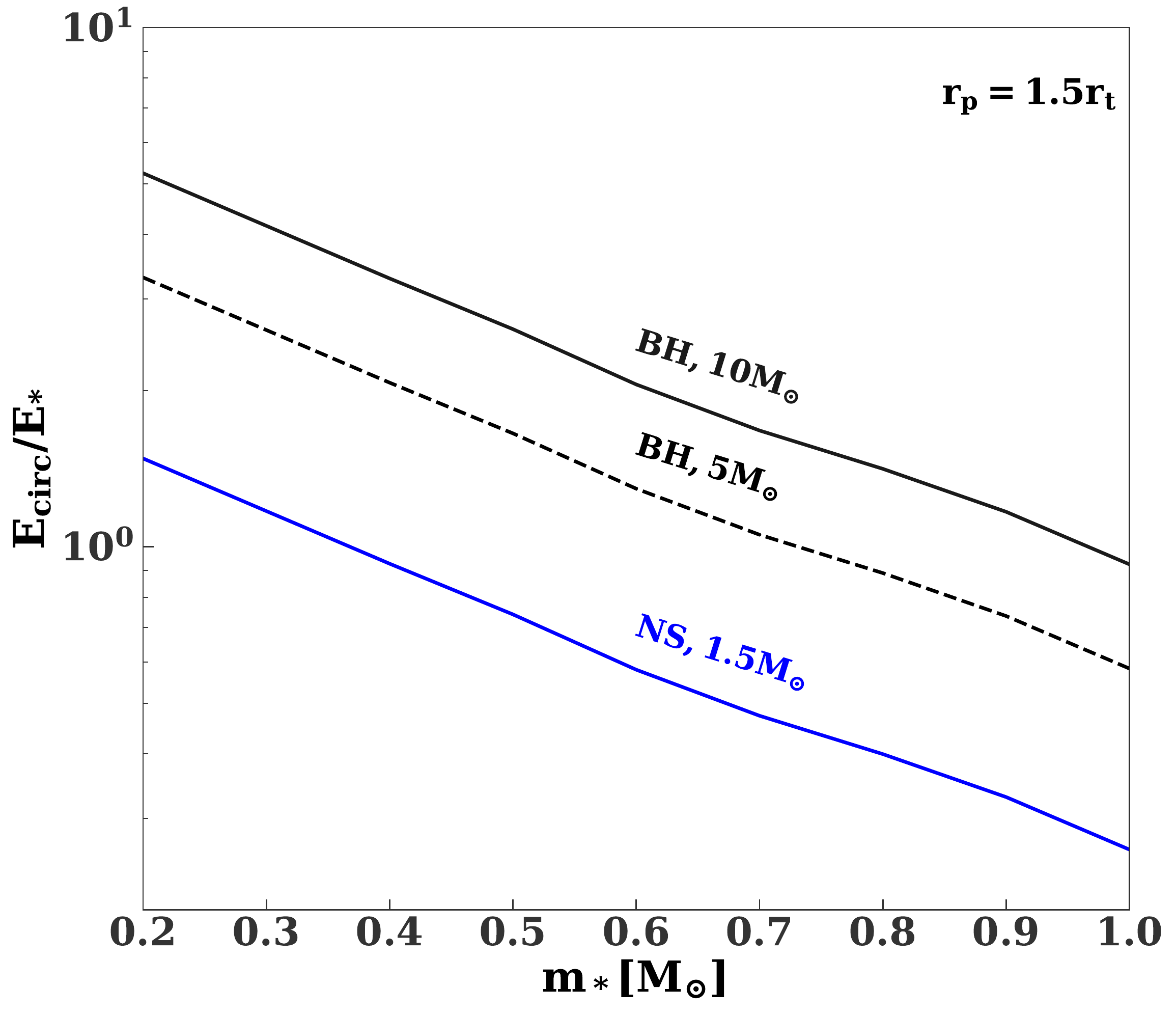}
\caption{\label{fig:ncirc}  Ratio of the
required to circularize the star into a binary with a NS or BH ($E_{\rm
circ}$) to the total (internal + gravitational binding) energy of the star
($E_{\star}$) as a function of stellar mass. Although $E_{\rm circ} \gtrsim
E_{\star}$ across much of the parameter space, tidal capture is not
necessarily fatal for the star because the circularization energy can be
deposited by modes primarily in the outer layers of the star, where it is
likely to drive non- destructive mass loss.}
 \end{figure}

Another hazard for a tidally captured star is a string of partial disruptions
due to the energy deposited by tides and tidal stripping near pericenter.
Complete destruction of the star is energetically allowed if the energy
released during circularization $E_{\rm circ}$ exceeds the total
(internal + gravitational binding) energy of the star $E_{\star}$.  As shown in
Fig.~\ref{fig:ncirc}, a star captured by a black hole necessarily has $E_{\rm
circ} \gtrsim E_{\star}$ (e.g. \citealt{kochanek1992,alexander&morris2003}).
The energy required for a star to circularize around a NS is smaller than the
BH case, but still can be comparable to the energy of a low mass star.

However, even if $E_{\rm circ} \gtrsim E_{\star}$, this does not necessarily
mean the star will be destroyed.  If a significant fraction of the mode energy
is thermalized near the stellar surface (e.g. as non-linear oscillations
steepen into shocks), then it could be carried outwards in a super-Eddington
wind \citep{fuller&lai2011,fuller&lai2012,wu2018}. Whatever remains of the
star following this process would then still circularize, albeit with a
lower mass and potentially higher entropy than its original state prior to
being captured.

The star will lose mass during the circularization process (either due to mode
dissipation or direct dynamical stripping at pericenter). The time-scale for
mass loss is shorter than the thermal time-scale of the star and its radius
will grow adiabatically \citep{linial&sari2017}. As the star grows tidal
dissipation becomes stronger, potentially leading to run-away heating and
disruption of the star \citep{kochanek1992}. If the mass loss occurs primarily
from the side of star closer to the compact object, the pericenter can grow
faster than the stellar radius averting the run-away. However, this effect
would only become important in nearly equal mass binaries in which the l=3
mode enhances (reduces) the displacement on the near (far) side of the star
\citep{manukian+2013}.

Another important issue is the time-scale over which the mode energy deposited
into the star is dissipated. \citet{mardling1995} has argued that a tidally
captured star necessarily undergoes a random walk in eccentricity, since mode
oscillations from consecutive pericenter passages would interfere with each
other, leading to chaotic exchange of energy between the orbit and the star
that would likely lead to the latter's disruption. However, this will not
occur if the mode energy is dissipated over the course of a single orbit. This
can plausibly occur via non-linear mode-mode couplings
\citep{kumar&goodman1996}, especially for the large amplitude modes that will
be excited by tidal capture in the GC, where the energy deposited into the
star on the first pericenter passage is $\gsim$ 10$^{47}$ erg. For example,
\citet{kumar&goodman1996}  show that the $f$-modes excited in low mass stars
can dissipate energy on a time-scale of 30 $\left(E/10^{45} {\rm erg}\right)$
days, where $E$ is the energy deposited into the star. This would be shorter
than the  orbital period of a captured low mass star ($\gsim 5$ days). Higher
mass stars  can dissipate energy even more efficiently by resonantly exciting
g-modes, and nonlinear oscillations may dissipate their energy even faster by
steepening into shocks.

The long-term evolution of a highly eccentric tidal capture binary remains an
open question, and its solution is beyond the scope of this work. For the
remainder of this paper, we assume that tidal capture binaries are in fact
able to circularize without being destroyed, so long as no more than 10\% of
the star's mass is lost on the first pericenter passage, but this assumption
must be examined in future modeling.

\subsection{X-ray Binary Formation and Evolution}
\label{sec:xray}

Once the star circularizes into an orbit around the compact object, the binary
semi-major axis $a$ will be roughly twice the pericenter radius of the captured star.\footnote{The relation $a=2r_{\rm p}$ will be exact if angular momentum and mass are conserved during circularization, except for a correction for stellar spin \citep{lee&ostriker1986}.} The orbit will then decay over long timescales due to gravitational wave emission\footnote{In principle spin-down by magnetic braking also contributes to angular momentum losses from the star. However, there is considerable uncertainty in the spin-down rate for high rotation speeds in contact binaries. Empirically, magnetic braking is sub-dominant to gravitational wave emission in BH binaries, as otherwise one predicts a
population of bright, persistent short period BH LMXBs that are not observed
(see \citealt{yungelson+2006,ivanova&kalogera2006}). By analogy with cataclysmic variables, magnetic braking is likely also sub-dominant in NS systems with periods $\lesssim$ 3 hours.}, such that $a$ decreases according to   
\begin{align}
\frac{\dot{a}}{a}=2 \frac{\dot{J}_{\rm GW}}{J} = -\frac{64}{5} \frac{G^3}{c^5}
\frac{m_\star       m_{\rm c} (m_\star+m_{\rm c})}{a^4}.   
\end{align}
Here $J$ is the circular orbit angular momentum, and $\dot{J}_{\rm GW}$ is the quadrupole-order rate of angular momentum radiation \citep{peters1964}.  Once the system enters Roche-lobe contact, the subsequent evolution of the semi-major axis and mass accretion rate onto the compact object obey (e.g. \citealt{frank+2002})
\begin{align}
  &\frac{\dot{m}_\star}{m_\star}= \frac{\dot{J}_{\rm GW}}{J}
  \left[1.2-\frac{m_\star}{m_c}\right]^{-1} \nonumber \\
  &\frac{\dot{a}}{a}=2 \frac{\dot{J}_{\rm GW}}{J}-2 \frac{\dot{m}_\star}{m_\star} \left(1-\frac{m_\star}{m_c}\right),
  \label{eq:binEvolve}
\end{align}
where we have assumed the star maintains thermal equilibrium, i.e.~that its radius follows the main-sequence, $r_{\star} \propto m_\star^{0.8}$.

\begin{figure}
\includegraphics[width=8.5cm]{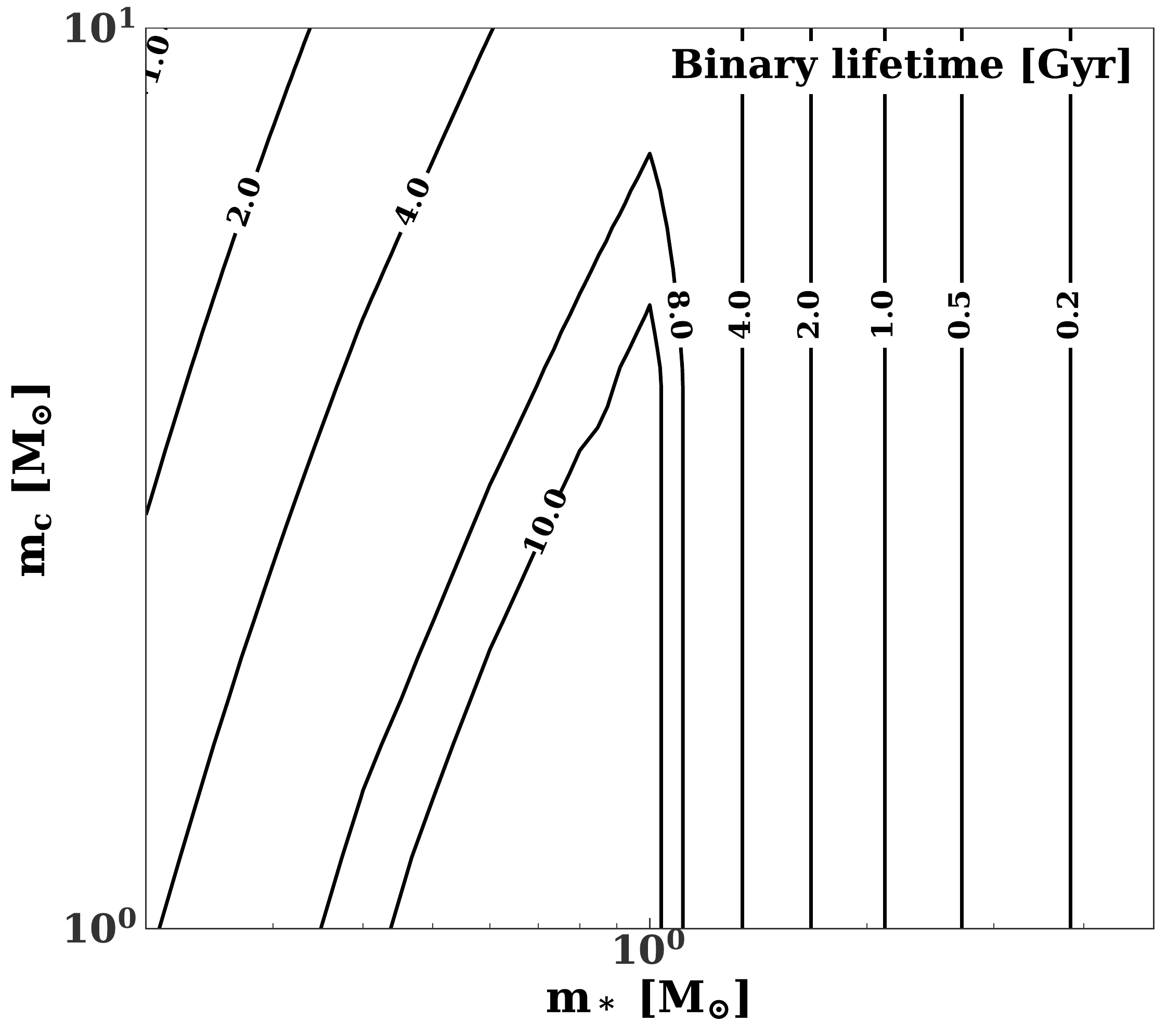}
\caption{\label{fig:lifetime} Binary lifetime after Roche-Lobe contact is
reached as a function of the stellar mass ($m_{\star}$) and compact object mass
($m_{\rm c}$). The binary lifetime is defined as the interval over which the
following criteria are met: (1) the companion mass still exceeds 0.1 $\Msun$
and (2) the star has not evolved off the main sequence.  We use equation 5
from \citet{hurley+2000} for the main sequence lifetime (and assume a solar
metallicity star).}
\end{figure}

Fig.~\ref{fig:lifetime} shows the binary lifetime after the star enters
Roche-Lobe contact as a function of the masses of the star and compact
remnant.  The lifetime is defined as the interval over which (1) the star has
not yet evolved off the main sequence and (2) the star's mass still exceeds
0.1$\Msun$.  The last condition is motivated by the fact that once $m_{\star}
\lesssim 0.1 \Msun$ the star's equation of state changes, resulting in a one
to two order of magnitude reduction in the mass-transfer rate (and an
undetectably dim X-ray source).  Likewise, if the star evolves off the main
sequence, the star and compact object may undergo a common envelope phase,
with an outcome that is uncertain theoretically.

Fig.~\ref{fig:lifetime} shows that the binary lifetime decreases for larger
compact object masses, due to more rapid evolution through gravitational
wave emission.  For low-mass stars ($m_\star \lsim 1 M_{\odot}$), the binary
lifetime also increases with $m_{\star}$ because the tidal radius (and thus
the initial separation) is larger for higher mass stars.  For massive stars
($m_\star \gtrsim 1 M_{\odot}$), the binary lifetime is instead limited by the
main-sequence lifetime.  

The present-day ($t = t_{\rm h} = 10$ Gyr) density of XRBs at radius $r$ is
approximately given by\footnote{Eq.~\eqref{eq:binNum} implicitly assumes
that binaries are visible as XRB at the radii where they are formed.  In
reality, binaries radially diffuse over time after forming, an effect we
quantify in Fig. \ref{fig:binNum}.}
\begin{align}
n_x(r) =  \int_{0}^{t_h}\int_{r_{\rm min}(t)}^{r_{\rm
    max}(t)} d\Gamma(r_p, r, t) d r_p dt, 
\label{eq:binNum}
\end{align}
where 
\begin{align}
&d\Gamma(r_p, r, t) = n_c(r, t) n_\star(r, t) \sigma  2 \pi r_p \times\nonumber\\ 
&\left[I_0(r_p, \sigma)+ \frac{G (m_c+m_\star)}{\sigma^2 r_p} I_1(r_p, \sigma)\right]\nonumber\\
&I_0(r_p, \sigma)= \int_{0}^{v_\infty(r_p)} \frac{v_\infty}{\sigma} f(v_\infty) d v_\infty \nonumber\\
&I_1(r_p, \sigma)= \int_{0}^{v_\infty(r_p)} \frac{\sigma}{v_\infty} f(v_\infty) d v_\infty,
\label{eq:rateCaptPeri}
\end{align} 
is the capture rate per unit pericenter, $v_{\rm \infty} (r_p)$ is the maximum
relative velocity that would result in a capture (the second term in the
brackets dominates). The limits of integration in eq.~\ref{eq:binNum} are the
minimum and maximum initial pericenters for which the binary would be active
today. 

For close pericenters, the star loses a significant fraction of its 
mass via direct tidal stripping, leading to the star's destruction in a series of partial disruptions. Quantitatively, \citet{ivanov&novikov2001}
find that an $n=3/2$ (3) polytrope would lose 10\% of its mass for 
a pericenter of 1.5 (1) $r_t$. Based on these results we also require
\begin{align} &r_p > \begin{cases} 1.5 r_t, & m_* \leq 0.7 \Msun\\
r_t, & m_* >0.7 \Msun \end{cases}.
\end{align} 
Modern hydrodynamic simulations \citep{mainetti+2017} find comparable results
with ten percent mass loss at $r_p/r_t\approx 1.6$ ($r_p/r_t\approx 0.95$) for
n=3/2 (n=3) polytropes. As we use the tidal coupling constants from
\citet{ivanov&novikov2001}, we also use their prescription for stellar
mass loss.

To accurately calculate the tidal capture rate at small Galactocentric radii
(where the rate becomes zero for stellar velocities equal to the local
velocity dispersion $\sigma$), we must integrate over the velocity
distribution.  For a Maxwellian velocity distribution, the integrals over
relative velocity in eq.~\eqref{eq:rateCaptPeri} can be evaluated analytically:
\begin{align} &I_0(r_p,\sigma)=\sqrt{\frac{2}{\pi }} \left[e^{-\frac{v_{\rm \infty}(r_p)^2}{2
\sigma^2}}    \left(-\frac{v_{\rm
\infty}(r_p)^2}{\sigma^2}-2\right)+2\right]\nonumber\\ &I_1(r_p,
\sigma)=\sqrt{\frac{2}{\pi}}\left(1-e^{-v_{\rm \infty}(r_p)^2/2
\sigma^2}\right). 
\label{eq:rateCaptPeri1} 
\end{align}

\begin{figure}
\includegraphics[width=8.5cm]{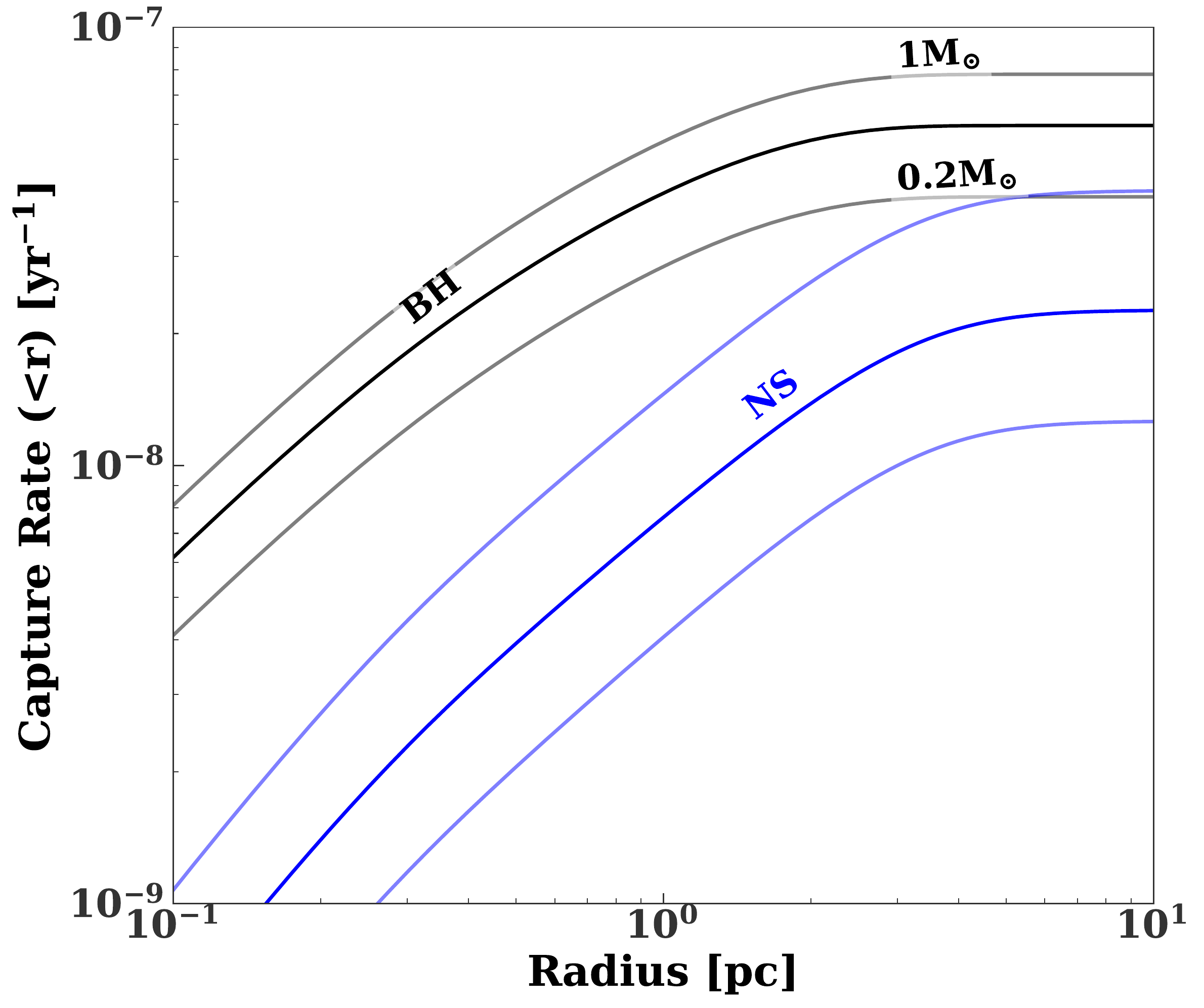}
\caption{\label{fig:captRateSingle} Present-day binary formation rate from
tidal captures of low mass stars by BHs (\emph{black lines}) and NSs
(\emph{blue lines}) interior to a given Galactocentric radius $r$, as
calculated using our Fiducial model for the population of stars and compact
remnants in the GC (Fig.~\ref{fig:stellarBackground2}). The thin lines show
models in which we have fixed the stellar density but consider a single-mass
population of stars with $m_{\star} = 0.2 \Msun$ or 1$\Msun$.  The thick lines
show the capture rate assuming a more realistic Kroupa mass function which
extends from 0.2 -1 $M_{\odot}$.}
\end{figure}

\begin{figure} \includegraphics[width=8.5cm]{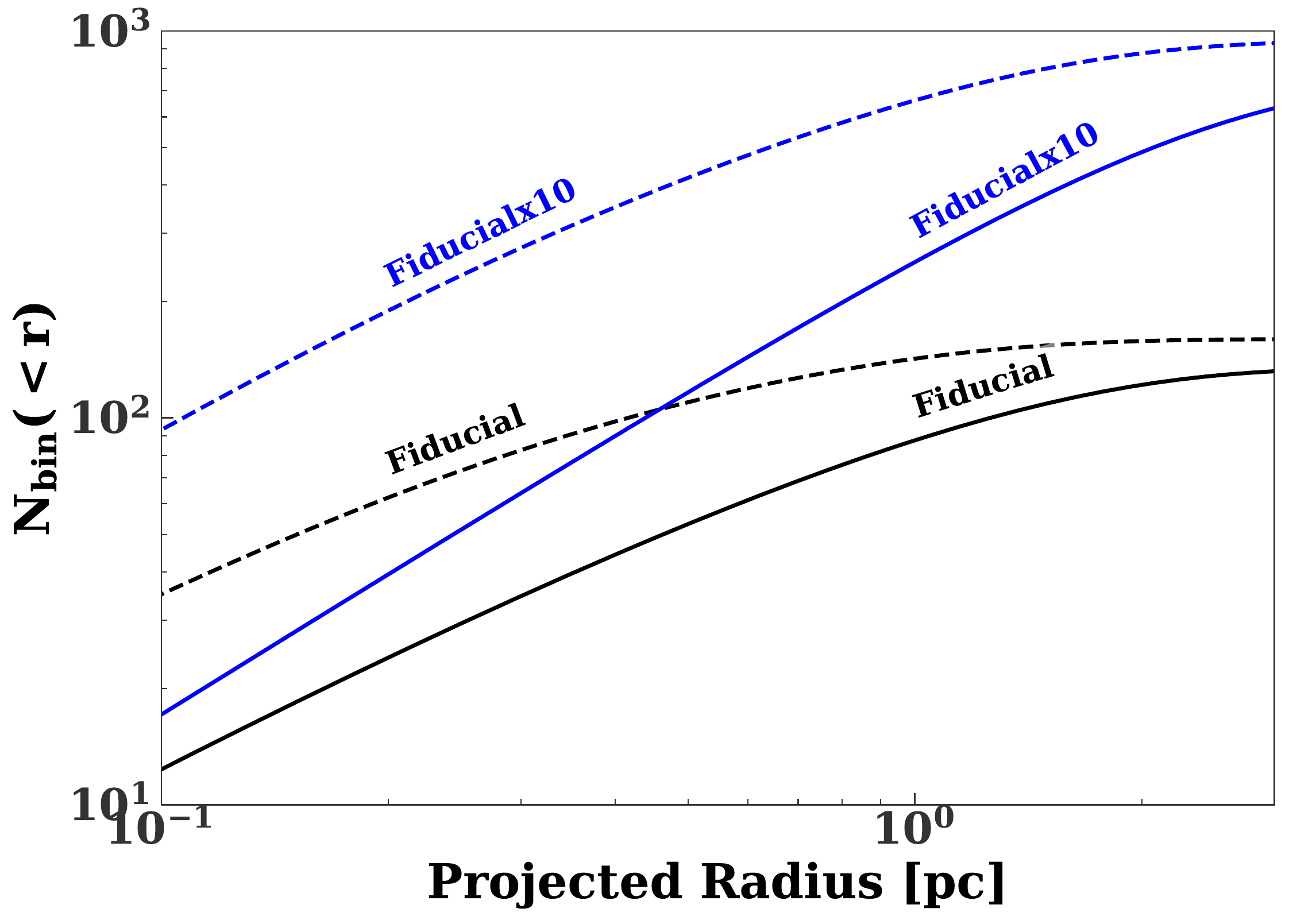}
\includegraphics[width=8.5cm]{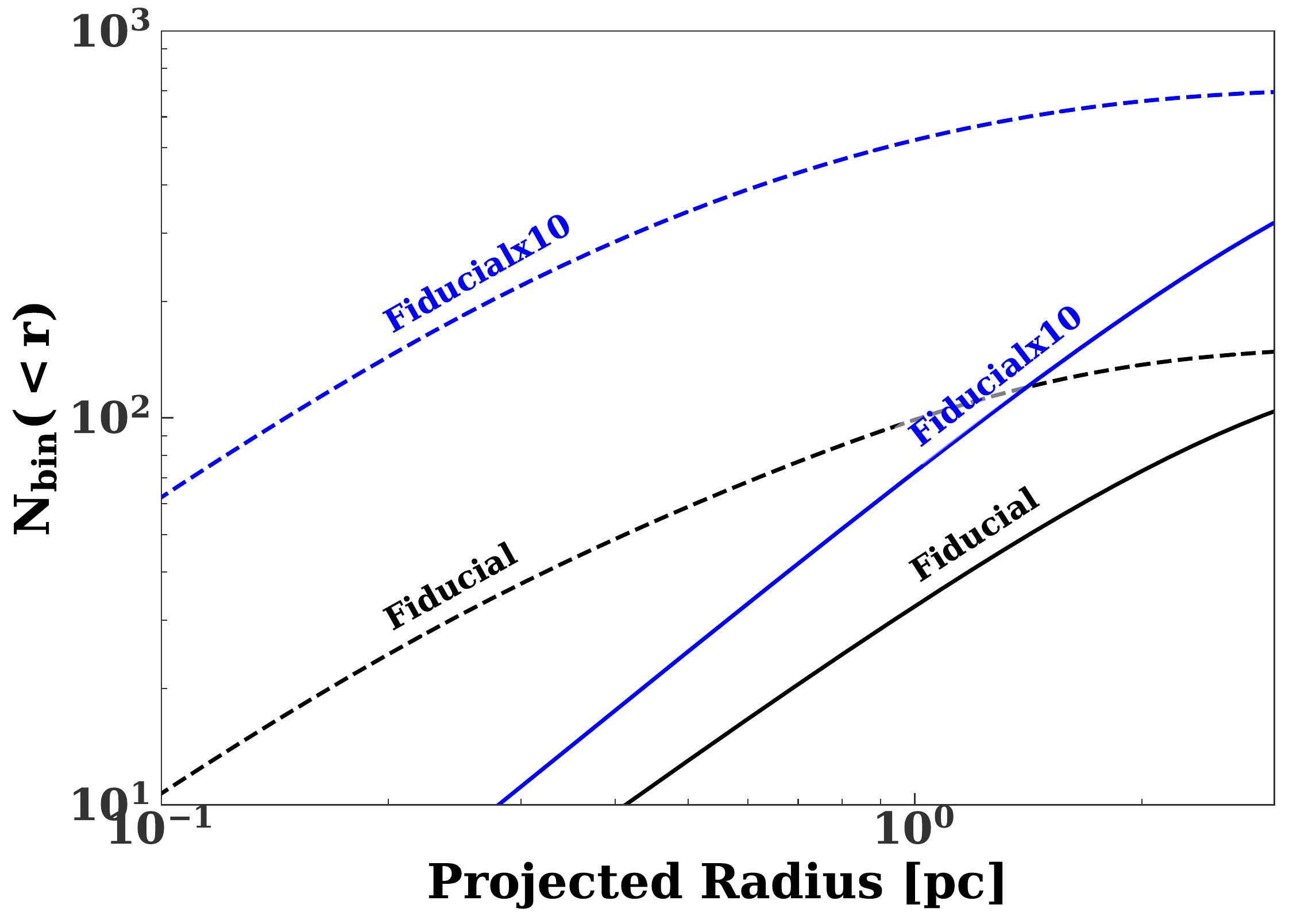} \caption{\label{fig:binNum}
Cumulative number of tidal capture BH-XRB (top panel) and NS-XRB (bottom
panel) predicted inside Galactocentric radius $r$ for our Fiducial and
Fiducial$\times10$ scenarios.  Dashed lines show the distribution of initially-
formed binaries, while solid lines show the final distribution after allowing
for dynamical relaxation of the binary population (these are calculated by
inserting a tracer population with formed distribution of the binaries into
the model snapshot corresponding the mean binary age).}
\end{figure}

Fig.~\ref{fig:captRateSingle} shows our calculation of the cumulative tidal
capture rate inside radius $r$, using our Fiducial model for the
time-dependent density profiles of BHs, NSs, and stars
(Fig.~\ref{fig:stellarBackground2}).  We explore the dependence of the capture
rate on stellar mass by fixing the number density of the stars, but varying
their mass $m_{\star}$.  The \textit{per star} capture rate is larger for
higher mass stars due to their larger tidal radii; however, lower mass stars
are more numerous for any realistic mass function and thus dominate the total
number of formed binaries.

Fig.~\ref{fig:binNum} shows our fiducial model predictions for the present-day
total number of accreting BH and NS XRBs interior to a given radius.  Dashed
lines show the initial radial distribution of the binaries just after forming,
while solid lines show the distribution, allowing for relaxation within the
cluster potential. To calculate the latter, we first find the
\textsc{PhaseFlow} snapshot with a look-back time equal to the mean
binary lifetime. Then, we insert a ``tracer" population of binaries with the
expected initial distribution, and evolve the system forward in time.

Table~\ref{tab:bh} summarizes the predictions of our fiducial models for the
number of tidally-captured XRBs in the central parsec of our GC.  The average
accretion rate for BH (NS) binaries is $10^{-10}$ (3$\times 10^{-11}$) $\Msun$
yr$^{-1}$, corresponding to $5\times 10^{-4}$ ($10^{-3}$) of the Eddington rate $\dot{M}_{\rm
Edd} = L_{\rm Edd}/0.1c^{2}$, where $L_{\rm Edd} = 1.3\times 10^{38}(m_{\rm
c}/M_{\odot})$ erg s$^{-1}$ is the Eddington luminosity. These accretion rates
are generally less than the theoretical critical threshold value below which the disk is thermally unstable,
\begin{align}
\dot{M}_{\rm crit}\approx& ~3.2\times 10^{-11}M_{\odot} \,{\rm yr^{-1}} \left(\frac{m_c}{\rm M_{\odot}}\right)^{0.5} \left(\frac{m_\star}{\rm M_{\odot}}\right)^{-0.2} \\
&\times \left(\frac{P}{1 \,{\rm hour}}\right)^{1.4} , \notag
\end{align}
where $m_c$, $m_\star$, and $P$ are the mass of the compact object, mass of the donor star, and period 
of the orbit, respectively \citep{dubus+1999}. Thus, we expect XRBs formed by tidal 
capture to be transient sources, with long quiescent periods interspersed with bright outbursts.

\section{X-ray observations}
\label{sec:obs}
\citet{hailey+2018} discovered twelve new non-thermal X-ray sources in the
central parsec of our galaxy. Of these, six are solid BH-XRB candidates,
while the identity of the remainder is less certain (they may be either
additional XRBs or MSPs). In principle, many more sources may be present with
luminosities below the \textit{Chandra} detection threshold of $L_x \approx
4\times 10^{31}$ erg s$^{-1}$. Indeed, field BH-XRBs are known with
luminosities as low as $L_{x} \approx 2 \times 10^{30}$ erg s$^{-1}$
(\citealt{armas-padilla+2014}). To estimate the total number of unobserved
XRBs lurking in the central parsec, \citet{hailey+2018} first estimate what
the flux of the minimum luminosity source from \citet{armas-padilla+2014}
would be if it were in the GC (accounting for absorption and instrumental
response). Then, extrapolating the field luminosity function ($N(>F)\propto
F^{-\alpha}$, $\alpha=1.4\pm 0.1$) to this flux, they conclude that the
total number of XRBs could be as high as 300$-$1000.

\subsection{Comparison to Tidal Capture Model} 
\begin{figure}  \includegraphics[width=8.5cm]{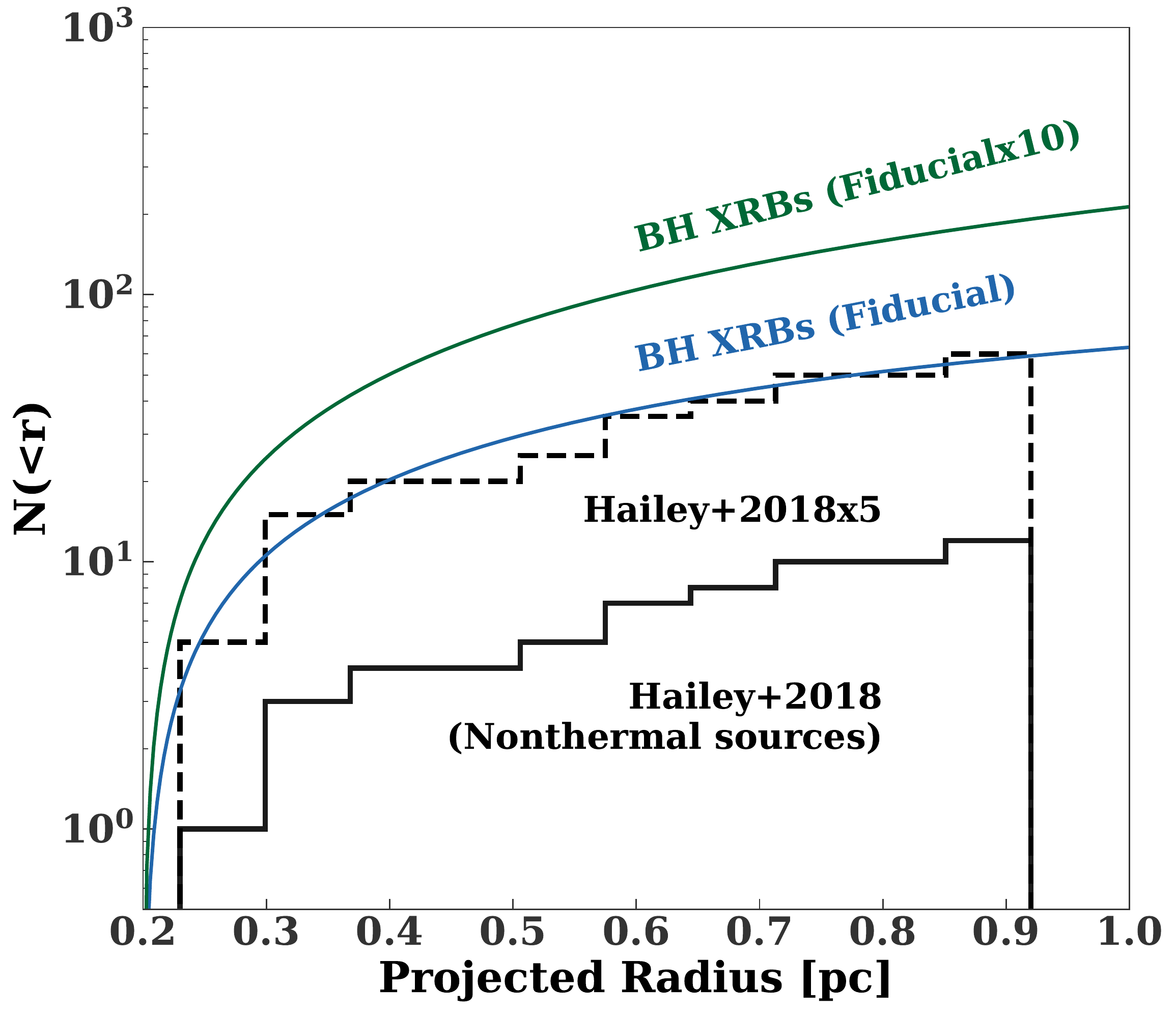}
\caption{\label{fig:surf} Cumulative number of BH XRBs inside projected
Galactocentric radius $r$ from our fiducial models compared with the non-
thermal sources identified by \citet{hailey+2018} (solid black line). We have
included the six sources that may be MSPs instead of BH-XRBs in the latter.
The dashed black line shows the distribution of sources scaled up to match the
normalization of the Fiducial model. The region inside of 0.2 pc is not
included as the population of non-thermal sources is not observationally
constrained there.} \end{figure}

For our fiducial models, we
predict a total of 60$-$200 BH XRBs, which is comparable to the total number
inferred  from observations (Table~\ref{tab:bh}). These numbers would require
the luminosity function to extend a factor of $\sim 3-15$ below the detection
threshold. Fig.~\ref{fig:surf} shows that cumulative radial distribution of
XRBs from our models at radii $\gtrsim 0.2$ pc also agrees well with the
distribution measured by \citet{hailey+2018}. (We only consider binaries
outside of 0.2 pc, as observational limits prevent the identification of
individual sources inside this radius). Specifically, we predict average XRB
surface density profiles $\Sigma \propto r^{-1.4}$ and $\propto r^{-0.9}$ in
our Fiducial and Fiducial$\times$ 10 models, respectively.  These slopes are
close to the measured surface density profile of high S/N sources: $\Sigma
\propto r^{-1.5 \pm 0.3}$ in the radial range $0.2$ pc $\lesssim r \lesssim 1$
pc \citep{hailey+2018}. There are no strong detections ($>100$ counts) outside
of the central parsec, but there are an additional $40$ lower significance
detections ($>50$) counts between 1 and 3.5 pc (though some of these may be
due background contamination).

The numbers we give assume that the stars lose a small fraction of their mass
as their orbits circularize. As previously discussed, the final outcome of
this process is uncertain, and it is possible that some stars may be
completely destroyed. Alternatively, some stars may lose an order unity
fraction of their mass even if they are not destroyed. We re-did our
calculations, assuming that every star loses half of its mass in the
circularization process. The number of BH-XRBs is reduced by a factor of three
under this assumption.


\begin{table*}
\begin{threeparttable}
\begin{minipage}{10cm}
\centering{}
\caption{\label{tab:bh} Number of tidally-captured BH- and NS-XRBs in the GC
predicted for our fiducial scenarios as compared to the observed population.
The ``Observed'' XRBs corresponds to the population detected by
\citet{hailey+2018}, while the ``Extrapolated'' sources account for an
(uncertain) extrapolation of the X-ray luminosity function below the {\it
Chandra} detection threshold (see text for details).}
\begin{tabular}{ccccc}
  Scenario &  BH-XRB  & BH-XRB & NS-XRB & NS-XRB \\ 
  & ($r\le 1$  pc) &  ($r \le 3.5$  pc) &  ($r \le 1$  pc)&   ($r \le 3.5$  pc)\\
  \hline 
  Fiducial &  64 &  110 & 29 & 110 \\
  Fiducialx10 &  210 & 640 & 67 &  370\\
  \hline 
  Observed &  6$-$12 & $\lsim 50$ & 1$-$3 (LMXB), $\leq$6 (MSP) & 3-6 (LMXBs), $\lesssim$ 50 (MSP) \\
  Extrapolated & 300$-$1000 &  & $\lesssim$200 (MSP) & $\lesssim 1000$  (MSP) \\
\end{tabular}
\end{minipage}
\end{threeparttable}
\end{table*}


\subsection{Neutron Stars} 
The number of NS-XRBs formed in the central parsec for our Fiducial and
Fiducial$\times10$ models are $N_{\rm ns} \approx 30$ and $\approx 70$,
respectively.  These numbers, which are a factor of $\approx 2-3$ times lower
than the predicted number of tidal capture BH-XRBs in this region, and
significantly exceed the $\leq$ 3 NS-XRBs observed thus far.

What might suppress the NS population?  First, as in the BH case,
not all NS binaries manifest as luminous XRBs.  Furthermore, some NS-XRBs may
evolve into millisecond pulsars later in their evolution.  Only 3\% of the
known population of MSP have properties which would make them detectable in
the GC \citep{perez+2015}; given that up to six of the observed X-ray sources
in the central parsec could be MSPs, as many as $\sim$200 MSPs could exist in
this region.

The relative number of tidally captured BHs versus NSs binaries also depends
on the fate of massive stars.  Although we have assumed that stars of ZAMS
mass $\gtrsim 25\Msun$ become BHs, in reality there is not a single mass
separating BH and NS progenitors \citep{sukhbold+2016}, and the fraction of
O/B stars that evolve into NSs (as opposed to BHs) may differ between the
field and the GC.  

The NS population could also be reduced by supernova kicks, which would eject
$\sim$40\% of isolated NS formed in the central disk for a Maxwellian kick
velocity distribution, with $\sigma$=265 km s$^{-1}$ \citep{hobbs+2005}.
However, $\sim 40\%$ (10\%) of the NS binaries in our Fiducial
(Fiducial$\times 10$) model come from NSs associated with the old stellar
population, calibrated to the \citet{ivanova+2008} model of globular clusters
(and this already accounts for supernova kicks). Overall supernova kicks would
reduce the number of neutron star binaries by $\sim 25\%$ ($\sim 40\%$) in our
Fiducial (Fiducial$\times 10$) model.

Finally, we note the population of NS XRBs is more sensitive to the initial
conditions than BH XRBs. The progenitors of NS XRBs in our models typically
formed $\sim$7-8 Gyr ago (in comparison to $\sim 4$ Gyr ago for BH XRBs). The cluster
expands over time, so our models assume the NSC was initially more compact
than it is today. However, a small uncertainty in the present day density
translates into a large uncertainty in the initial density. We redid our
calculation for the number of NS XRBs holding the stellar density fixed to the
present day profile. We find that the number of NS XRBs is reduced by a factor
of $\sim$2, while the number of BH XRBs is only reduced by 30\%.

\section{Predictions and implications of our models} 
\label{sec:discuss} 
In this section we summarize various implications of our models, including
properties of binaries and rates of various electromagnetic transients
(including tidal disruption events and stellar collisions). We also estimate
the formation rate of BH-BH binaries due to bound-free gravitational wave
emission. Table~\ref{tab:exotica} summarizes the rates of these processes in
our GC models.

\subsection{Properties of binaries}
\begin{figure}
\includegraphics[width=8.5cm]{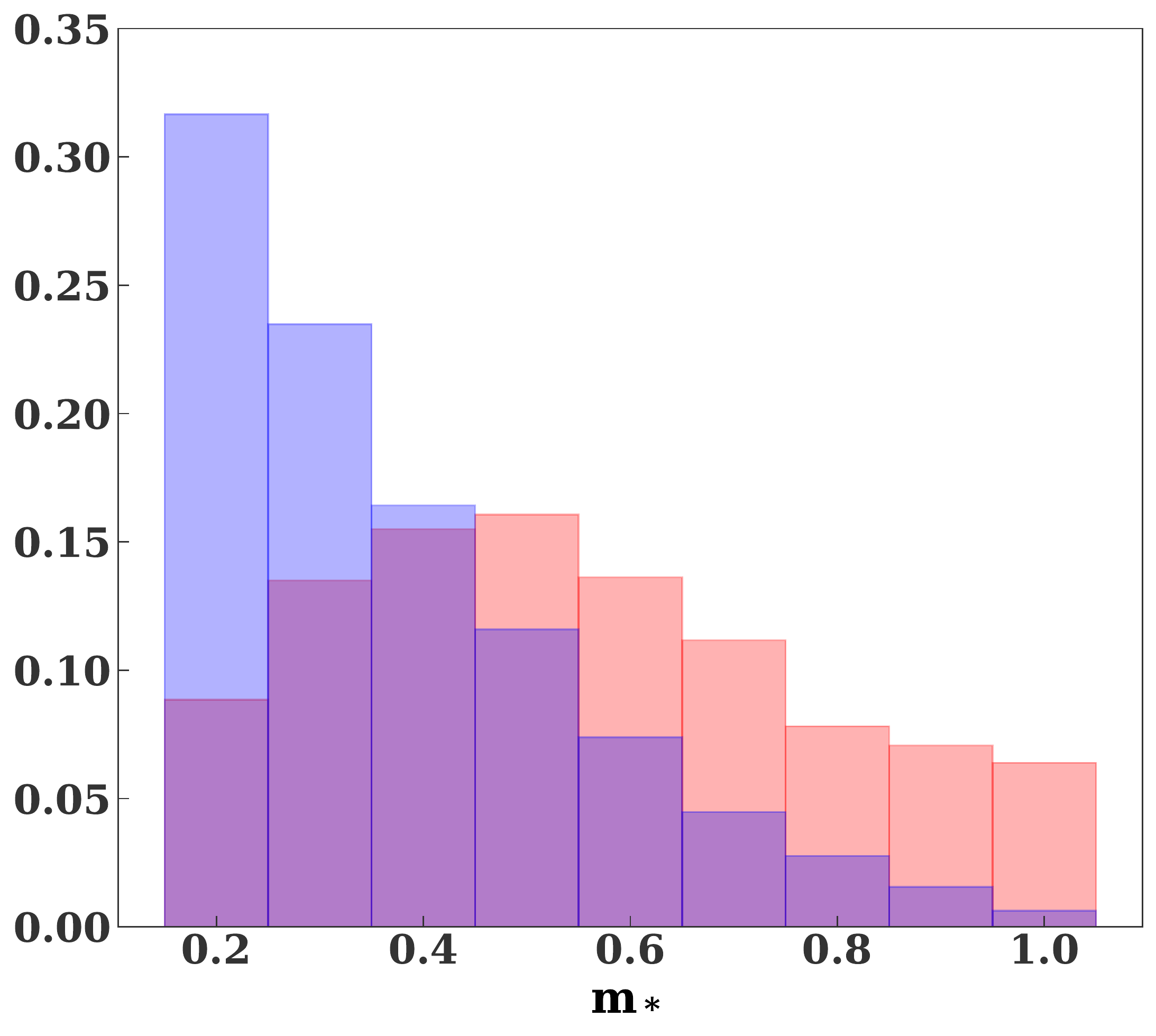}
\caption{\label{fig:mhist} Histogram of companion 
masses of present-day BH XRBs (\textit{blue}) and companion masses of their
progenitors (\textit{red}). The difference between the final and initial
masses is due entirely to post-circularization Roche-Lobe overflow accretion;
additional mass may be lost during the circulization process itself.}
\end{figure}

XRBs formed by tidal capture are necessarily short period systems. The
binaries in our models have main sequence companions with periods of $\lesssim
10$ hours (with a median period of $\sim$3.6 hours). Any future periodicity
identified in the quiescent population would be a powerful discriminant
between tidal capture and other channels (e.g. binary exchange) that can form
long period XRBs. We show a histogram of the companion masses of present day 
BH XRBs in our Fiducial model in Fig.~\ref{fig:mhist}.

In the field such short-period XRBs possess low luminosities of $\lesssim
$10$^{31}$ erg s$^{-1}$ \citep{armas-padilla+2014} which are below the
detection threshold of \citet{hailey+2018} and thus could not be contributing
to the observed population. However, the current sample of short period
BH-XRBs is small (only four are known a with period of less than six hours).

\subsection{Tidal disruptions by the central SMBH} 
Stars may also be tidally disrupted by the central SMBH \citep{hills1975}.
Such tidal disruption events (TDEs) can produce bright electromagnetic flares
\citep{rees1988}. Many candidate  flares have now been detected in optical/UV
\citep{gezari+2006,gezari+2008,van-velzen+2011,gezari+2012,chornock+2014,holoien+2014,arcavi+2014,vinko+2015,holoien+2016,holoien+2016a,blagorodnova+2017},
and X-ray wavelengths (see \citealt{auchettl+2017} and the references
therein).

The total TDE rate due to two-body relaxation has been estimated for a large
{\it Hubble Space Telescope} ({\it HST}) sample of nearby galactic nuclei
\citep{wang&merritt2004, stone&metzger2016}. These authors find that the
average per-galaxy disruption rate is $\sim1-10 \times10^{-4}$ per year.  This
range appears discrepant with observationally inferred TDE rate estimates,
which are often $\sim 10^{-5}~{\rm galaxy}^{-1}~{\rm yr}^{-1}$
\citep{donley+2002, van-velzen&farrar2014}.  While recent work has suggested
that properly accounting for the broad TDE luminosity function \citep{van-velzen2018} may bring observational TDE rates into agreement with theory,
it is worth considering one limitation of the theoretical estimates: in the
smallest galaxies, even {\it HST} observations underresolve the SMBH influence
radius (from which most TDEs are sourced), and moderate inward extrapolation
is needed to calibrate theoretical models \citep{stone&metzger2016}.  The TDE
rates predicted by our Fokker-Planck models have been calibrated off scales
far smaller than the Sgr A$^\star$ influence radius, and are thus a useful
sanity check on TDE rate calculations in general.

\begin{figure}
\includegraphics[width=8.5cm]{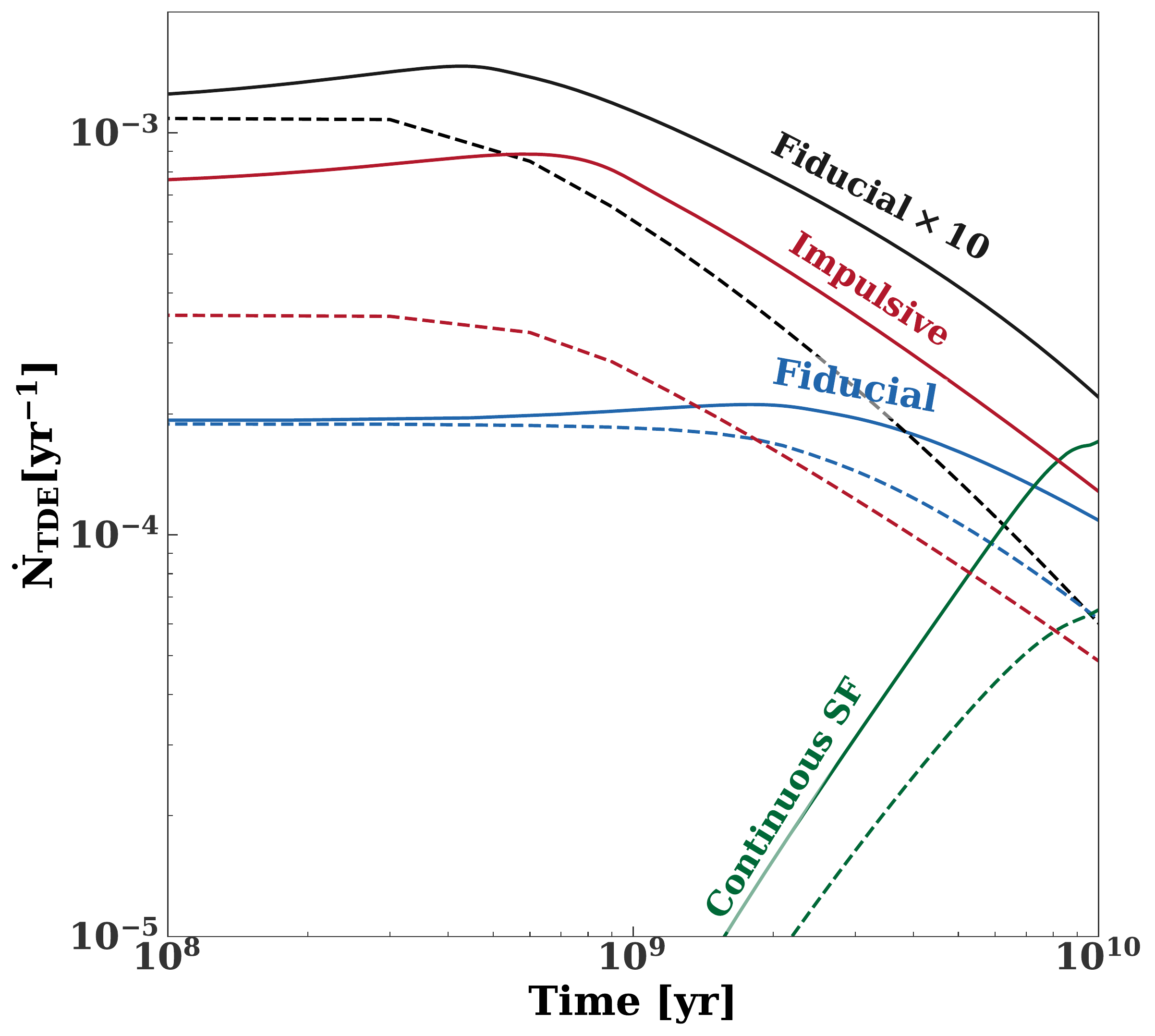}
\caption{\label{fig:tdeRate} Rate of tidal disruption by the central
  SMBH as a function of time for our GC models (Fiducial and Fiducial$\times
  10$). We also show hypothetical models with continuous star formation
  (\emph{green line}), and a single population formed $10^{10}$ years ago
  (\emph{red lines}) (see $\S~\ref{sec:nonfiducial}$). The dashed lines show
  what the disruption rate would be without stellar mass BHs (calculated by
  excluding the compact objects from the angular momentum diffusion
  coefficients).}
\end{figure}

Fig.~\ref{fig:tdeRate} shows the TDE rate for a few different models for the
GC.  The present-day TDE rate in each is $\sim 10^{-4}$ stars per year
($3\times 10^{-5} \Msun$ yr$^{-1}$), similar to previous theoretical estimates
for SMBHs of similar size \citep{wang&merritt2004,stone&metzger2016}.
Unsurprisingly, the \emph{present-day} disruption rate is similar for
different models as they are all tuned to reproduce the present-day observed
stellar density profile.  However, different star formation histories lead to
very different temporal behavior in TDE rates (see also
\citealt{aharon+2016}). In our models of the GC (Fiducial and
Fiducial$\times$10), all of the lower main sequence stars formed impulsively
in the distant past, and the star cluster expands over time. Therefore, the
TDE rate decreases at late times.  In contrast, the TDE rate monotonically
increases in a galactic nucleus that is continuously forming stars (see the
dashed gray line in Fig.~\ref{fig:tdeRate}).

Our current sample of (thermal) TDEs is limited to the low-redshift universe,
but {\it LSST} and {\it eROSITA} are expected to find TDEs out to $z \approx
1$.  The rates of high-$z$ tidal disruption that these surveys find will
therefore carry information on the growth history of nuclear star clusters
\citep{aharon+2016}.\footnote{Although other factors, such as the evolution of the
SMBH mass function, will also contribute - see e.g. \citealt{kochanek2016}).}

The SMBH can accumulate a substantial fraction of its mass by disrupting stars
and accreting compact objects. After a TDE, half of the disrupted star is
bound to the SMBH. If the SMBH consumed half of each disrupted star it would
grow by $3\times 10^5$ (1.4$\times 10^6$) $\Msun$ ($\sim 8-40\%$ of its
present day mass). However, a significant fraction of the initially bound
debris may be lost in outflows, so the mass accreted may be $\lesssim10\%$ of
the disrupted star's mass \citep{metzger&stone2016}. If the SMBH accretes ten
percent of each disrupted star it would grow by $10^5$ ($8\times 10^5$)
$\Msun$. In the last case, most of this mass ($5\times 10^5
\Msun$) comes from consumption of compact objects. For simplicity, we fix the
mass of the SMBH to $4\times 10^6 M_{\odot}$ in our fiducial models.

\subsection{Tidal disruptions by stellar mass compact objects} Stars that
enter the tidal radius of a stellar compact object are also tidally disrupted, powering a
transient flare of electromagnetic emission. We calculate the total rate of
such ``micro-TDEs'' in our Fiducial model to be $\sim 6\times 10^{-7}$ per year
(see Fig.~\ref{fig:colRateDist}). Thus, the micro-TDE rate in the GC is
comparable to the rate from globular clusters, perturbations of wide binaries
in the field, and disruptions induced by natal kicks \citep{perets+2016}.
Because the resulting flare is short-lived (\citealt{perets+2016} estimate the
the viscous time-scale of the debris to be less than a day), it is highly
unlikely any such disruption events would be observable in our own GC today.
However, such events in other galactic nuclei might produce rare short-lived transients
detectable at cosmological distances - for example, ``ultra long'' gamma ray
bursts \citep[GRBs;][]{levan+2014}. Taking into account selection effects, the total
rate ultra-long GRBs may be comparable to the rate of classic long GRBs: $\sim$
10$^{-6}$ per galaxy per year (at $z=0$) after beaming corrections \citep{guetta+2005}.
Interestingly, this is comparable to the micro-TDE rate. However, we note that ultra long
GRBs can also be explained by the core collapse of massive stars \citep{greiner+2015}.

At very small Galactocentric radii, these micro-TDEs may occur without
producing observable accretion flares.  This will occur if the relative
velocity $v_\infty$ between the star and the compact object is too large for
any of the tidal debris to remain bound, i.e. if $v_\infty^2/2 > (m_{\rm
c}/m_\star)^{1/3} Gm_\star / r_\star$ \citep{hayasaki+2018}.  We have excluded
such hyperbolic micro-TDEs from our rate estimates.

The small mass ratio between NSs and main sequence stars means that many
``micro-TDEs'' involving NSs will actually be direct physical collisions,
where a Thorne-Zytkow object may be formed (although the stability of 
such objects remains uncertain).

\subsection{Red giant depletion}
As pointed out by \citet{genzel+1996}, there is a dearth of bright red
giants ($K<10.5$) within $\sim 0.2$ pc of the GC.  There
is a similar dearth of intermediate luminosity ($10.5<K<12$) giants
within $\sim 0.08$ pc. The distribution of fainter stars, on the
other hand, is smooth, and has no holes on small scales.

It has been suggested that collisions of red giants with main sequence stars
and BHs \citep[D09 hereafter]{dale+2009} could cause the observed holes in the
red giant population. D09 find that stripping is only effective in reducing
the brightness of giants in the RGB phase (and has little effect on AGB and
horizontal branch stars). Furthermore, only close pericenters ($r_p\lesssim 15
R_{\odot}$ for a solar type giant) will remove enough material to
significantly alter the evolution of the giant (see also
\citealt{leigh+2016}). They conclude that 2$\times 10^4$ BHs inside of 0.1 pc
are required to explain the observed dearth of intermediate luminosity giants.
The gap in the bright giants is harder to explain, as it would require even
larger numbers of BHs that would make the gap in the intermediate
luminosity giants too large.

\begin{table*}
\begin{threeparttable}
\begin{minipage}{10cm}
\centering
\caption{\label{tab:exotica} Present-day rates of various ``exotic'' collisional
stellar interactions in our GC models.  From top to bottom: physical collisions 
between ordinary stars, close encounters between BHs and red giants that would remove a significant fraction of the latter's 
envelope ($r_p\lsim 15 R_{\odot}$), disruptions of ordinary stars by the central SMBH and by smaller mass remnants, BH-BH binary formation by bound-free gravitational wave emission, ejection of stars from the GC in strong scatterings with BHs.}
\begin{tabular}{l|l|l}
  Interaction &  Fiducial & Fiducial$\times$ 10\\ 
  \hline 
  Star-star collisions [yr$^{-1}$] & $7\times 10^{-6}$ &  $7\times 10^{-6}$ \\
  BH-Red giant collisions [yr$^{-1}$ giant$^{-1}$ at 0.1 pc] & 5$\times 10^{-11}$  & 1.5$\times 10^{-10}$ \\
  micro-TDEs (BH) [yr$^{-1}$]& 6$\times 10^{-7}$ & $2\times 10^{-6}$\\
  micro-TDEs (NS) [yr$^{-1}$]& $9\times 10^{-8}$ & $3\times 10^{-7}$\\
  TDEs (SMBH) [yr$^{-1}$] & $10^{-4}$ & 2$\times 10^{-4}$\\
  GW bound-free captures (BH-BH) [yr$^{-1}$] & 1.4$\times 10^{-10} $ & 1.1$\times 10^{-9}$\\
  Ejection of stars by strong scattering [yr$^{-1}$] & 3$\times 10^{-4}$ &  $10^{-3}$
\end{tabular}
\end{minipage}
\end{threeparttable}
\end{table*}

In our Fiducial (Fiducial$\times 10$) model the number of BHs inside 0.1 pc is
1200 (3600), much smaller than the number required to explain the depleted
giants.
The intermediate luminosity  giants are $\sim$2-3 solar mass stars that spend
$\lsim$100 Myr on the red giant branch. The time-scale for close encounters only
becomes comparable to the giant lifetime inside of $\sim$0.01 pc. We conclude
that it is difficult to account for the depletion of red giants by collisions
with BHs alone.  However, there are many alternative explanations for the dearth of red
giants in the literature. For example, red giants may be destroyed by collision
with a clumpy gas disk \citep{amaro-seoane&chen2014, kieffer&bogdanovic2016}.

Ordinary stars may also collide with each other. We calculate the present-day
rate of star-star collisions outside of 0.1 pc to be 7$\times 10^{-6}$ per
year.

\subsection{Two body BH-BH binary formation} 
Close encounters between BHs can result in the formation of close binaries,
either via three-body interactions or two-body gravitational wave bound-free
emission \citep{antonini&rasio2016}. GW capture is generally sub-dominant to
three-body processes, but it is one of the few ways to produce LIGO
sources with a non-negligible eccentricity. All else being equal, eccentric
sources are louder and would be detectable to larger distances. Additionally,
eccentric sources can sometimes provide more stringent tests of strong field gravity, as
a larger fraction of the energy is emitted when the source is moving at high
velocities \citep{loutrel+2014}. The maximum impact parameter that results in
binary formation is
\begin{align}
&b_{\rm gw}= \left(\frac{340 \pi}{3}\right)^{1/7} \frac{G m_{\rm tot}}{c^2} \eta^{1/7} \left(\frac{v_\infty}{c}\right)^{-9/7}\nonumber\\
&\eta = \frac{m_1 m_2}{(m_1+m_2)^2},
\end{align}
as in equation 17 of \citet{oleary+2009}. The total rate of GW captures in our
Fiducial (Fiducial$\times$ 10) model is $\sim 10^{-10}~{\rm yr}^{-1}$
($10^{-9}~{\rm yr}^{-1}$; see also Table~\ref{tab:exotica}), within the 
range of estimates from \citet{oleary+2009}.

An estimate of the total rate of double compact object binary formation, including
three-body processes, is beyond the scope of this paper, and we leave this to
future work.

\section{Summary and Conclusions} \label{sec:conc} \citet{hailey+2018} have
recently identified 6-12 quiescent BH-LMXB candidates within one parsec of the
Galactic Center, and infer that there may be hundreds of fainter systems in
the same region. This means that the GC is three orders of magnitude more
efficient than the field at producing BH-XRBs, recalling the analogous massive
overproduction of NS-XRBs in a different dense environment (globular
clusters).  While suggestive, this analogy is incomplete: NS-XRBs are dynamically
manufactured in globulars by exchange interactions (e.g. binary-single
scatterings), but this mechanism is disfavored in the GC's high velocity
dispersion environment, which only permits the survival of the hardest main
sequence binaries.

We instead propose that the observed LMXBs are formed via tidal capture of low
mass stars by BHs.  We estimated the distribution of stars and compact
remnants in the GC using time-dependent Fokker-Planck models that predict
close encounter rates.  Taken at face value, tidal capture can explain the observed
(and extrapolated) inventory of BH-XRBs in the GC.  Our primary results are
summarized as follows:

\begin{enumerate} 
\item We calculated the rate at which low mass stars are tidally captured by
BHs and NS as a function of time, and used this to predict that there should
be $\sim$60-200 accreting BH-XRBs in the central parsec today. The number and
radial distribution of these binaries is consistent with the quiescent BH-XRB
population identified by \citet{hailey+2018}, given reasonable extrapolation
below the {\it Chandra} detection threshold.

\item Our models also produced a substantial number of NS-XRBs (far more than are currently
observed). However, there are several candidate mechanisms for suppressing our
predicted NS-XRB population. Alternatively, evolved NS binaries may also
manifest as MSPs, whose population is poorly constrained in the GC.

\item The compact object source terms in our Fokker-Planck models were
calibrated from the observed number of massive stars in the GC. Most of the stellar mass
BHs in the GC may originate in star forming disks with a top heavy IMF, like
the one currently observed at $\sim 10^{18}$ cm
\citep{krabbe+1995,paumard+2006,bartko+2010,lu+2013}.  In our models, {\it in
situ} star formation in these disks has left between $10^4$ and $4\times 10^4$ BHs
within the central parsec, at $z=0$.  Much smaller numbers of BHs would fail
to explain the observed BH-XRB population, yielding the first quantitative
constraints on the long-theorized ``dark cusp'' in the GC.

\item We also estimated the rates of other exotic dynamical interactions
between stars and compact objects. For example, we found that the rate of
disruption of stars by stellar mass BHs (``micro-tidal disruption'') in the
Galactic Center is $\sim 10^{-6}$ per year--comparable to previous estimates
of the total rate in the field and globular clusters \citep{perets+2016}, as
well as the rate of ultra-long GRBs \citep{levan+2014}.  The present-day TDE
rate from Sgr A$^\star$ is $\sim 1- 3\times 10^{-4}~{\rm yr}^{-1}$, similar to
other SMBHs of its mass.

\end{enumerate}  

The largest theoretical uncertainty in our model is the assumption that main
sequence stars tidally captured by stellar mass BHs are able to circularize
and settle into stable Roche-lobe overflow.  Such an outcome is not
energetically guaranteed, and it is likely that BHs above a certain mass will
rapidly destroy tidally captured stars by thermalizing too much mode energy
inside them, leading to super-Eddington accretion in a string of partial tidal
disruptions.  The precise BH mass threshold above which tidal capture becomes
catastrophic is an open question that we hope to address in future work.  A
second concern is that tidal capture binaries have periods $\lsim 10$ hours.
In the field, such systems have low X-ray luminosities, and, if placed in the
GC, would fall below the Chandra detection threshold. However, short period
field XRBs likely have a different formation mechanism, and it is not clear if
tidal capture XRBs would inherit their luminosity function.

While we have focused on tidal capture in an isotropized population of stars
and compact objects, it may be fruitful to examine high mass XRB formation
within star-forming disks. The small number of high mass stars in the Galactic
Center and their short lifetimes suggest that the number of high mass XRBs
would be small. However, the capture rate of disk stars is enhanced by (i)
their larger cross-section (which scales linearly with the star's mass if the
star is more massive the the BH) (ii) their larger escape speeds (iii) the
enhanced stellar densities within the star-forming disk. We would 
expect $\sim 0.3$ HMXBs from BHs interacting with the disk of young stars 
in our Fiducial model. This number would increase by up to two
orders of magnitude if the stellar mass BHs are aligned with the stellar disks
as in \citet{szoelgyen&kocsis2018}. It is also possible that stellar mass
objects migrating within a gaseous disk may smoothly capture into binaries due
to gas dissipation alone, even in the absence of strong tidal coupling. While
there are no HMXBs present in the GC today \citep{hailey+2018}, past or
extragalactic populations of nuclear HMXBs are potentially interesting as
progenitors of LIGO-band GW sources.

Our tidal capture model was motivated by surprising discoveries in the MW
Center, but it carries major implications for extragalactic NSCs as well.  If
the GC's inventory of XRBs is representative, it may complicate X-ray
searches for low-luminosity intermediate mass black hole AGN in dwarf
galaxies.  The unresolved, integrated X-ray luminosity from a large XRB
population represents a durable if dim contaminant; a single BH-XRB in
outburst would represent a more dangerous contaminant for single-epoch
searches.  The existence of dark cusps in galactic nuclei also carries major
implications for the highly uncertain rates of extreme mass ratio inspirals,
one of the primary scientific targets for future space-based GW laser
interferometers (e.g. {\it eLISA}).  Future dynamical modeling of XRB
formation in the GC may yield more sophisticated constraints on the radial
profile of our dark cusp, a local laboratory with which we may calibrate our
expectations for stellar dynamics in distant galactic nuclei.

\section*{Acknowledgments}
We thank the referee for helpful and insightful comments. We are happy to
acknowledge useful discussions with Fabio Antonini, Josh Grindlay, Nathan
Leigh, Nevin Weinberg, Doug Lin, Eugene Vasiliev, Jessica Lu and Jules
Halpern. We are especially grateful to Charles Hailey for discussions of his
recent observational findings. NCS received financial support from NASA
through Einstein Postdoctoral Fellowship Award Number PF5-160145, and thanks
the Aspen Center for Physics for its hospitality during the completion of this
work. BDM and AG acknowledge support from NSF Astronomy and Astrophysics
grants AST-1410950, AST-1615084; NASA Astrophysics Theory Program grants
NNX16AB30G, NNX17AK43G; and Hubble Space Telescope Grant HST-GO-14785.004-A.

\clearpage
  \appendix
\section{Tidal coupling constants}
\label{app:coupling}
The energy deposited into a star of mass $m_{\star}$ after a close encounter
with a compact object of mass $m_{\rm c}$ is given by
\begin{align}
&\Delta E =\frac{G m_\star^2}{r_\star} \left(\frac{m_c}{m_\star}\right)^2
\sum_{l=2}^{\infty} T_{l}(\eta) \left(\frac{r_\star}{r_p}\right)^{(2l+2)}\nonumber\\ 
&\eta \equiv \left(\frac{m_\star}{m_\star+m_c}\right)^{1/2}
\left(\frac{r_\star}{r_p}\right)^{-3/2},
\label{eq:dE}
\end{align}
where $r_p$ is the pericenter distance of the encounter, $r_\star$ is the
radius of the star, and $T_l$ is the tidal coupling constant of multipole order $l$, which depends on
the stellar structure and orbit.

For fixed stellar structure, the tidal coupling constant is a function of the
ratio $\eta$ of the star's dynamical time to the time spent near pericenter.
For the dominant $l=2$ modes the energy deposited in the star is
\begin{align}
\Delta E &= T_2(\eta) \left(\frac{r_p}{r_t}\right)^{-6} \frac{G m_\star^2}{r_\star},
\label{eq:enTide}
\end{align}
where $r_t = r_{\star}(m_{\rm c}/m_{\star})^{1/3}$ is the tidal radius.

\begin{figure}
  \includegraphics[width=8.5cm]{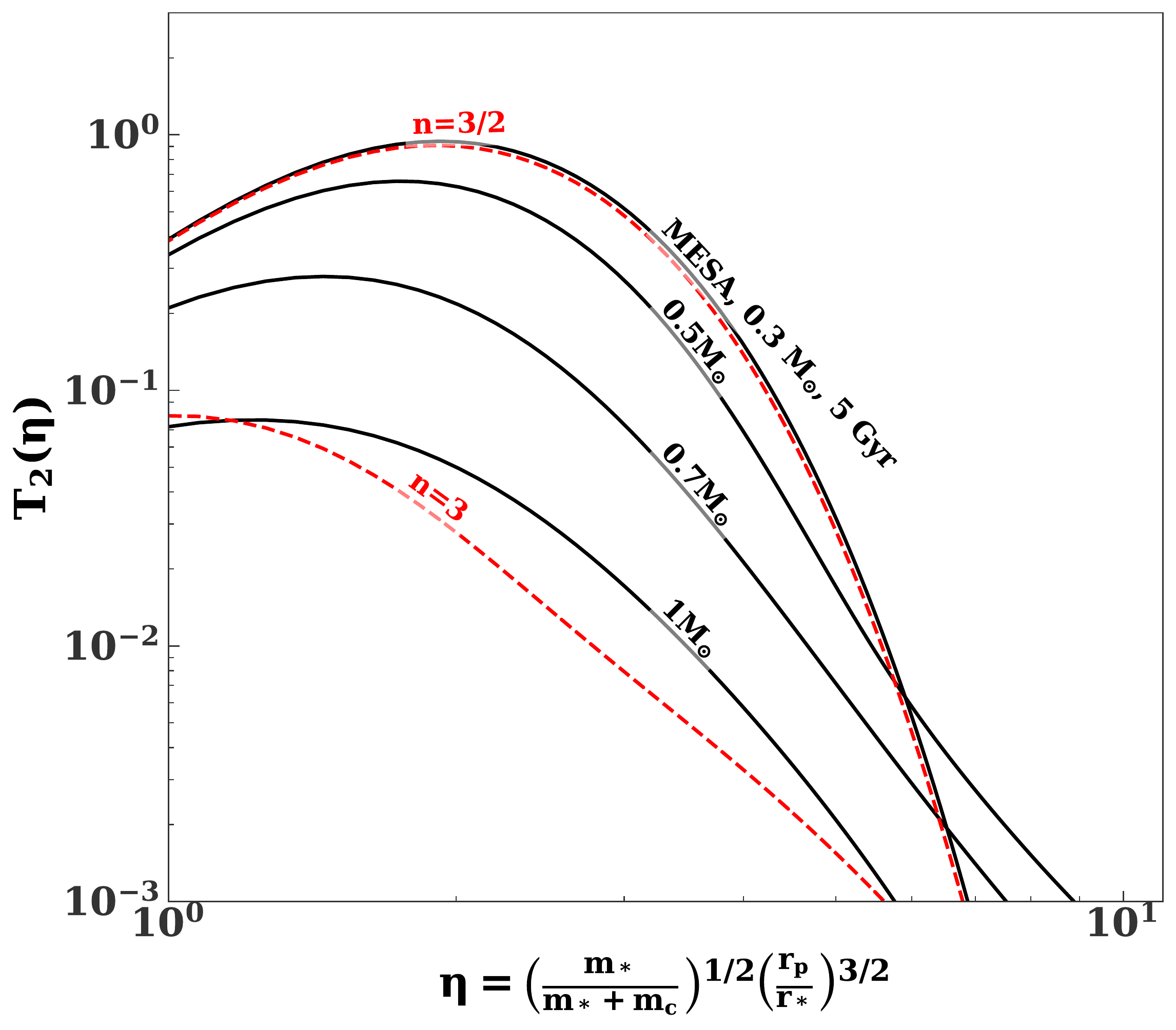}
  \caption{\label{fig:tidalCoupling} Comparison of tidal coupling
    constant as a function of $\eta$ (eq.~\ref{eq:dE}) for different stellar
    models as labeled. The dashed, red lines show the tidal coupling constants for polytropic stellar models. We have assumed a parabolic orbit.}
\end{figure}

 Fig.~\ref{fig:tidalCoupling} compares the $l=2$ tidal coupling constants for
both polytropic and MESA stellar models
\citep{paxton+2011,paxton+2013}.\footnote{http://mesa.sourceforge.net, version
9575} Calculating the tidal coupling constant requires a summation over
discrete stellar eigenmodes, which we calculate with GYRE
\citep{townsend+2013}.\footnote
{https://bitbucket.org/rhdtownsend/gyre/wiki/Home, version 5.  We assume
adiabatic oscillations.} Following \citet{lee&ostriker1986}, we include the
f-mode, the five lowest order p-modes, and the eighteen lowest order g-modes
(if they exist) in the summation. 

\begin{figure}
\includegraphics[width=8.5cm]{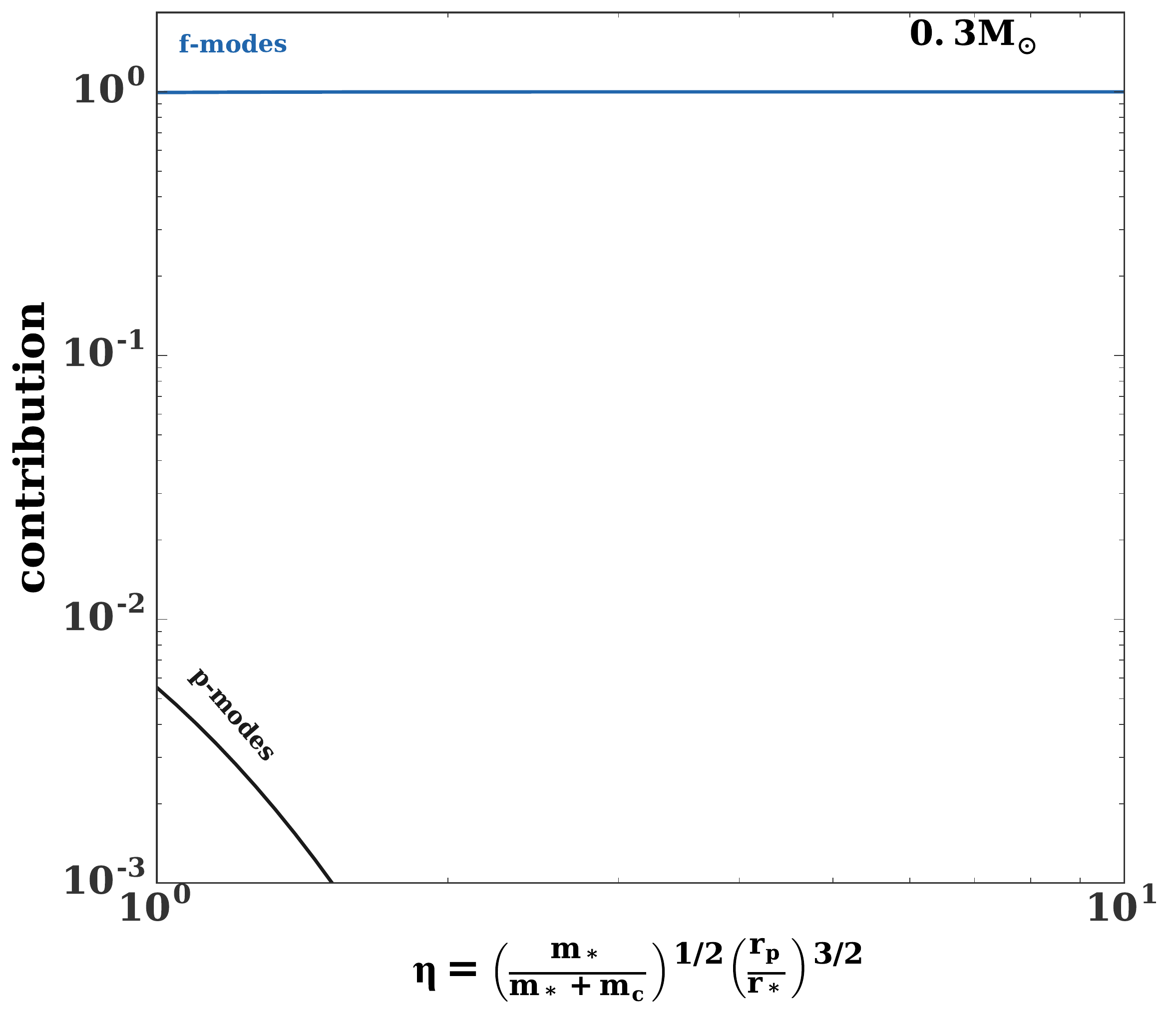}
\includegraphics[width=8.5cm]{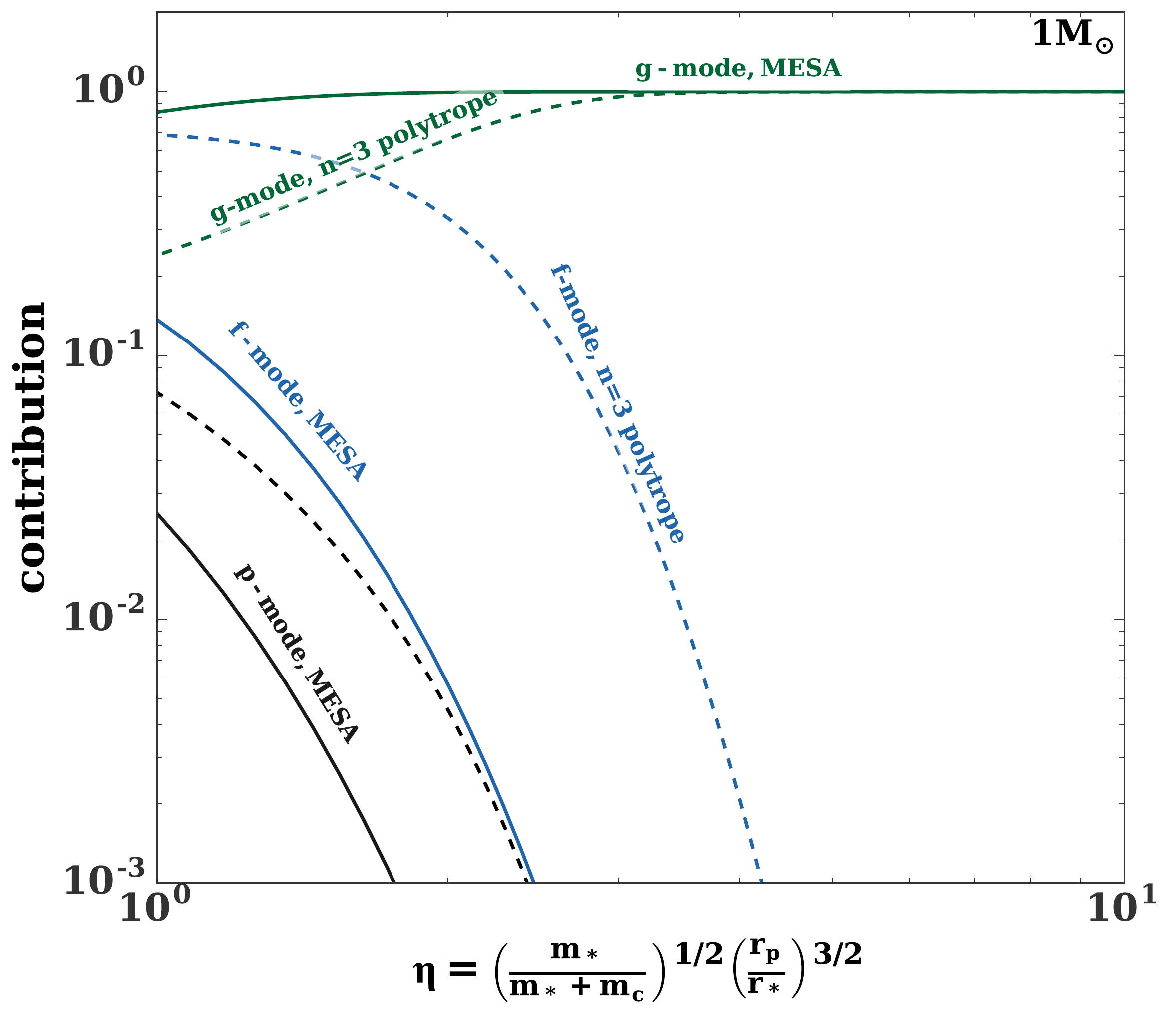}
\caption{\label{fig:modeSpec} \emph{Top panel:} Fraction of oscillation energy
deposited into p-, f-, and g-modes for a star of mass 0.3$M_{\odot}$.  For this
calculation we use a MESA model evolved for $5$ Gyr, but the results are indistinguishable from that of an $n=3/2$ polytrope. The g-modes do not contribute. \emph{Bottom panel:} Same as the top panel, but for a star of mass 1 $M_{\odot}$.  The mode decomposition is not accurately reproduced by a polytropic model, as can be seen by comparing the solid and dashed lines.}
\end{figure}

For stars of mass $\lsim$0.3 $\Msun$, g-modes are not excited at all and most
of the energy is deposited into the f-mode. Larger stellar masses and larger values of
$r_p$ result in greater energy transfer into g-modes, while p-modes are always
subdominant. Fig.~\ref{fig:modeSpec} shows the fraction of energy placed into
different modes as a function of pericenter for $m_\star=0.3M\odot$ and $m_\star=1 M_{\odot}$ stars. An
$n=3/2$ polytropic model accurately reproduces the mode spectrum of the low
mass star. However, the mode spectrum of the solar type star is poorly
aproximated by a polytropic model: the $n=3$ polytropic model underestimates
the energy in g-modes, and overestimates that in the f-mode, for small pericenter distances.

The tidal coupling constant of low mass stars ($m_\star \lsim 0.5 M_{\odot}$),
is close to that of an $n=3/2$ polytrope. The tidal coupling constant
approaches that of an $n=3$ polytrope as the stellar mass aproaches $1
M_{\odot}$.

\section{Corrections for non-linear effects}   
\label{app:nonlinear}
Linear theory underestimates the energy deposited in the star by a factor of a few for the close
pericenters of interest. Non-linear corrections have been calculated by
\citet{ivanov&novikov2001} for polytropic stellar models. We adopt their
prescriptions for the tidal coupling constant for close pericenters.

\begin{figure}
\includegraphics[width=8.5cm]{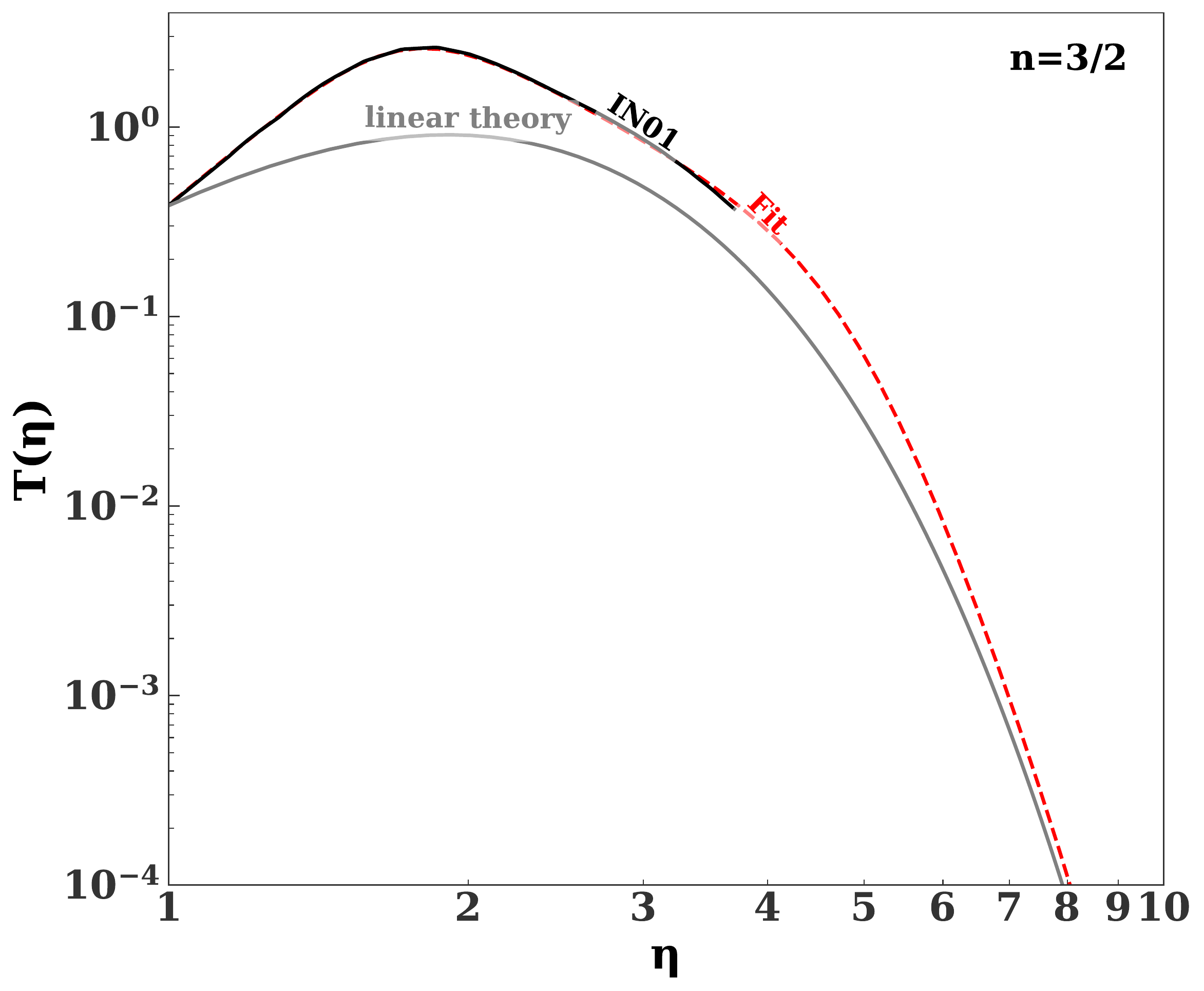}
\includegraphics[width=8.5cm]{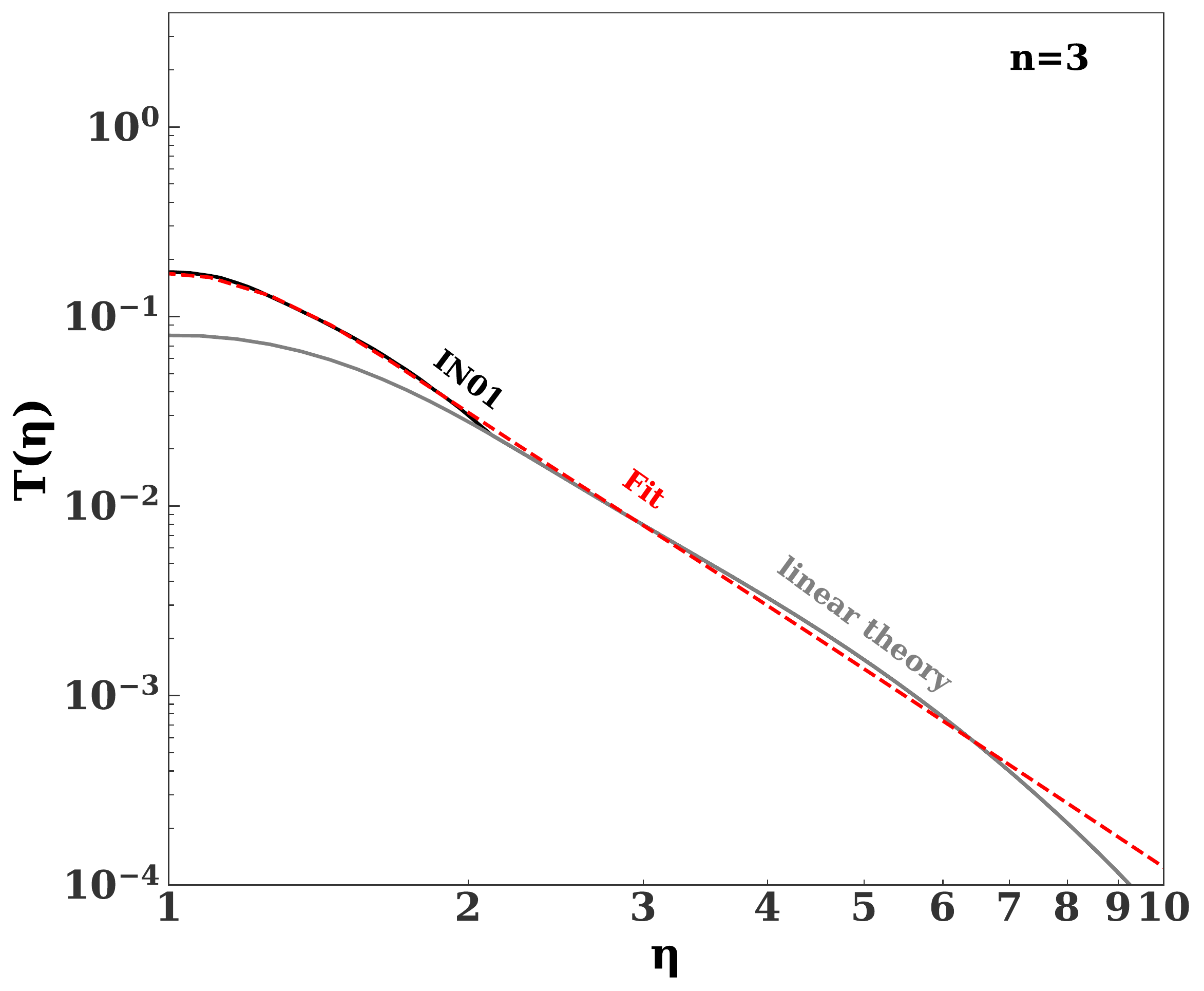}
\caption{Fitted tidal coupling constants $T(\eta)$ from the nonlinear results of \citet{ivanov&novikov2001}.  Results are shown for $n=3/2$ (top panel) and $n=3$ (bottom panel) polytropes.  \label{fig:nonLin}}
\end{figure}

Fig.~\ref{fig:nonLin} compares tidal coupling constants for polytropic models
from linear theory and from \citet{ivanov&novikov2001} (see also their Figures
13 and 15). The following expressions reproduce tidal coupling constants from
\citet{ivanov&novikov2001} for small pericenters, while approaching the
results of linear theory at large pericenters.

\noindent \textit{n=3/2 polytrope:}
\begin{align}
&T(\eta) = C 2^{(b-g)/s}\left(\frac{\eta}{\eta_o}\right)^{-g} \left(1+\left(\frac{\eta}{\eta_o}\right)^s\right)^{(g-b)/s} \nonumber\\
& \times \left(\frac{1}{2}-\frac{1}{2}\tanh\left[k\left(\frac{r}{r_1}-1\right)\right]\right)\nonumber \\
&C=2.58, \eta_0=1.73, g=-4.36, b=2.82, s=9.91,\nonumber \\
& r_1=4.5, k=4 
\label{eq:in1}
\end{align}

\noindent \textit{n=3 polytrope:}
\begin{align}
&T(\eta) = C 2^{(b-g)/s}\left(\frac{\eta}{\eta_o}\right)^{-g} \left(1+\left(\frac{\eta}{\eta_o}\right)^s\right)^{(g-b)/s} \nonumber \\
& \times \left(1+\left(\frac{\eta}{\eta_1}\right)^{s_2}\right)^{(b-b_2)/s_2}\nonumber\\
&C=0.17,\eta_o=1.07, \eta_1=1.92, g=-3.83, \nonumber \\
&b=5.5, b_2=3.49, s=3.59, s_2=6.68,
\label{eq:in2}
\end{align}
where we have adopted eq.~\eqref{eq:in1} for low mass stars with $m_\star \leq0.7 \Msun$, and eq.~\eqref{eq:in2} for higher stellar masses.

\section{Binary exchange interactions}
\label{app:binFrac}

\subsection{Binary fraction} When \textit{soft} binaries interact with field
stars in the GC they gain energy, become more loosely bound, and eventually
dissociate \citep{heggie1975,binney&tremaine1987}. A binary is soft if its
binding energy is less than the kinetic energy of a typical field star, i.e.
if,

\begin{equation}
\frac{G m_1 m_2}{a \langle m\rangle} < \sigma^{2},
\label{eq:hs}
\end{equation}
where $m_1$ and $m_2$ are the masses of the binary components, $a$ is the
semi-major axis, $\langle m\rangle$ is the mean stellar mass, and $\sigma$ is
the 1D velocity dispersion.  Binaries that do not satisfy eq.~\eqref{eq:hs}
are {\emph{hard}}. Interactions with field stars shrink
the separation of a hard binary over time, making it a smaller target. Thus, it
is much easier (and faster) to dissolve a soft binary than to push a hard binary to
coalescence.

\begin{figure} \includegraphics[width=8.5cm]{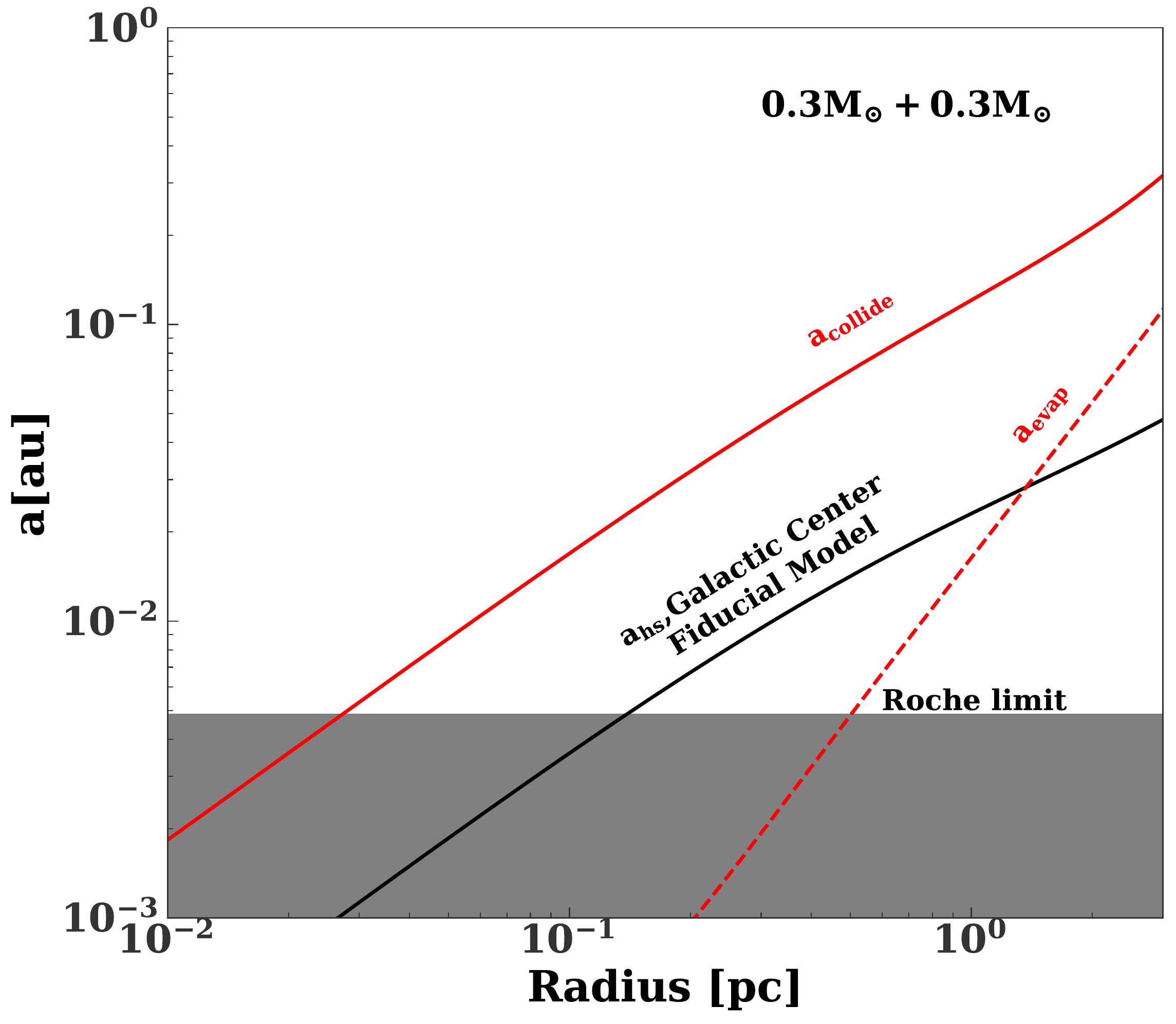}
\includegraphics[width=8.5cm]{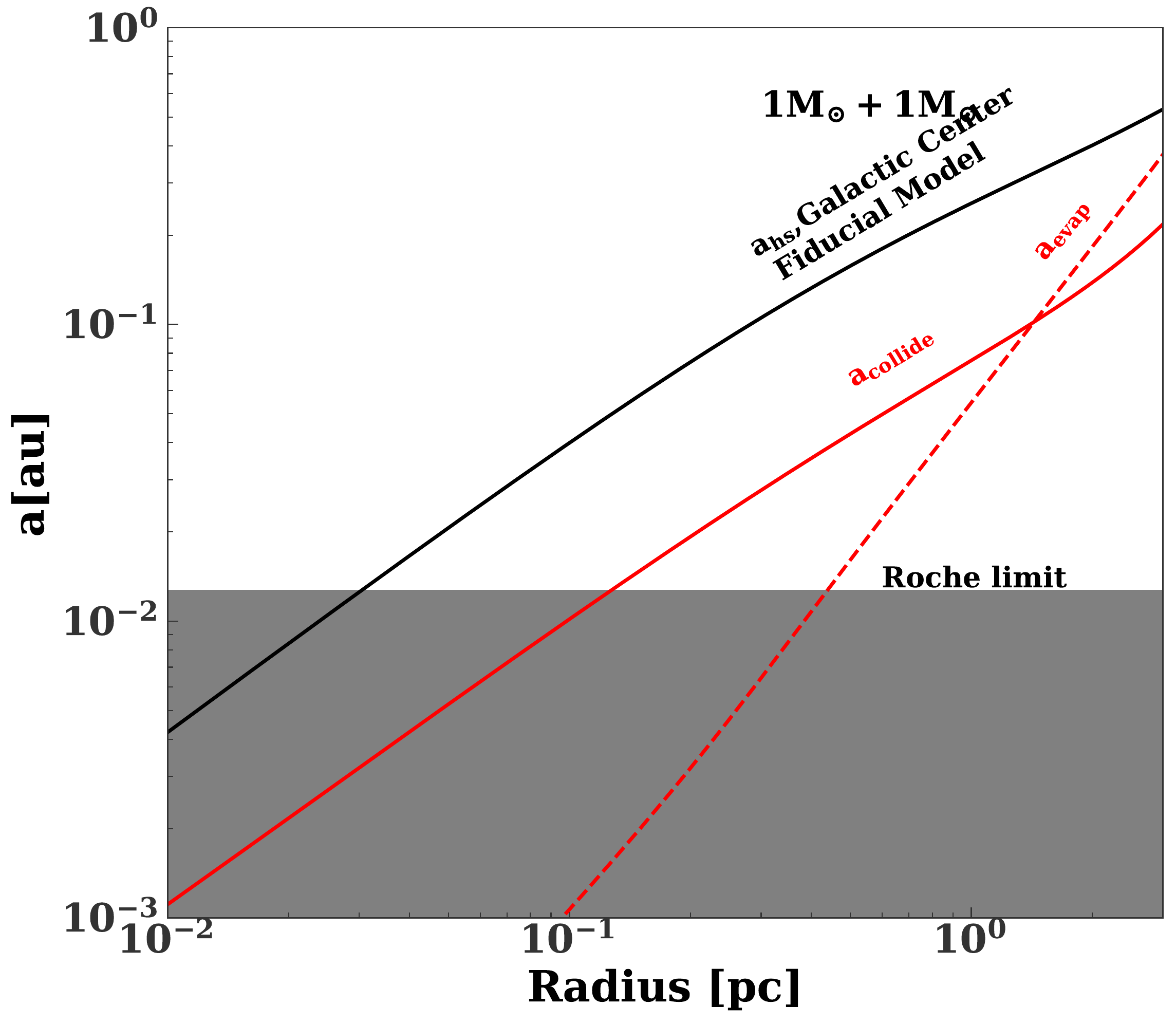} \caption{\label{fig:evap}  Hard-soft
boundary for 1+1 $\Msun$ (\emph{top panel}) and 0.3+0.3 $\Msun$ (\emph{bottom
panel}) in our fiducial model for the GC. The red lines show the semi-major axis
for which the time-scale for direct collisions (eq.~\ref{eq:tcoll}) and
evaporation (eq.~\ref{eq:evap}) is $10^{10}$ years. Binaries in the gray
region are either contact binaries or unphysical as the semi-major axis  of
the binary would be smaller than the Roche limit.} \end{figure}

The black lines in Fig.~\ref{fig:evap} shows the hard-soft boundary in our
Fiducial  model of the Galactic Center for two different binary masses. For a
binary distribution that is flat in $\log(a)$, from the semi-major axis of
Roche-Lobe contact $a_{\rm roche} \approx r_t \sim R_{\star}$ to $a = 900$ AU,
we find that $\sim$73\% (87\%) of binaries with two solar mass (0.3 solar mass
stars) are soft at 1 pc.  By contrast, in a globular cluster with $\sigma\sim
10$ km s$^{-1}$, only 40$-$50\% of the primordial binaries are soft
\citep{ivanova+2005}.

Soft binaries can be ionized in two different ways:
\begin{enumerate}[a]
\item Direct collisions with field stars, as occurs on a timescale
\begin{equation}
\tau_{\rm collide}=\frac{1}{\pi n_\star \sigma a^2 \left(1+\frac{2 G(m_1+m_2)}{\sigma^2 a}\right)}.
\label{eq:tcoll}
\end{equation}
In our fiducial models, the collision rate of binaries with stars exceeds  the
collision rate of binaries with compact objects. \item ``Evaporation" due to
perturbations from distant field stars. For an equal mass binary this occurs
on a timescale (\citealt{alexander&pfuhl2014}; their eq.~3)

\begin{equation}
\tau_{\rm evap}\approx 0.07 \frac{(m_{1}+m_2) \sigma}{ G n \langle m^2 \rangle \ln \Lambda},
\label{eq:evap}
\end{equation}
where $m_{bin}$ is the total mass of the binary, $n$ is the number density of
perturbers, $\langle m^2\rangle$ is the second moment of the mass function,
$\sigma$ is the 1D velocity dispersion, and $\ln \Lambda \approx 15$ is the
Coulomb logarithm. \end{enumerate}

The red lines in Fig.~\ref{fig:evap} show the semi-major axes for which 
the collision and evaporation times are equal to $10^{10}$ years.
Any primordial binaries with semi-major axes $\gtrsim 0.1$ AU within the
central parsec would be evaporated on a timescale of $\lesssim 10^{10}$ yr.

On the other hand, binaries with particularly {\it small} semi-major axes can
be destroyed by magnetic braking.  Following \citet{ivanova&kalogera2006}
(their eq.~4), we find that two stars of mass $m_{\star}$ with semi-major axes
obeying
\begin{equation}
a < 3 \left(\frac{m_\star}{M_{\odot}}\right)^{0.16} \left(\frac{t}{10\, {\rm Gyr}}\right)^{0.41} a_{\rm roche},
\label{eq:amin}
\end{equation}
are brought into Roche-Lobe contact after time $t$.  Solar mass stars in a two
day orbit would thus come into Roche-Lobe contact within $\lesssim$ 5 Gyr (see
also \citealt{andronov+2006}).

\begin{figure}
\includegraphics[width=8.5cm]{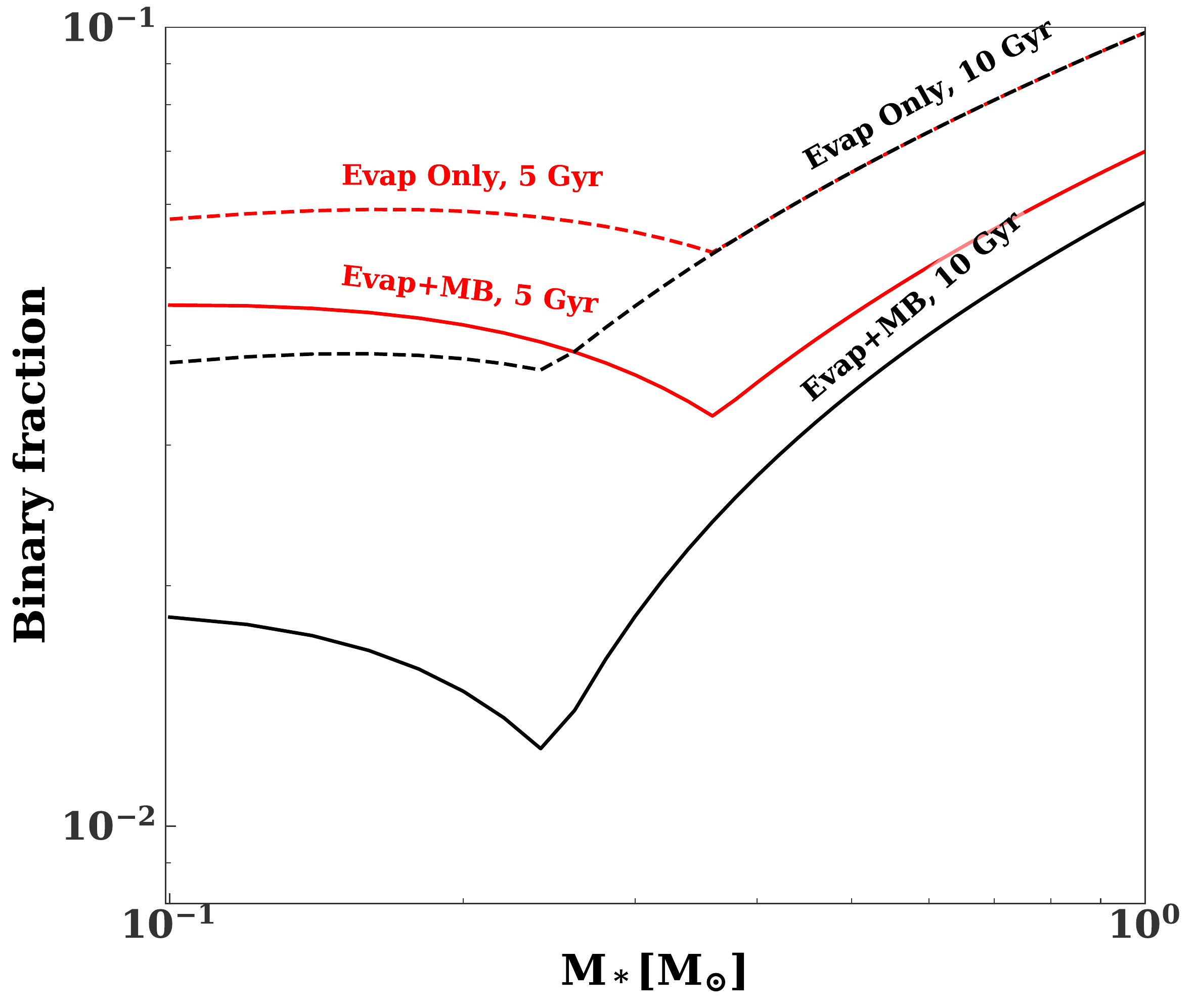}
\caption{\label{fig:binFrac} Binary fraction at 1 pc after 5 Gyr ({\emph{red
lines}}) and 10 (\emph{black lines}) Gyr, calculated for an assumed initial binary fraction
of 50\%. For the dashed lines we only account for evaporation of soft
binaries, while for the solid lines we account for the destruction of hard
binaries via magnetic braking. The x-axis indicates the mass of each of the stars in the binary.}
\end{figure}

The binary fraction is 
\begin{equation}
  f_b\equiv\frac{N_b}{N_s+N_b}=\frac{(1-f_d) f_{\rm b,o}}{1+f_d f_{\rm b,o}}
\end{equation}
where $N_b$ and $N_s$ are the numbers of single stars and binaries
respectively, $f_{\rm b,o}$ is the initial binary fraction and $f_d$ is the
fraction that are destroyed due to the effects  of evaporation and/or magnetic
braking. Figure~\ref{fig:binFrac} shows the expected binary fraction at 1 pc
after 5 and 10 Gyr, as a function of stellar mass (assuming equal mass
binaries). Weighting each mass bin by a Kroupa PDMF, we find that the binary
fraction is $\sim$4\% (3$\%$) after 5 (10) Gyr. Our estimate for the binary
fraction of solar mass stars accounting evaporation alone ($\sim$ 10\%) is
comparable to previous estimates \citep{hopman2009}.

Kozai-Lidov (KL) oscillations induced by the central SMBH can turn some soft binaries
into hard binaries, effectively increasing the binary fraction. In particular, KL
oscillations can excite binaries to very large eccentricities. Tides can then
dissipate energy, creating a tight stellar binary
\citep{antonini&perets2012,stephan+2016}. In practice, for the Galactocentric
radii of interest ($\sim 1$ pc), the time-scale to excite the binary to very
large eccentricities (the \textit{octupole} Kozai time scale) is generally
longer than the evaporation time-scale. Additionally, for a $1 \Msun$ binary,
GR precession will damp KL oscillations for binary separations

\begin{align}
a_{1} < 2 {\, \rm au} \left(\frac{a_2}{1 {\rm pc}}\right)^{3/4} \frac{(1-e_2^2)^{3/8}}{(1-e_1^2)^{1/4}}, 
\end{align}
where $e_1$ is the eccentricity of the inner binary orbit, while $a_2$  and
$e_2$ are the semi-major axis and eccentricity of the binary's orbit around
the SMBH (see equation 59 in \citealt{naoz2016}).

\subsection{Binary exchange rates}
Finally, the rate of compact objects exchanging into existing stellar binaries is
\begin{align}
\dot{n}_{2+1} =\int_{0}^{\infty} n_c f_b n_\star \Sigma v_\infty f(v_\infty) dv_\infty,
\end{align}
where $n_c$ and $n_\star$ are the densities of compact objects (BHs or NSs)
and stars, respectively,  and $\Sigma$ is the total cross-section for the
compact object to be captured into a binary with an ordinary star. This may
either occur via a \textit{prompt exchange} or a \textit{resonant capture}. In
the former case the exchange occurs quickly, while in the latter case a
metastable triple system is formed first. The cross-sections for these
processes have been calibrated from binary-single scattering experiments as \citep{valtonen&karttunen2006}
\begin{align}
&\Sigma_{\rm ex}\approx 0.51  \frac{2 m_c}{m_\star} \frac{\pi a^2}{v^2} (1-P_{\rm c})\\
&\Sigma_{\rm cap}\approx 1.18 (n-1) (1-v^2)^{n-2} \frac{2 m_c}{m_\star} \frac{\pi a^2}{v^2}(1-P_{\rm s})\\
&P_{\rm c} \approx 0.25(n-1)(1-v^2)^{n-2} \\
&P_{\rm s} = \frac{m_c^{-q}}{2m_\star^{-q}+m_{\rm c}^{-q}} \\
&v^2= \frac{2 m_c}{M} \frac{v_\infty^2 a_o}{G m_\star},
\end{align}
where $m_c$, $m_\star$, and $M$ are the masses of the compact object, the
stars in the binary (assumed to be equal in mass), and the three-body system, respectively. The cross-sections go to 0
for $v\gsim 1$. The power law index $n$ ($q$) depends on the angular momentum of the system,
and is expected to vary between $4.5$ and $3$ ($1$ and $3$) as v goes from 0 to 1. We choose
$n=q=3$, but the results are not very sensitive to this choice.

Using densities profiles of our Fiducial model, the rate of 2+1 encounters per unit volume at radii $\lesssim 1$ pc is approximately given by
\begin{align}
\dot{n}_{\rm bh, 2+1}= &5 \times 10^{-11} \left(\frac{r}{1 {\rm pc}}\right)^{-2.5} 
\left(\frac{f_b}{0.01}\right) \left(\frac{m_\star}{\bar{m}_*}\right) {\rm pc^{-3} yr^{-1}} \nonumber\\
\dot{n}_{\rm ns, 2+1}= &3 \times 10^{-11} \left(\frac{r}{1 {\rm pc}}\right)^{-1.9} 
\left(\frac{f_b}{0.01}\right) \left(\frac{m_\star}{\bar{m}_*}\right) {\rm pc^{-3} yr^{-1}}
\end{align}
where $\bar{m}_\star=0.3 \Msun$. Integrating over volume and a Kroupa PDMF
($m_{\star} = 0.2-1 \Msun$), we find that the total rate of 2+1 encounters
inside of 1 pc is
\begin{align}
&\dot{N}_{\rm bh, 2+1}= 8\times 10^{-10} \left(\frac{f_b}{0.01}\right) {\rm yr^{-1}} \nonumber\\
&\dot{N}_{\rm ns, 2+1}= 4\times 10^{-10}  \left(\frac{f_b}{0.01}\right) {\rm yr^{-1}},  
\end{align}
where we have truncated the volume integral where $a_{\rm hs}$ equals the
stellar radius. Comparing to the tidal capture rates
(Fig.~\ref{fig:captRateSingle}), we see the rate of 2+1 encounters is
sub-dominant for binary fractions of $\lesssim$50\% for BHs and 15\% for NSs,
as expected in the GC from the above considerations.  We stress that these
calculations are generous to the 2+1 formation channel, as we have assumed
that every exchange interaction involving a main sequence binary and a compact
object will lead to XRB formation, while in reality this is only true for a
subset of these interactions. For example, three-body interactions can result
in a physical stellar collision \citep{fregeau+2004}. Thus, for the low binary
fractions expected in the GC, binary-single exchange interactions should be
highly sub-dominant to tidal capture in the formation of XRBs.

\footnotesize{ 
\bibliographystyle{mnras} 
\bibliography{master} 
}

\end{document}